\documentclass[12pt]{article}
\makeatletter
\newcommand{\singlespacing}{\let\CS=\@currsize\renewcommand{\baselinestretch}{1}\tiny\CS}
\usepackage{epsfig}
\usepackage{graphicx}
\usepackage{verbatim}   % useful for program listings
\usepackage{color}      % use if color is used in text
\usepackage{subfigure}  % use for side-by-side figures
\usepackage{amsfonts}

\begin{document}
\baselineskip=24pt
%\singlespacing
%\doublespacing
\parskip = 10pt
\def \qed {\hfill \vrule height7pt width 5pt depth 0pt}
\newcommand{\ve}[1]{\mbox{\boldmath$#1$}}
\newcommand{\IR}{\mbox{$I\!\!R$}}
\newcommand{\1}{\Rightarrow}
\newcommand{\bs}{\baselineskip}
\newcommand{\esp}{\end{sloppypar}}
\newcommand{\be}{\begin{equation}}
\newcommand{\ee}{\end{equation}}
\newcommand{\beanno}{\begin{eqnarray*}}
\newcommand{\inp}[2]{\left( {#1} ,\,{#2} \right)}
\newcommand{\eeanno}{\end{eqnarray*}}
\newcommand{\bea}{\begin{eqnarray}}
\newcommand{\eea}{\end{eqnarray}}
\newcommand{\ba}{\begin{array}}
\newcommand{\ea}{\end{array}}
\newcommand{\nno}{\nonumber}
\newcommand{\dou}{\partial}
\newcommand{\bc}{\begin{center}}
\newcommand{\ec}{\end{center}}
\newcommand{\2}{\subseteq}
\newcommand{\cl}{\centerline}
\newcommand{\ds}{\displaystyle}
\newcommand{\mr}{\mathbb{R}}
\newcommand{\ul}{\underline}
\def\refhg{\hangindent=20pt\hangafter=1}
\def\refmark{\par\vskip 2.50mm\noindent\refhg}

\title{\sc Birnbaum-Saunders Distribution: A Review of Models, Analysis and Applications}

\author{\sc N. Balakrishnan{\footnote{Department of Mathematics and Statistic, McMaster University, Canada. Corresponding author.}} and 
Debasis Kundu{\footnote{Department of Mathematics and Statistics, Indian Institute of
Technology Kanpur, Pin 208016, India.}}}

\date{}
\maketitle

\begin{abstract}

Birnbaum and Saunders \cite{BS:1969a, BS:1969b} introduced a two-parameter lifetime distribution to model 
fatigue life of a metal, subject to cyclic stress.  Since then, extensive work has been
done on this model providing different interpretations, constructions, generalizations, inferential methods, and 
extensions to bivariate, multivariate and
matrix-variate cases.  More than two hundred papers and one research monograph have already appeared describing 
all these aspects and developments.  
In this paper, we provide a detailed review of all these developments and at the same time indicate several open problems 
that could be considered for further research.

\end{abstract}


%\newpage

\section{\sc Introduction}

Among all statistical distributions,  the normal distribution is undoubtedly the most used one  
in practice.  Several new distributions have been developed by employing some  
transformations on the normal distribution.  Two-parameter Birnbaum-Saunders (BS) distribution is one such 
distribution which has been developed by making a monotone transformation on the standard normal random 
variable.  
The BS distribution has appeared in several different contexts, with varying derivations.  It was
given by Fletcher \cite{Fletcher:1911}, and according to Schr{\"o}edinger \cite{Schr:1915}, it was originally obtained by 
Konstantinowsky \cite{Konst:1914}.  Subsequently, it was obtained by Freudenthal and Shinozuka \cite{FS:1961} as a model that will 
be useful in life-testing.  But,
it was the derivation of Birnbaum and Saunders \cite{BS:1969a} that brought the usefulness of this distribution 
into a clear focus.  Birnbaum and Saunders \cite{BS:1969a} introduced the distribution, that has come to bear their
names, specifically for the purpose of modeling fatigue life of metals subject to periodic stress; 
consequently, the distribution is also sometimes referred to as the fatigue-life distribution.

Since the re-introduction of the model in 1969, extensive work has been done on this specific 
distribution.  Even though Birnbaum and Saunders \cite{BS:1969a} provided a natural physical justification
of this model through fatigue failure caused under cyclic loading, Desmond \cite{Desmond:1985} presented a more
general derivation based on a biological model.  He also strengthened the physical 
justification for the use of this distribution by relaxing some of the original assumptions made by Birnbaum and 
Saunders.  Bhattacharyya and Fries \cite{BF:1982} proved that a BS distribution
can be obtained as an approximation of an  inverse Gaussian (IG) distribution.  Desmond \cite{Desmond:1986} observed an interesting 
feature that a  BS distribution can be viewed as an equal mixture of an IG distribution and its reciprocal.  This mixing property becomes useful in deriving
several properties of the BS distribution using well-known properties of the IG
distribution, and it also becomes useful for estimation purposes as well.  Balakrishnan et al. \cite{BGKLS:2010}
used this mixture property effectively to define mixtures of BS distributions, and then establish various 
properties and inferential methods for such a mixture BS distribution.

Many different properties of BS distribution have been discussed by a number of  
authors.  It has been observed that the probability density function (PDF) of the BS  
distribution is unimodal.  The shape of the hazard function (HF) plays an important role
in lifetime data analysis.  In this regard, Mann et al. (\cite{MSS:1974}, page 155) first conjectured that the 
HF is not an increasing function, but the average HF is nearly a non-decreasing 
function.  Three decades later, Kundu et al. \cite{KKB:2008} and Bebbington et al. \cite{BLZ:2008} 
proved formally that the HF of BS distribution is an unimodal function; see also Gupta and Akman \cite{GA:1997} 
in this respect.  In many real life  
situations, the HF may not be monotone and that it increases up to a point and then 
decreases.  For example, in the study of recovery from the breast cancer, it has been observed by 
Langlands et al. \cite{LPKG:1979} that the maximum mortality occurs about three years after the 
disease and then
it decreases slowly over a fixed period of time.  In a situation like this, BS  
distribution can be used quite effectively for modeling.  Moreover, in this aspect, log-normal
and BS distributions behave very similarly.  Although the HF of the log-normal 
distribution tends to zero, as $t \rightarrow \infty$, the hazard function of the BS distribution  
tends to a constant.  Hence, in the long run, the BS behaves like an exponential distribution.
Cheng and Tang \cite{CT:1993, CT:1994a} developed several
reliability bounds and tolerance limits for BS distribution.  Cheng and Tang \cite{CT:1994b}
also developed a random number generator for the BS distribution, and a comparison of the performance of
different BS generators can be found in Rieck \cite{Rieck:2003}.

The maximum likelihood (ML) estimators of the shape and scale parameters based on a complete sample were discussed originally by 
Birnbaum and Saunders \cite{BS:1969b}, and their asymptotic distributions were obtained by Engelhardt 
et al. \cite{EBW:1981}.  The existence and uniqueness of the ML estimators have been formally established by 
Balakrishnan and Zhu \cite{BZ:2014a}
who have shown that the ML estimators of the unknown parameters cannot be obtained in an explicit form and need to be 
obtained by solving a non-linear equation.  Extensive work has been done on developing 
the point and interval estimation of the parameters in the case of complete as well as 
censored samples.  Rieck \cite{Rieck:1995} first developed the estimators of the unknown parameters in the case
of a censored sample.  Ng et al. \cite{NKB:2003} provided modified moment (MM) estimators of the 
parameters in the case of a complete sample which are in explicit simple form, and they later extended 
to the case of a Type-II censored sample; see Ng et al. \cite{NKB:2006}.  From and Li \cite{FL:2006} suggested
four different estimation techniques for both complete and censored samples.  Several other estimation
procedures for complete or censored samples, mainly from the frequentist point of view, have been proposed  
by Dupuis and Mills \cite{DM:1998},  Wang et al. \cite{WDL:2006}, Lemonte et al. \cite{LNV:2007, LSC:2007},
Desmond et al. \cite{DRL:2008}, Ahmed et al. \cite{ABLV:2008}, 
 Balakrishnan et al. \cite{BLSC:2009}, and Balakrishnan and Zhu \cite{BZ:2014b}.  Padgett \cite{Padgett:1982} first considered  
Bayesian inference for the scale parameter $\beta$ of the model assuming the shape parameter $\alpha$ to be known, with 
the use of a non-informative prior.   Achcar \cite{Achcar:1993} considered the same problem when both parameters are
unknown and Achcar and Moala \cite{AM:2010} addressed the problem in the presence of censored data and covariates.  
Recently, Wang et al. \cite{WSP:2016} provided Bayes estimates and associated credible intervals for  
the parameters of BS distribution under a general set of priors.

Several other models associated with BS distribution, and their properties and statistical inferential 
methods, have been discussed in the literature.  For example, Rieck and Nedelman \cite{RN:1991} considered the 
log-linear model for the BS distribution. Desmond et al. \cite{DRL:2008} considered the BS
regression model.  The logarithmic version of BS (Log-BS) regression model has been discussed by 
Galea et al. \cite{GLP:2004} and Leiva et al. \cite{LBPG:2007}.  Lemonte and Cordeiro \cite{LC:2009} considered 
the BS non-linear regression model.  A length-biased version of the BS 
distribution and its different applications have been provided by Leiva et al. \cite{LSA:2009}.  
Balakrishnan et al. \cite{BGKLS:2010} considered three different mixture models associated with BS 
distribution and discussed their fitting methods and applications.  Santos-Neto et al. \cite{SCLB:2016} 
proposed a reparameterized BS regression model with varying precision and discussed associated  
inferential issues.  Recently, Bourguingon et al. \cite{BLLS:2017} and Desousa et al. \cite{DSLS:2017} proposed
transmuted BS and tobit-BS models, respectively, and discussed several inferential issues for these models.  For a review of
the BS model, we refer the readers to the monograph by Leiva \cite{Leiva:2016}; see also Johnson et al. 
\cite{JKB:1995}, Leiva and Saunders \cite{LS:2015} and Leiva and Vivanco \cite{LV:2016}
in this regard.

D{\'i}az-Garc{\'i}a and  Leiva \cite{DL:2005} introduced the generalized BS  distribution  by replacing the 
normal kernel in (\ref{cdf-bs}) with elliptically symmetric kernels.  Since then, quite a bit of
work has been done on this generalized BS distribution.  Leiva et al. \cite{LSSP:2008} 
developed a procedure for generating random samples from the generalized BS distribution, while 
Leiva et al. \cite{LRBS:2008} discussed lifetime analysis based on generalized BS  
distribution.  A comprehensive discussion on generalized BS distributions has been provided by 
Sanhueza et al. \cite{SLB:2008}.  Kundu et al. \cite{KBJ:2010} introduced the bivariate BS  
distribution and studied some of its properties and characteristics.  The multivariate 
generalized BS distribution has been introduced, by replacing the normal kernel by an  
elliptically symmetric kernel, by Kundu et al. \cite{KBJ:2010}.  Caro-Lopera et al. \cite{CLB:2012} developed subsequently  
the generalized matrix-variate BS distribution and discussed its properties.

The main aim of this paper is to provide a  review of all these developments on BS  
distribution and also to suggest some open problems along
the way.  The rest of this paper is organized as follows.  In Section 2, we define the BS distribution and   
provide physical interpretations and 
 some basic properties of the model.  Point and interval 
estimation methods based on complete samples are detailed in Sections 3 and 4, respectively.  Bayesian 
inference is discussed next in Section 5.  In Section 6, we discuss the point and interval estimation 
of the model parameters based on censored samples.  Some of the other univariate issues are discussed in Section 7.  
Bivariate and multivariate BS 
distributions are described in Sections 8 and 9., respectively.  In Sections 10 and 11, we describe several 
related models and the BS regression model, 
respectively.  Different generalizations are described in Section 12, and several illustrative examples are
presented in Section 13.  Finally, some concluding remarks are made in Section 14.

\section{\sc Definition \& Some Basic Properties}

In this section, we provide the definition, give some physical interpretations and discuss 
several properties of the two-parameter BS distribution.

\subsection{\sc Cumulative Distribution Function, Probability Density Function \& Hazard Function}

The cumulative distribution function (CDF) of a two-parameter BS random variable 
$T$ can be written as
\be
F_T(t; \alpha, \beta) = \Phi \left [ \frac{1}{\alpha} \left \{ \left ( \frac{t}{\beta} \right )^{1/2} - 
\left ( \frac{\beta}{t} \right )^{1/2} \right \} \right ], \ \ \ 0 < t < \infty, \ \ \alpha, \beta > 0, 
    \label{cdf-bs}
\ee
where $\Phi(\cdot)$ is the standard normal CDF.  The parameters $\alpha$ and $\beta$ in (\ref{cdf-bs}) are the shape and 
scale parameters, respectively.

\begin{figure}[t]
\begin{center}
\subfigure[]{
\includegraphics[height=4.5cm,width=4.5cm]{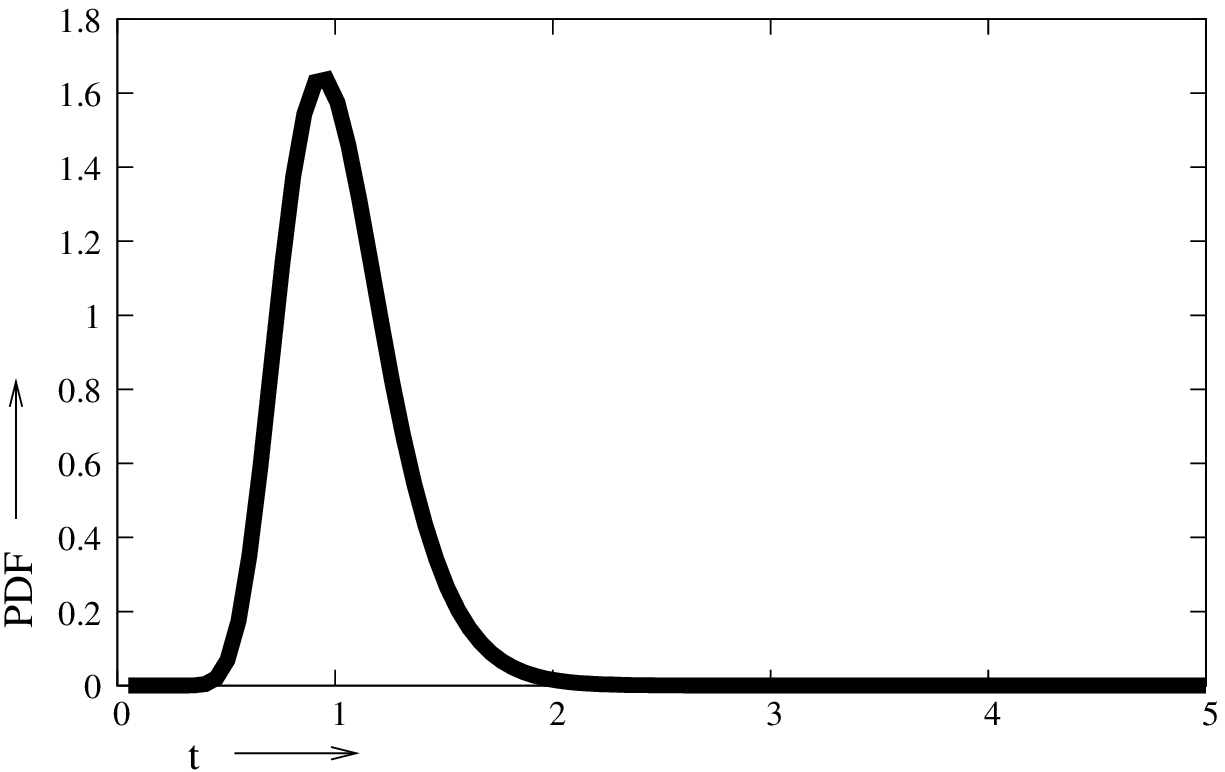}}
\subfigure[]{
\includegraphics[height=4.5cm,width=4.5cm]{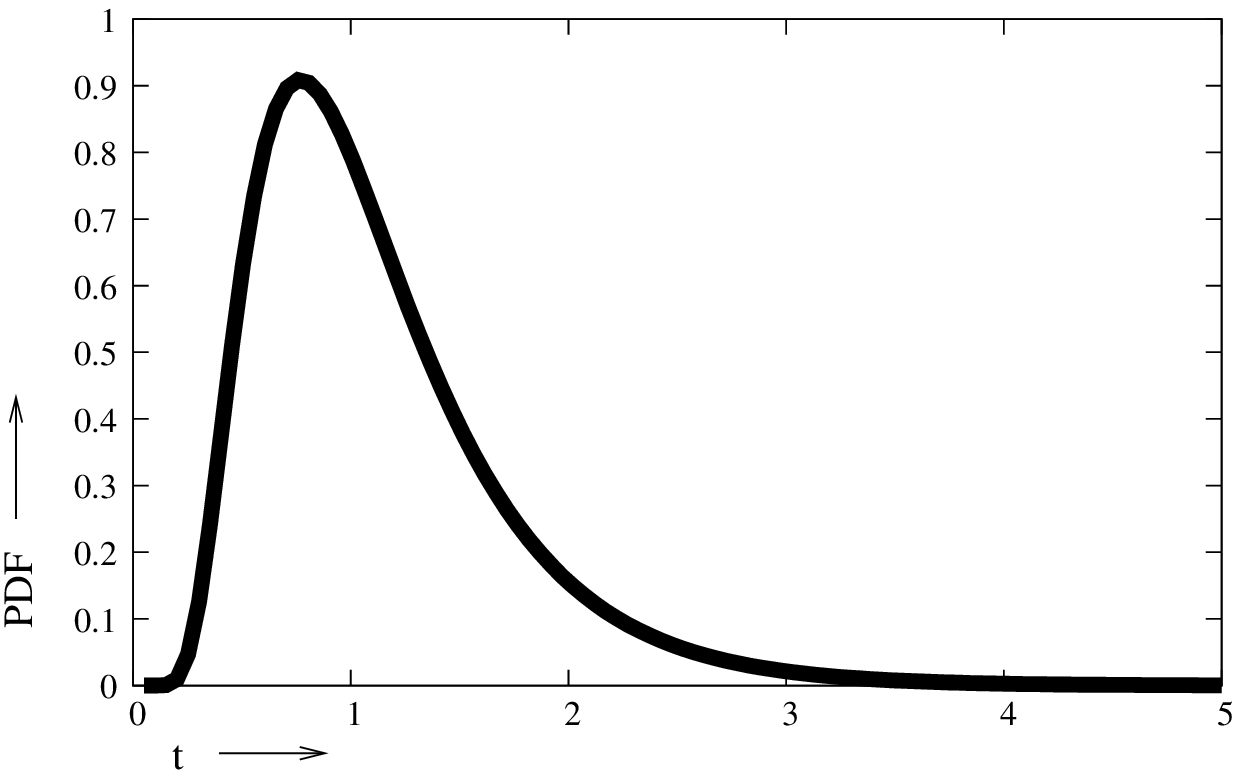}}

\subfigure[]{
\includegraphics[height=4.5cm,width=4.5cm]{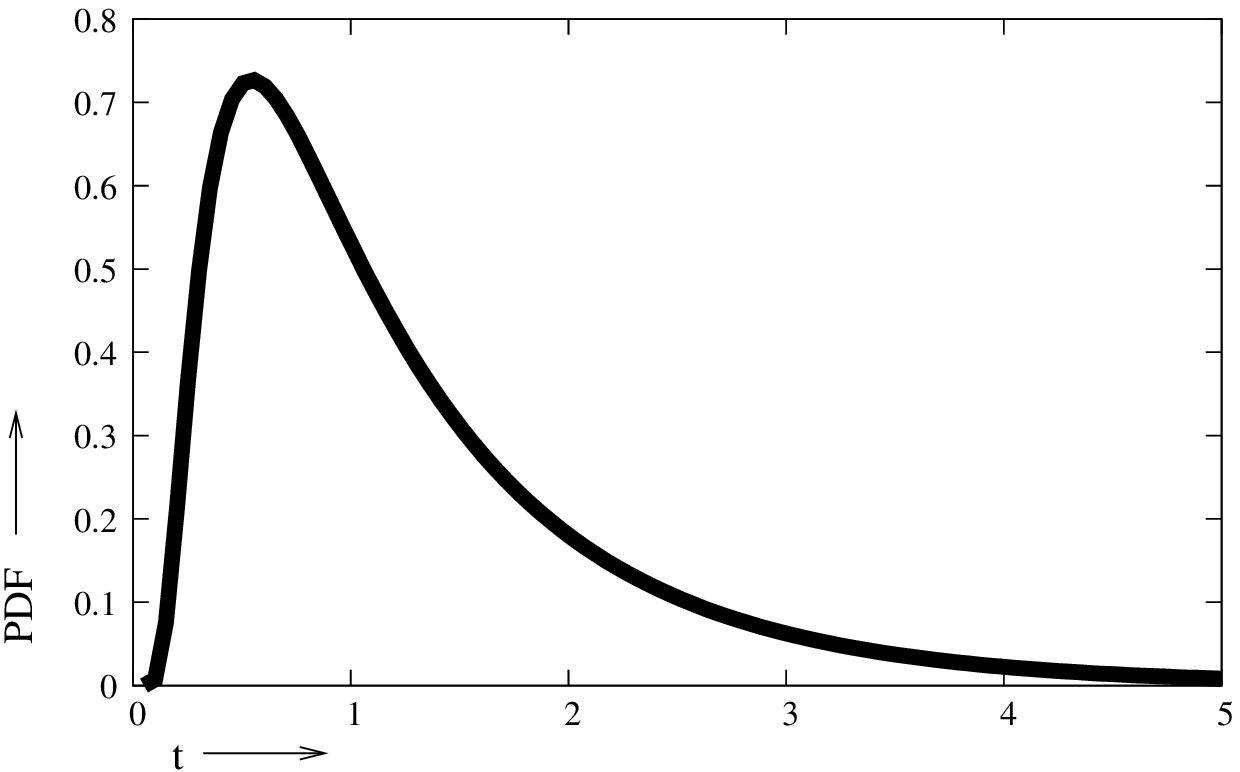}}
\subfigure[]{
\includegraphics[height=4.5cm,width=4.5cm]{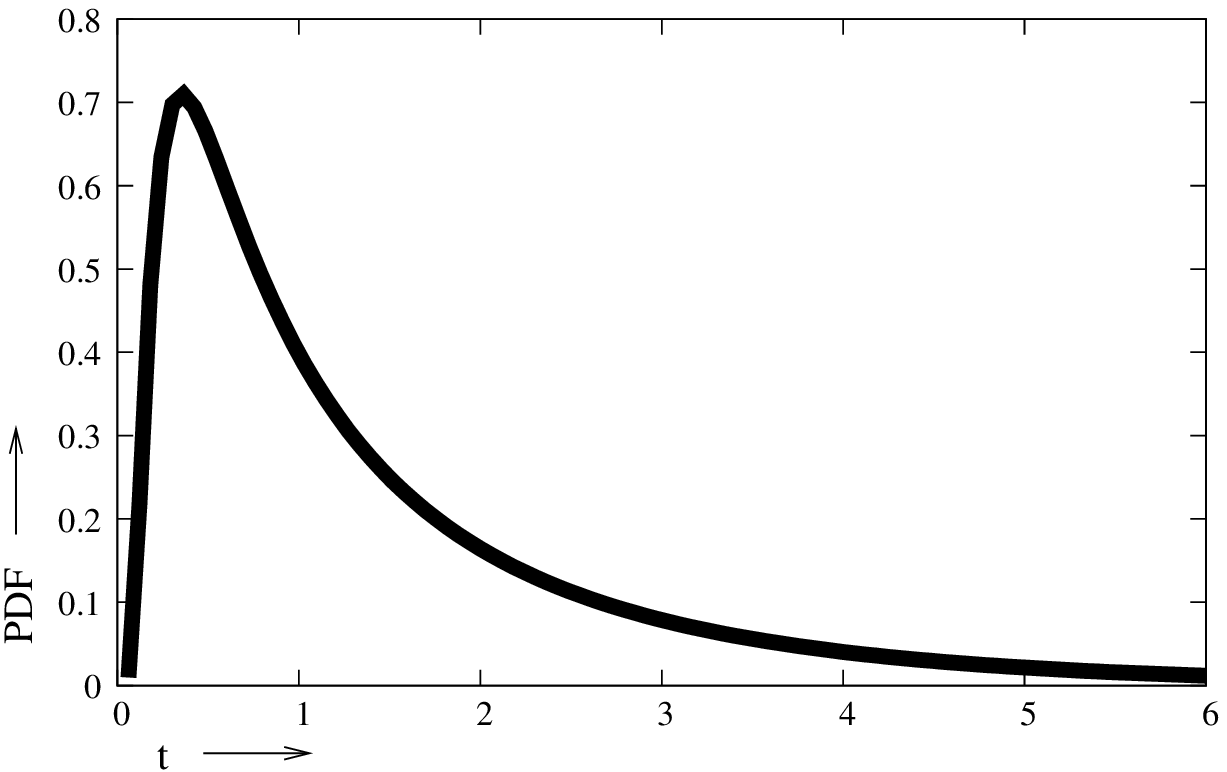}}

\caption{Plots of the PDF of BS$(\alpha, \beta)$ for different values of $\alpha$ and when $\beta$ = 1: \ 
(a) $\alpha$ = 0.25, (b) $\alpha$ = 0.5 \ (c) $\alpha$ = 0.75, \ (d) $\alpha$ = 1.0.}
\label{pdf-plot}
\end{center}
\end{figure}

If the random variable $T$ has the BS distribution function in (\ref{cdf-bs}), then the 
corresponding probability density function (PDF) is 
\be
f_T(t; \alpha, \beta) = \frac{1}{2 \sqrt{2 \pi} \alpha \beta} \left [ \left ( \frac{\beta}{t} \right )^{1/2}
+ \left ( \frac{\beta}{t} \right )^{3/2} \right ] \exp \left [ - \frac{1}{2 \alpha^2} 
\left ( \frac{t}{\beta} + \frac{\beta}{t} - 2 \right ) \right ], \ \ \  t > 0, \alpha > 0, \beta > 0.  \label{bs-pdf}
\ee
From here on, a random variable with PDF in (\ref{bs-pdf}) will be simply denoted by BS$(\alpha, \beta)$.  It is clear that 
$\beta$ is the scale parameter while $\alpha$ is the shape parameter.  It can be seen that 
 the PDF of BS$(\alpha, \beta)$ goes to 0 when $t \rightarrow 0$ as well as when $t \rightarrow \infty$.  For 
all values of $\alpha$ and $\beta$, the PDF is unimodal.    The plots of the PDF of BS$(\alpha, \beta)$  in (\ref{bs-pdf}), 
for different values of $\alpha$, are presented in Figure \ref{pdf-plot}.

The mode of BS$(\alpha, \beta)$ cannot be obtained in an explicit form. But, the 
mode of the PDF of BS$(\alpha, 1)$, say $m_{\alpha}$, can be obtained as the solution of the 
following cubic equation [see Kundu et al. \cite{KKB:2008}]:
\be
t^3 + t^2 (\alpha^2 + 1) + t(3 \alpha^2 - 1) - 1 = 0.
\ee
It can be easily shown that the cubic equation in (\ref{cubic-eq}) has a unique solution.
The mode of BS$(\alpha, \beta)$ can then be obtained simply as $\beta \ m_{\alpha}$.  It has been observed that 
$m_\alpha$ is an 
increasing function of $\alpha$.  

If $T \sim$ BS$(\alpha, \beta)$, consider the following transformation 
\be
Z = \frac{1}{\alpha} \left [ \left ( \frac{T}{\beta} \right )^{1/2} - \left ( \frac{T}{\beta} \right )^{-1/2}
\right ],    \label{relx}
\ee
or equivalently
\be
T = \frac{\beta}{4}\left [\alpha Z + \sqrt{(\alpha Z)^2 + 4} \right ]^2.  \label{relt}
\ee
It then readily follows from (\ref{cdf-bs}) that $Z$ is a normal random variable with mean zero 
and variance one. 
From (\ref{relx}), it is immediate that 
\be
W = Z^2 = \frac{1}{\alpha^2} \left [ \frac{T}{\beta} + \frac{\beta}{T} - 2 \right ] \sim \chi^2_1.
\label{square-rel}
\ee
Eq. (\ref{relt}) can be used effectively to generate BS random variables
from  standard normal random variables; see, for example, Rieck \cite{Rieck:2003}.  Alternatively, the relation in  
(\ref{square-rel}) can also be used to generate BS random variables; see Chang and Tang \cite{CT:1994b}.
Rieck \cite{Rieck:2003} showed the generator based on (\ref{relt}) to be more efficient than the one based on (\ref{square-rel}).

Evidently, the $q$-th quantile of BS$(\alpha, \beta)$ random 
variable is 
\be
\frac{\beta}{4} \left [ \alpha z_q + \sqrt{(\alpha z_q)^2 + 4} \right ]^2,
\ee
with $0 < q < 1$, where $z_q$ is the $q$-th quantile of the standard normal random variable.  From
(\ref{relx}), it is also immediate that $\beta$ is the median of $T$.

\begin{figure}[t]
\begin{center}
\subfigure[]{
\includegraphics[height=4.5cm,width=4.5cm]{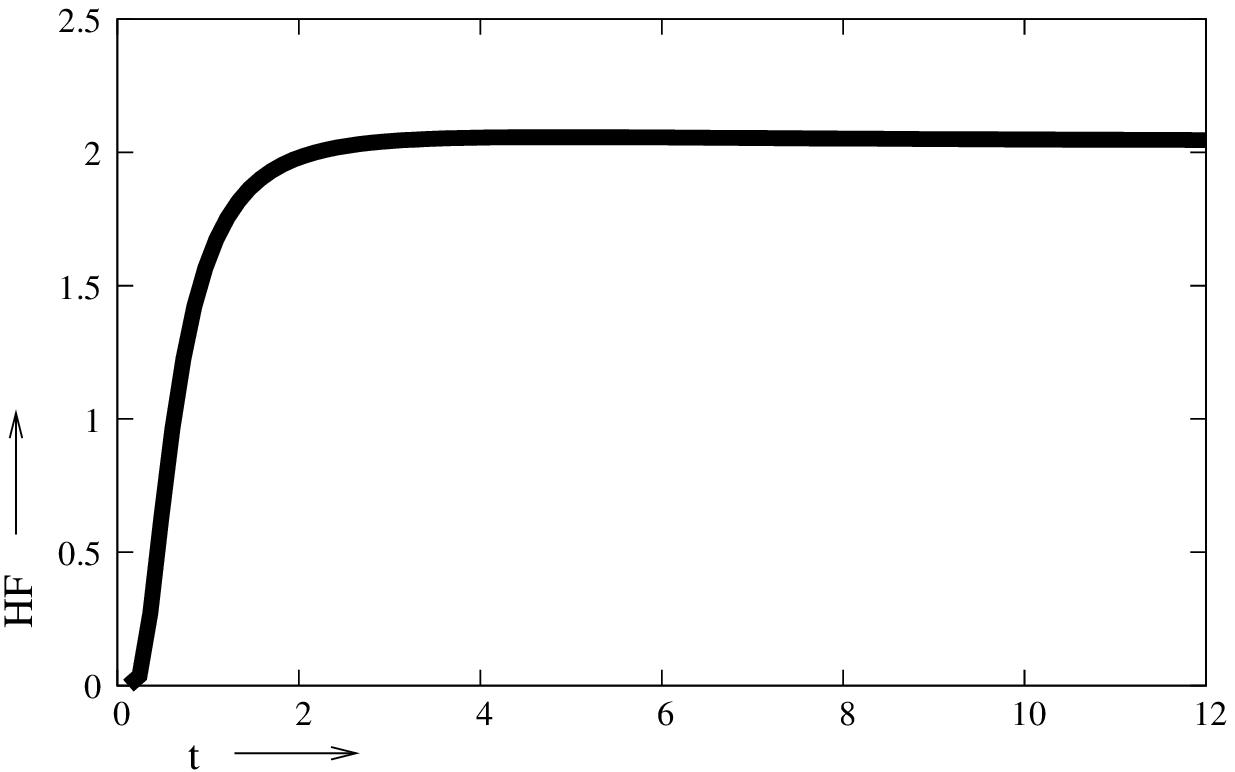}}
\subfigure[]{
\includegraphics[height=4.5cm,width=4.5cm]{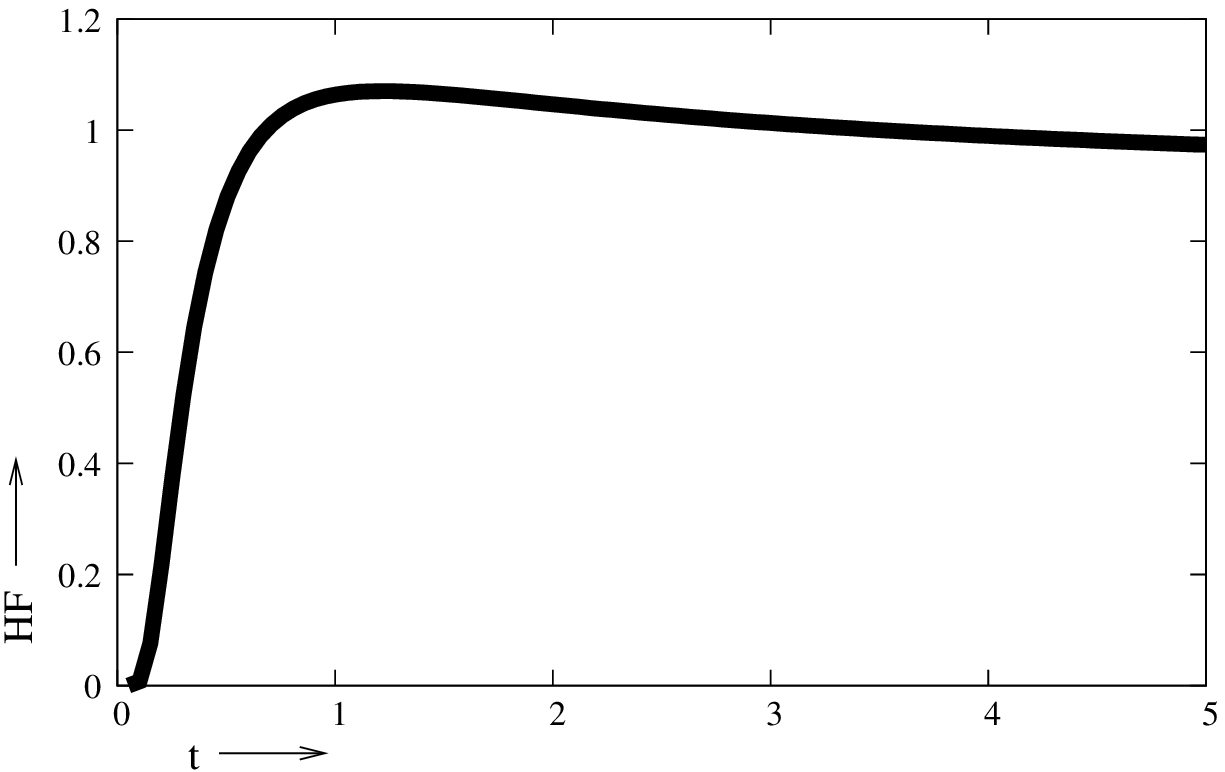}}

\subfigure[]{
\includegraphics[height=4.5cm,width=4.5cm]{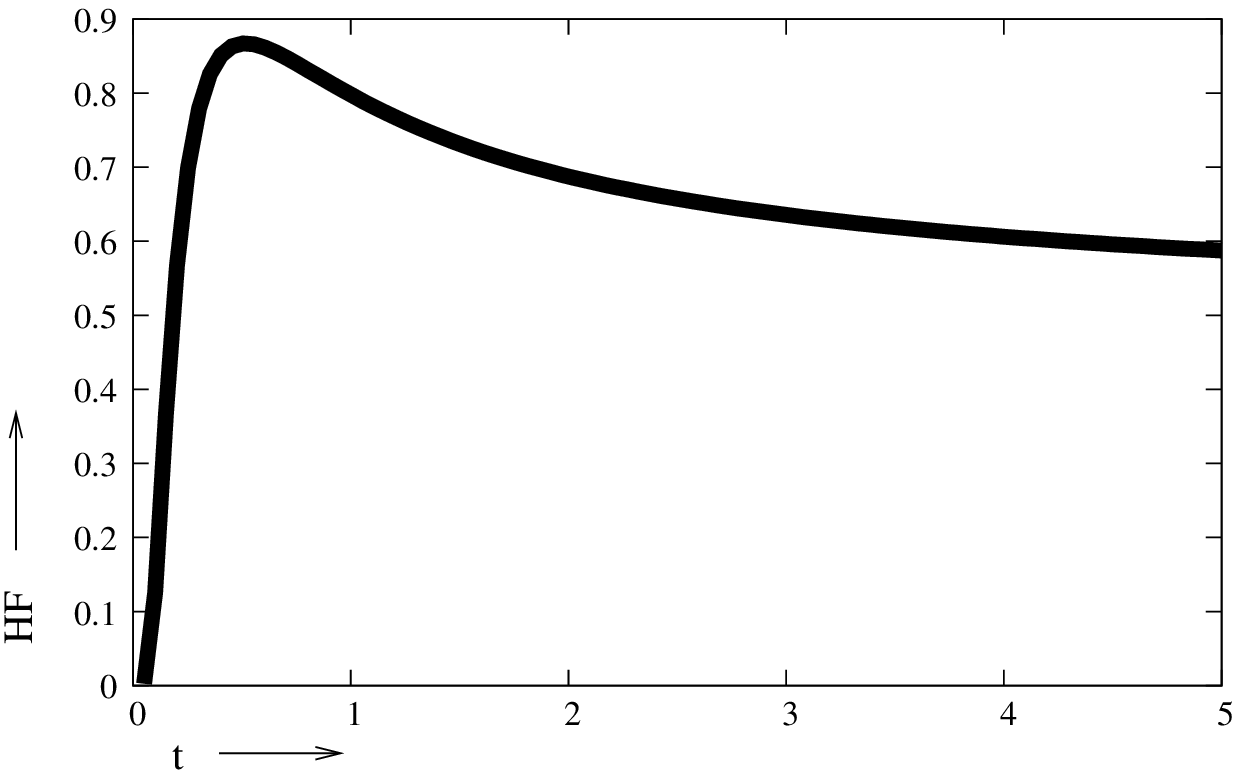}}
\subfigure[]{
\includegraphics[height=4.5cm,width=4.5cm]{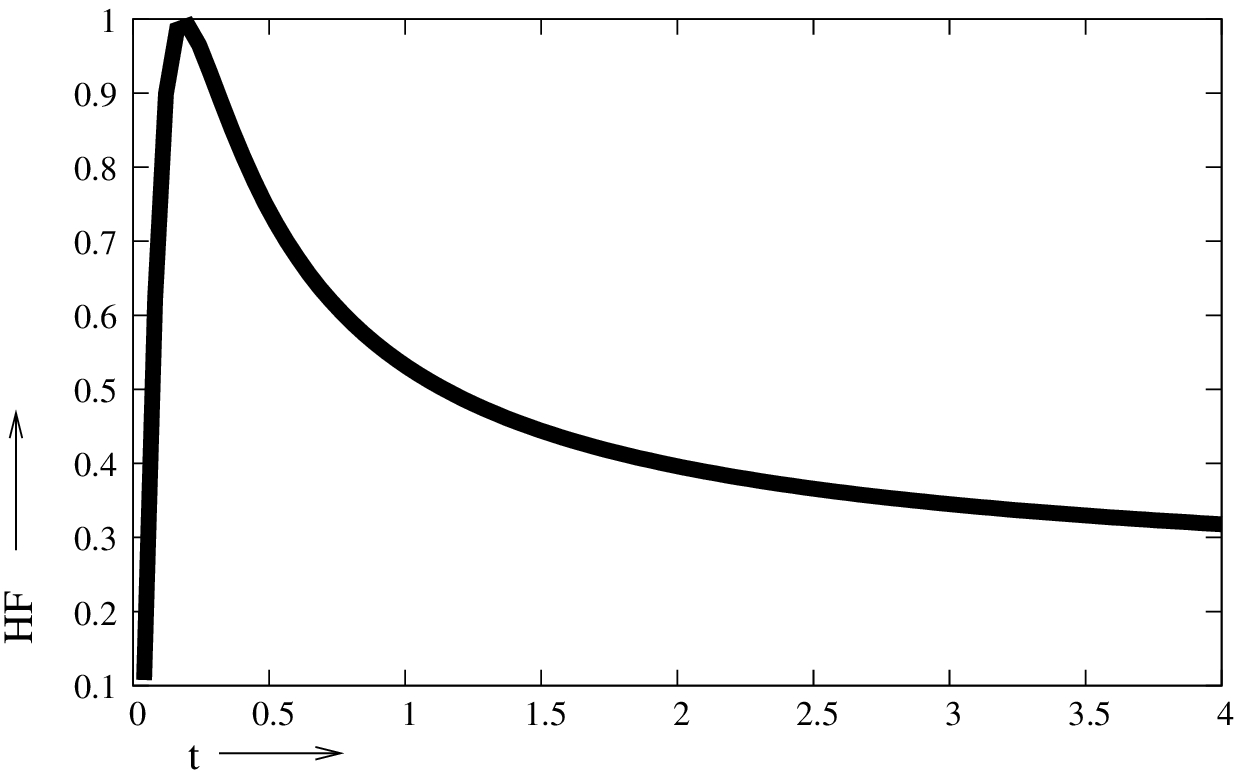}}

\caption{Plots of the HF of BS$(\alpha, \beta)$ for different values of $\alpha$ and when $\beta$ = 1: \ 
(a) $\alpha$ = 0.50, (b) $\alpha$ = 0.75 \ (c) $\alpha$ = 1.0, \ (d) $\alpha$ = 1.5.}
\label{hf-plot}
\end{center}
\end{figure}

We now discuss the shape characteristics of the HF of a BS random variable.  With $T \sim$ 
BS($\alpha, \beta$), the HF of $T$ is given by 
\be
h_T(t; \alpha, \beta) = \frac{f_T(t; \alpha, \beta)}{1 - F_T(t; \alpha, \beta)}, \ \ \ t > 0.
\label{hazard}
\ee
Since the shape of the HF does not depend on the scale parameter $\beta$, we may take $\beta = 1$ without 
loss of generality.  In this case, the HF in (\ref{hazard}) takes the form
\be
h_T(t; \alpha, 1) = \frac{\frac{1}{\sqrt{2 \pi} \alpha} \epsilon'(t) e^{-\frac{1}{2 \alpha^2} \epsilon^2(t)}}
{\Phi \left ( - \frac{\epsilon(t)}{\alpha} \right )},  \label{hazard-alt}
\ee
where
\be
\epsilon(t) = t^{1/2} - t^{-1/2}, \ \ \ \epsilon'(t) = \frac{1}{2t} (t^{1/2} + t^{-1/2}) \ \ \hbox{and} \ \ 
\epsilon''(t) = - \frac{1}{4t^2} (t^{1/2} + 3 t^{-1/2}).    \label{cubic-eq}
\ee
Kundu et al. \cite{KKB:2008} then  showed that the HF in (\ref{hazard-alt}), and hence the one in (\ref{hazard}), is always 
unimodal.  The plots of the HF of BS$(\alpha, \beta)$ in (\ref{hazard-alt}) for different values of $\alpha$, are presented 
in Figure \ref{hf-plot}.  From 
(\ref{hazard-alt}), it can be shown that $\ln (h_T(t; \alpha, 1)) \rightarrow 1/(2\alpha^2)$ as $t \rightarrow \infty$.

Kundu et al. \cite{KKB:2008} have shown that the change point, $c_{\alpha}$, of the hazard 
function of BS$(\alpha, 1)$ in (\ref{hazard-alt}) can be obtained as a solution of the non-linear equation
\be
\Phi \left ( - \frac{1}{\alpha} \epsilon(t) \right ) \left \{ - (\epsilon'(t))^2 \epsilon(t) 
+ \alpha^2 \epsilon''(t) \right \} + \alpha \phi \left ( - \frac{1}{\alpha} \epsilon(t) \right )
( \epsilon'(t))^2 = 0.
\ee
Since $\beta$ is the scale parameter,
the change point, $c_{\alpha, \beta}$, of the HF of BS$(\alpha,\beta)$ in (\ref{hazard}) can be obtained immediately as 
 $c_{\alpha, \beta} = \beta \ c_{\alpha}$.  Values of $c_{\alpha}$, for different choices of $\alpha$, have been computed by 
Kundu et al. \cite{KKB:2008}.  They have also observed that for $\alpha > 0.25$, the following
approximation of $c_{\alpha}$ can be useful: 
\be
\widetilde{c_{\alpha}} = \frac{1}{(-0.4604 + 1.8417 \alpha)^2},   \label{cp-app}
\ee
and that this approximation becomes quite accurate when $\alpha > 0.6$.  

\subsection{\sc Physical Interpretations}

It is well known that for the analysis of fatigue data, any of the two-parameter families of distributions, 
such as Weibull, log-normal or gamma, could be used to model effectively the region of central tendency.  Moreover,
since we usually have only small sample sizes in practice, none of them may get rejected by a goodness-of fit test, say 
Chi-square test or Kolmogorov-Smirnov test, for a given data.  However, when the goal is to predict the safe life, 
say the one-thousandth percentile, we would expect to have a wide discrepancy in the results obtained from these models.

For this reason, Birnbaum and Saunders \cite{BS:1969a} proposed the fatigue failure life distribution based
on a physical consideration of the fatigue process, rather than using an arbitrary parametric family of distributions.
They obtained a two-parameter family of non-negative random 
variables as an idealization for the number of cycles necessary to force a fatigue crack to grow to a critical 
value, based on the following assumptions:

\noindent 1. Fatigue failure is due to repeated application of a common cyclic stress pattern;  

\noindent 2. Due to the influence of this cyclic stress, a dominate crack in the material grows until
a critical size $w$ is reached at which point fatigue failure occurs;

\noindent 3.  The crack extension in each cycle is a random variable with the same mean and the 
same variance;

\noindent 4.  The crack extensions in cycles are statistically independent.

\noindent 5.  The total extension of the crack, after a large number of cycles, is approximately 
normally distributed, justified by the use of central limit theorem (CLT).

Specifically, at the $j$-th cycle, the crack extension $X_j$ is assumed to be a random variable with mean $\mu_0$ and variance
$\sigma_0^2$.  Then, due to CLT, $\ds \sum_{i=1}^n X_i$ is approximately normally distributed with mean 
$n \mu_0$ and variance $n \sigma_0^2$, and consequently the probability that the crack does not exceed a critical length 
$\omega$, say, is given by 
\be
\Phi \left ( \frac{\omega - n \mu_0}{\sigma_0 \sqrt{n}} \right ) = \Phi \left ( \frac{\omega}{\sigma_0 \sqrt{n}}
- \frac{\mu_0 \sqrt{n}}{\sigma_0} \right ).
\ee
It is further assumed that the failure occurs when the crack length exceeds $\omega$.  If $T$ denotes 
the lifetime (in number of cycles) until failure, then the CDF of $T$ is approximately
\be
P(T \le t) = F_T(t) \approx 1 - \Phi \left ( \frac{\omega}{\sigma_0 \sqrt{t}}
- \frac{\mu_0 \sqrt{t}}{\sigma_0} \right ) = \Phi \left (\frac{\mu_0 \sqrt{t}}{\sigma_0} -
 \frac{\omega}{\sigma_0 \sqrt{t}}\right ).   \label{approx}
\ee
Note that in deriving (\ref{approx}), it is being assumed that the probability that $X_j$'s take negative values is negligible. 
If (\ref{approx}) is assumed to be the exact lifetime model, then it is evident from (\ref{cdf-bs}) that $T$ follows a BS 
distribution with CDF in (\ref{cdf-bs}), where 
$$
\beta = \frac{\omega}{\mu_0}, \ \ \ \ \hbox{and} \ \ \ \ \alpha = \frac{\sigma_0}{\sqrt{\omega \mu_0}}.
$$
%\noindent Comments:  Need to add something from Desmond's paper.

\subsection{\sc Moments and Inverse Moments}

Using the relation in (\ref{relt}), and different moments of the standard normal random variable, 
moments of $T$, for integer $r$, can be obtained as follows:
\be
\hbox{E}(T^r) = \beta^r \sum_{j=0}^r {{2r}\choose{2j}} \sum_{i=0}^j {i\choose j} \frac{(2r-2j+2i)!}{2^{r-j+i}
(r-j+i)!} \left ( \frac{\alpha}{2} \right )^{2r - 2j + 2i};    \label{moment-1}
\ee
see Leiva et al. \cite{LSA:2009}.
Rieck \cite{Rieck:1999} also obtained $\hbox{E}(T^r)$, for fractional values of $r$, in terms of Bessel function, from the moment 
generating function of $\hbox{E}(\ln (T)$).

From (\ref{moment-1}), the mean and variance are obtained as  
\be
\hbox{E}(T) = \frac{\beta}{2} (\alpha^2 + 2) \ \ \ \ \ \hbox{and} \ \ \ \ \hbox{Var}(T) = \frac{\beta^2}{4}(5 \alpha^4 +
4 \alpha^2),   \label{mean-var}
\ee
while the coefficients of variation ($\gamma$), skewness ($\delta$) and kurtosis ($\kappa$) are found to be 
\be
\gamma = \frac{\sqrt{5 \alpha^4 + 4 \alpha^2}}{\alpha^2 + 2}, \ \ 
\delta = \frac{44 \alpha^3 + 24 \alpha}{(5 \alpha^2 + 4)^{3/2}} \ \ \ \hbox{and} \ \ \ 
\kappa = 3 + \frac{558 \alpha^4 + 240 \alpha^2}{(5\alpha^2 + 4)^2}, \label{cv-sk-kur}
\ee
respectively.  It is clear that as $\alpha \rightarrow 0$, the coefficient of kurtosis approaches 3, and the behavior of
the BS distribution appears to be approximately normal with mean and variance being approximately $\beta$ and $\beta^2 \alpha^2$, 
respectively.  Hence, as $\alpha \rightarrow 0$, the BS distribution becomes degenerate
at $\beta$.  On the other hand, when $\alpha \rightarrow \infty$, both mean and variance diverge, while the 
coefficients of variation, skewness and kurtosis all converge to some fixed constants.  

If $T \sim$ BS$(\alpha, \beta)$, it can be easily observed from (\ref{bs-pdf}) that $T^{-1} \sim$ 
BS$(\alpha, \beta^{-1})$.  Therefore, for integer $r$, we readily obtain from (\ref{moment-1}) that
\be
\hbox{E}(T^{-r}) = \beta^{-r} \sum_{j=0}^r {{2r}\choose{2j}} \sum_{i=0}^j {i\choose j} \frac{(2r-2j+2i)!}{2^{r-j+i}
(r-j+i)!} \left ( \frac{\alpha}{2} \right )^{2r - 2j + 2i}.    \label{moment}
\ee
Also, the mean and variance of $T^{-1}$ are obtained, from (\ref{mean-var}), to be 
\be
\hbox{E}(T^{-1}) = \frac{1}{2 \beta} (\alpha^2 + 2) \ \ \ \ \ \hbox{and} \ \ \ \ \hbox{Var}(T^{-1}) = 
\frac{1}{4 \beta^2}(5 \alpha^4 + 4 \alpha^2), 
\ee
while the corresponding $\gamma$, $\delta$ and $\kappa$ remain the same as presented in (\ref{cv-sk-kur}). 

\subsection{\sc Relationships with Inverse Gaussian Distribution}   \label{section-ig}

A random variable $X$ is said to have an inverse Gaussian (IG) distribution, with parameters $\mu > 0$  
and $\lambda > 0$, if $X$ has the PDF 
\be
f_X(t; \mu, \lambda) = \left ( \frac{\lambda}{2 \pi t^3} \right )^{1/2} 
\exp \left \{- \frac{\lambda}{2 \mu^2 t}(t-\mu)^2 \right \}, \ \ \  t > 0.
\label{pdf-ig}
\ee 
An IG random variable, with PDF in (\ref{pdf-ig}), will be denoted by 
IG$(\mu, \lambda)$.  The IG distribution has several interesting statistical and probabilistic
properties, of which one of the most intriguing one is in the presence of $\chi^2$ and $F$ distributions in 
associated inferential methods.  Book length accounts of IG distribution can be found in Chhikara and Folks \cite{CF:1989} and Seshadri 
\cite{Sesh:1993, Sesh:1999}.

Desmond \cite{Desmond:1986} first observed a relationship between BS and IG distributions.  Consider
two random variables $X_1$ and $X_2$ with $X_1 \sim$ IG$(\mu, \lambda)$ and $X_2^{-1} \sim$ IG$(\mu^{-1},
\lambda \mu^2)$.  Now, let us define a random variable $T$ as 
\be
T = \left \{ \matrix{X_1 & \hbox{with probability} & 1/2,  \cr 
&  &  \cr X_2 & \hbox{with probability} & 1/2.
\cr} \right .   \label{mix-random}
\ee 
Then, the PDF of $T$ becomes
\be
f_T(t; \mu, \lambda) = \frac{1}{2} f_{X_1}(t; \mu, \lambda) + \frac{1}{2} f_{X_2}(t; \mu, \lambda),
\label{mix-rep}
\ee
where $f_{X_1}(t; \mu, \lambda)$ has the same form as in (\ref{pdf-ig}) and 
$\ds f_{X_2}(t; \mu, \lambda) =  t f_{X_1}(t; \mu, \lambda)/\mu$.  A simple algebraic calculation shows that the random 
variable $T$ in (\ref{mix-random}) has BS$(\alpha, \beta)$, where $\ds \alpha = \sqrt{\mu/\lambda}$ and 
$\ds \beta = \mu$.  Using the mixture representation in (\ref{mix-rep}), many properties of the IG  
distribution can be readily transformed to those of the BS distribution.  For example, by  using
the moment generating functions (MGF) of the IG distribution and the length-biased IG distribution, the 
MGFs of the BS distribution can be obtained easily.  Specifically, if $T \sim$ BS$(\alpha, \beta)$, then the MGF of 
$T$ is 
$$
M_T(t) = \frac{1}{2} \left [ M_{X_1}(t) + M_{X_2}(t) \right ],
$$ 
where $M_{X_1}(t)$ and $M_{X_2}(t)$ are the MGFs of $X_1$ and $X_2$, respectively.   Since it is known that  
[see Jorgensen et al. \cite{JSW:1991}]
\beanno
M_{X_1}(t) & = & \exp \left \{(\lambda/\mu - ((\lambda/\mu)^2 - 2\lambda t)^{1/2}) \right \},  \\  
M_{X_2}(t) & = & \exp \left \{(\lambda/\mu - ((\lambda/\mu)^2 - 2\lambda t)^{1/2}) \right \},  
(1 - 2 \mu^2 t/ \lambda)^{-1/2}
\eeanno
 we readily find  
$$
M_T(t) = \frac{1}{2} \exp \left \{(\lambda/\mu - ((\lambda/\mu)^2 - 2\lambda t)^{1/2}) \right \} 
(1 + (1 - 2 \mu^2 t/ \lambda)^{-1/2}). 
$$
This mixture representation becomes useful while estimating the parameters of  
BS$(\alpha, \beta)$ distribution through an EM algorithm, and also in estimating the parameters
of a mixture of BS distributions.  These details are presented in later sections.  

\section{\sc Point Estimation: Complete Sample}  \label{section-point-estimation}

In this section, we discuss different methods of estimation of the parameters $\alpha$ and $\beta$ based on a random sample
$\{T_1, \cdots, T_n\}$ with observations (data) $\{t_1, \cdots, t_n\}$ of size $n$ from BS$(\alpha, \beta)$.

\subsection{\sc ML Estimators}

Birnbaum and Saunders \cite{BS:1969b} considered the ML estimators of parameters $\alpha$ and $\beta$ 
based on $\{t_1, \cdots, t_n\}$.  The log-likelihood function, without the additive constant, 
is given by  
\be
l(\alpha, \beta| \hbox{data}) = - n \ln (\alpha) - n \ln (\beta) + 
\sum_{i=1}^n \ln \left [ \left (\frac{\beta}{t_i} \right )^{1/2} + 
\left (\frac{\beta}{t_i} \right )^{3/2} \right ] - \frac{1}{2\alpha^2} \sum_{i=1}^n  \left (
\frac{t_i}{\beta} + \frac{\beta}{t_i} - 2 \right ).   \label{ll-comp}
\ee 
Based on the observed sample, let us define the sample arithmetic and harmonic means as
$$
s = \frac{1}{n} \sum_{i=1}^n t_i \ \ \ \ \hbox{and} \ \ \ \
r = \left [ \frac{1}{n} \sum_{i=1}^n t_i^{-1} \right ]^{-1},
$$
respectively.  Differentiating (\ref{ll-comp}) with respect to $\alpha$ and equating it to zero, we obtain
\be
\alpha^2  = \left [ \frac{s}{\beta} + \frac{\beta}{r} - 2 \right ].    \label{ne-alpha-1}
\ee
Next, differentiating (\ref{ll-comp}) with respect to $\beta$ and equating it to zero and after substituting 
$\alpha^2$ from (\ref{ne-alpha-1}), the following non-linear equation is obtained:
\be
\beta^2 - \beta ( 2 r + K(\beta)) + r(s + K(\beta)) = 0,   \label{ne-beta}
\ee
where 
$$
K(x) = \left [ \frac{1}{n} \sum_{i=1}^n (x+t_i)^{-1} \right ]^{-1} \ \ \ \ \hbox{for} \ x \ge 0.
$$
The ML estimator of $\beta$, say $\widehat{\beta}$, is then the positive root of 
(\ref{ne-beta}).  Birnbaum and Saunders \cite{BS:1969b} showed that $\widehat{\beta}$ is the unique positive root
of (\ref{ne-beta}) and furthermore, $r < \widehat{\beta} < s$.  A numerical iterative procedure is needed to 
solve the non-linear equation in (\ref{ne-beta}).  In their work, Birnbaum and Saunders \cite{BS:1969b} 
proposed two different iterative methods to compute $\widehat{\beta}$ and showed that with any 
arbitrary initial guess value between $r$ and $s$, both methods do converge to $\widehat{\beta}$.  
Once the ML estimator of $\beta$ is obtained, the ML estimator  of $\alpha$ can then be obtained as 
\be
\widehat{\alpha}  = \left [ \frac{s}{\widehat{\beta}} + \frac{\widehat{\beta}}{r} - 2 \right ]^{1/2}.   
\ee
Birnbaum and Saunders \cite{BS:1969b} also proved that the ML estimators are consistent estimates of the parameters.
Recently, Balakrishnan and Zhu \cite{BZ:2014a} showed that if $n$ = 1, the ML estimators of $\alpha$ and $\beta$ do not
exist, but when $n > 1$, the ML estimators always exist and are unique.

Engelhardt et al. \cite{EBW:1981} showed that it is a regular family of distributions and that the   
Fisher information matrix is given by 
\be
{\ve I}(\alpha,\beta) = - \left [ \matrix{\frac{2n}{\alpha^2} & 0 \cr 0 & n \left [ \frac{1}{\alpha^2 \beta^2}
+ E \left ( \frac{1}{(T + \beta)^2} \right ) \right ]\cr} \right ] = 
- \left [ \matrix{\frac{2n}{\alpha^2} & 0 \cr 0 & \frac{n}{\alpha^2 \beta^2} 
(1 + \alpha(2 \pi)^{-1/2} h(\alpha)), 
\cr} \right ],    \label{fish-info}
\ee
where
\be
h(\alpha) = \alpha \sqrt{\pi/2} - \pi e^{2/\alpha^2}[ 1 - \Phi(2/\alpha)].   \label{h-func}
\ee
By simple calculations, it has been shown by Engelhardt 
et al. \cite{EBW:1981} that the joint distribution of $\widehat{\alpha}$ and $\widehat{\beta}$ is 
 bivariate normal, i.e., 
\be
\left ( \matrix{\widehat{\alpha} \cr   \cr \widehat{\beta} \cr} \right ) \sim
\hbox{N} \left [ \left ( \matrix{\alpha \cr  \cr \beta \cr} \right ), 
\left ( \matrix{\frac{\alpha^2}{2n} & 0 \cr   &  \cr 
0 & \frac{\beta^2}{n(0.25 + \alpha^{-2} + I(\alpha))}}
\right ) \right ],   \label{asymp-dist-mle}
\ee
where 
\be
I(\alpha) = 2 \int_0^{\infty} \left \{(1 + g(\alpha x))^{-1} - 1/2 \right \}^2 \hbox{d} \Phi(x) 
\label{i-func}
\ee
with
$$
g(y) = 1 + \frac{y^2}{2} + y \left ( 1 + \frac{y^2}{4} \right )^{1/2}.
$$
It is of interest to note that $\widehat{\alpha}$ and $\widehat{\beta}$ are asymptotically 
independent.  An asymptotic confidence interval for $\alpha$ can be easily obtained 
from (\ref{asymp-dist-mle}), and an asymptotic confidence interval for $\beta$, for a given $\alpha$,
can also be obtained from (\ref{asymp-dist-mle}).

\subsection{\sc Moment and Modified Moment Estimators}

The moment estimators of $\alpha$ and $\beta$ can be obtained by equating the sample mean and sample 
variance, respectively,  with $\hbox{E}(T)$ and $\hbox{Var}(T)$, provided in (\ref{mean-var}).  Thus, if $s$ and $v$ denote the 
sample mean and sample variance, respectively, then the moment estimators of $\alpha$ and $\beta$ are obtained 
by solving 
\be
s = \frac{\beta}{2}(\alpha^2 + 2), \ \ \  v = \frac{\beta^2}{4}(5 \alpha^4 + 4 \alpha^2).
\ee
It can be easily seen that the moment estimator of $\alpha$ can be obtained as the root of the 
non-linear equation
\be
\alpha^4(5 s^2 - v) + 4 \alpha^2(s^2 - 4) - 4 v = 0.
\ee
It is clear that if the sample coefficient of variation is less than $\sqrt{5}$, then the moment estimator
of $\alpha$ is
\be
\widehat{\widehat{\alpha}} = \left [ \frac{-2(s^2-4) + 2\sqrt{(s^2-4)^2 + v(5s^2-v)}}{(5s^2 - v)} \right ]^{1/2}  
\ee
and the moment estimator of $\beta$ is then
\be
\widehat{\widehat{\beta}} = \frac{2s}{\widehat{\widehat{\alpha}}^2+2}.
\ee
However, if the sample coefficient of variation is greater than $\sqrt{5}$, then the moment estimator
of $\alpha$ may not exist.

Since the ML estimators cannot be obtained in explicit form and that the moment estimators may not always exist, 
Ng et al. \cite{NKB:2003} suggested MM estimators of the parameters $\alpha$ and $\beta$ by 
utilizing the fact that if $T \sim$ BS($\alpha, \beta$), then $T^{-1} \sim$ BS$(\alpha, \beta^{-1})$.  Therefore, 
equating the sample arithmetic and harmonic means, $s$ and $r$, respectively, with the corresponding population versions, we obtain
\be
s = \beta \left (1 + \frac{1}{2} \alpha^2 \right ) \ \ \ \ \hbox{and} \ \ \ \
r^{-1} = \beta^{-1} \left ( 1 + \frac{1}{2} \alpha^2 \right ).
\ee
Thus, the MM estimators of $\alpha$ and $\beta$ are obtained as 
\be
\widetilde{\alpha} = \left \{ 2 \left [ \left ( \frac{s}{r} \right )^{1/2} - 1 \right ] \right \}^{1/2} \ \ \
\hbox{and} \ \ \  \widetilde{\beta} = (s r)^{1/2}.
\ee
The interesting point about the MM estimators is that they always exist unlike the moment estimators, and moreover they have simple  
explicit forms.  Using the CLT on $\ds (1/n) \sum_{i=1}^n T_i$ and
$\ds (1/n) \sum_{i=1}^n T_i^{-1}$, when $T_i$'s are independent and identically distributed ($i.i.d.$)
 BS$(\alpha, \beta)$, Ng et al. \cite{NKB:2003}
proved that the joint asymptotic distribution of $\widetilde{\alpha}$ and $\widetilde{\beta}$ is bivariate 
normal,i.e., 
\be
\left ( \matrix{\widetilde{\alpha} \cr   \cr \widetilde{\beta} \cr} \right ) \sim
\hbox{N} \left [ \left ( \matrix{\alpha \cr  \cr \beta \cr} \right ), 
\left ( \matrix{\frac{\alpha^2}{2n} & 0 \cr   &  \cr 
0 & \frac{\alpha\beta^2}{n} \left ( \frac{1 + \frac{3}{4} \alpha^2}{(1 + \frac{1}{2} \alpha^2)^2}
\right ) \cr} \right )\right ].   \label{asymp-dist-mme}
\ee
Interestingly, in this case also, we observe that the MM estimators are asymptotically independent.

Ng et al. \cite{NKB:2003} performed extensive Monte Carlo simulations to compare the performances of the ML 
estimators and MM estimators 
for different sample sizes and for different choices of parameter values.  They then observed that the performances of 
ML estimators and MM estimators are almost identical for any sample size when the shape parameter $\alpha$ is not
large (say, $< 0.5$).  For small sample sizes, both estimators are highly biased if $\alpha$ is large.
It has been observed that
\be
\hbox{Bias}(\widehat{\alpha}) \approx \hbox{Bias}(\widetilde{\alpha}) \approx - \frac{\alpha}{n} \ \ \  \hbox{and} \ \ \ 
\hbox{Bias}(\widehat{\beta}) \approx \hbox{Bias}(\widetilde{\beta}) \approx - \frac{\alpha^2}{4n}.
   \label{bias-mle-mme}
\ee
So, the bias-corrected ML estimators and MM estimators perform quite well.  Alternatively, jackknifing [see Efron \cite{Efron:1982}]  
can also 
be performed to determine bias-corrected ML estimators  and MM estimators, and these estimators have also been observed to perform 
quite well.  

\subsection{\sc From and Li Estimators}

From and Li \cite{FL:2006} proposed four other estimators of $\alpha$ and $\beta$, which are as follows.

\noindent From and Li Estimator-1:  If $T \sim$ BS$(\alpha, \beta)$, then $\ds Z$ 
defined in (\ref{relx}) follows the standard normal distribution and based on this fact, an estimator of $(\alpha, \beta)$
can be obtained by solving the following two equations:
\be
\frac{1}{n} \sum_{i=1}^n  \frac{1}{\alpha} \left [ \left ( \frac{t_i}{\beta} \right )^{1/2} - 
\left ( \frac{t_i}{\beta} \right )^{-1/2} \right ] = 0 \ \ \  \hbox{and} \ \ \
\frac{1}{n} \sum_{i=1}^n  \frac{1}{\alpha^2} \left [ \left ( \frac{t_i}{\beta} \right ) +
\left ( \frac{t_i}{\beta} \right )^{-1} - 2 \right ] = 1,    \label{fl-ne-1}
\ee
giving rise to the estimators  
$$
\widehat{\beta}_{FL,1} = \frac{\sum_{i=1}^n t_i^{1/2}}{\sum_{i=1}^n t_i^{-1/2}} \ \ \ \hbox{and} \ \ \ 
\widehat{\alpha}_{FL,1} = \left ( \sum_{i=1}^n  \frac{1}{n} \left [ \left ( \frac{t_i}{\widehat{\beta}_{FL,1}} 
\right ) +
\left ( \frac{t_i}{\widehat{\beta}_{FL,1}} \right )^{-1} - 2 \right ] \right )^{1/2}.
$$
These can be seen to be variations of the moment estimators.

\noindent From and Li Estimator-2:  Since $\beta$
is the median of $T$ irrespective of $\alpha$, From and Li \cite{FL:2006} obtained estimators by equating the 
sample median and sample variance to their corresponding population versions.  Thus, the estimator 
of $\beta$ in this case is simply  
$$
\widehat{\beta}_{FL,2} = \hbox{median}\{t_1, \cdots, t_n\},
$$
while an estimator of $\alpha$ can be obtained from the equation  
$$
v = \frac{\widehat{\beta}_{FL,2}}{4} \left ( 5 \alpha^4 + 4 \alpha^2 \right ),
$$
for which the solution is 
$$
\widehat{\alpha}_{FL,2} = \sqrt{\frac{-2 + 2 \sqrt{1+5v/\widehat{\beta}_{FL,2}}}{5}}.
$$
Here, $v$ is the sample variance as before.

\noindent From and Li Estimator-3:  If we denote $t_{1:n} < \cdots < t_{n:n}$ as the ordered sample of
$t_1, \cdots, t_n$, then take $\widehat{\beta}_{FL,3}$ same as $\widehat{\beta}_{FL,2}$, and solve for 
$\alpha$ from 
$$
F_T(t_{i:n}; \alpha, \widehat{\beta}_{FL,3}) = \frac{i}{n+1}, \ \ \  i = 1, \cdots, n,
$$
$i.e.$, 
$$
\widehat{\alpha}(i) = \frac{\epsilon(t_{i:n}/\widehat{\beta}_{FL,3})}{\Phi^{-1}(i/(n+1))}, \ \ \  
i = 1, \cdots, n.
$$
Here, $\epsilon(\cdot)$ is same as defined earlier in (\ref{cubic-eq}).
Finally, an estimator of $\alpha$ is proposed as
$$
\widehat{\alpha}_{FL,3} = \hbox{median}\{\widehat{\alpha}(1), \cdots, \widehat{\alpha}(n)\}.
$$

\noindent From \& Li Estimator-4:  Instead of using the whole data, From and Li \cite{FL:2006} proposed the 
following estimators of $\alpha$ and $\beta$ using only the middle portion of the data, which produces
more robust estimators.  For $1 \le n_1 < n_2 \le n$, these estimators are as follows:
\be
\widehat{\beta}_{FL,4} = \frac{\sum_{i=n_1}^{n_2} t_{i:n}}{\sum_{i=n_1}^{n_2} (1/\sqrt{t_{i:n}})} \ \ \ \
\hbox{and} \ \ \
\widehat{\alpha}_{FL,4} = \sqrt{\frac{\sum_{i=n_1}^{n_2} \epsilon^2(t_{i:n}/\widehat{\beta}_{FL,4})}
{\sum_{i = n_1}^{n_2} \left ( \Phi^{-1}(i/(n+1)) \right )^2}};  \label{form-li-4}
\ee
here, $n_1$ and $n_2$ are chosen such that $n_1/n < 0.5$ and $n_2/n > 0.5$.  Since the estimators in (\ref{form-li-4}) use only the middle
portion of the data, it is expected that the estimators will not be affected by the presence of outliers, but a 
loss in efficiency is to be expected in the case when there are no outliers in the data.

From and Li \cite{FL:2006} performed extensive Monte Carlo simulation experiments to compare the performance of ML estimators with those of their four 
estimators.  They observed that among their estimators, From and Li Method-1 performed nearly the same as  the 
ML estimators if the model is true and that there are no outliers in the data.  If some outliers
are present, then the  ML estimators do not perform well, but From and Li Method-4 performed the best in this case.

\subsection{\sc Balakrishnan and Zhu Estimator}

Balakrishnan and Zhu \cite{BZ:2014b} developed their estimators of $\alpha$ and $\beta$ based on some key properties of the 
BS distribution.  Suppose $T_1, \ldots, T_n$ 
are $i.i.d.$ BS$(\alpha, \beta)$ random variables.  Then, consider the following $\ds {n\choose{2}}$ pairs of random variables 
$(Z_{ij}, Z_{ji})$, where
$$
Z_{ij} = \frac{T_i}{T_j} \ \ \ \ \hbox{for} \ \ \  1 \le i \ne j \le n.
$$
Now, observe that 
$$
\hbox{E}(Z_{ij}) = \hbox{E} \left ( \frac{T_i}{T_j} \right ) = \hbox{E}(T_i) \hbox{E} \left ( \frac{1}{T_j} \right ) = \left (1 + \frac{\alpha^2}{2} \right )^2.
$$
Upon equating the sample mean of $Z_{ij}$ with its population mean, we obtain the equation 
\be
\bar{z} = \frac{1}{n(n-1)} \sum_{1 \le i \ne j \le n} z_{ij} = \left (1 + \frac{\alpha^2}{2} \right )^2.    \label{bala-zhu-me}
\ee
Here, $z_{ij}$ denotes the sample value of $Z_{ij}$.  Upon solving (\ref{bala-zhu-me}), an estimator of $\alpha$ is obtained as  
\be
\widehat{\alpha}_{BZ} = \left ( 2 ( \sqrt{\bar{z}} - 1) \right )^{1/2}.    \label{bz-est-alpha}
\ee
Balakrishnan and Zhu \cite{BZ:2014b} then proposed two different estimators of $\beta$ based on $\widehat{\alpha}_{BZ}$ in  
(\ref{bz-est-alpha}), as follows. 
Since
$$
\hbox{E} \left ( \frac{1}{n} \sum_{i=1}^n T_i \right )  = \beta \left ( 1 + \frac{1}{2} \alpha^2 \right ),
$$
an estimator $\beta$ can be obtained immediately as
$$
\widehat{\beta}_{BZ,1} = \frac{2 s}{(\widehat{\alpha}_{BZ})^2 + 2} = \frac{s}{\sqrt{\bar{z}}};  %\label{bz-est-beta}
$$
here, $s$ is the sample mean of the observed $t_1, \ldots, t_n$.  Moreover, based on the fact that
$$
\frac{1}{n} \sum_{i=1}^n \hbox{E} \left ( \frac{1}{T_i} \right ) = \frac{1}{\beta} \left ( 1 + \frac{\alpha^2}{2} \right ),
$$
Balakrishnan and Zhu \cite{BZ:2014b} proposed another estimator of $\beta$ as
$$
\widehat{\beta}_{BZ,2} = r \left (1 + \frac{1}{2}(\widehat{\alpha}_{BZ})^2 \right ) = r \sqrt{\bar{z}};  %\label{bz-est-beta-2}
$$
here, $r$ is the harmonic mean of $t_1, \ldots, t_n$, as defined before.  

Theoretically, it has been shown by Balakrishnan and Zhu \cite{BZ:2014b} that $\widehat{\alpha}_{BZ}$, $\widehat{\beta}_{BZ,1}$ and 
$\widehat{\beta}_{BZ,2}$ always exist and that $\widehat{\alpha}_{BZ}$ is negatively biased.  Based on extensive simulations, they have 
 observed that
the performances of their estimators are very similar to those of the ML estimators and the MM estimators in terms of bias and MSE.

\subsection{\sc New Estimators}

We may propose the following estimators which are in explicit form.  First, we may take  
$\widehat{\beta}_{\hbox{new}}$ = $\hbox{median}\{t_1, \cdots, t_n\}$.  Then, we consider the new transformed variables
$$
\ds u_i = \left (\frac{t_i}{\widehat{\beta}_{\hbox{new}}} \right )^{1/2} - 
\left ( \frac{\widehat{\beta}_{\hbox{new}}}{t_i} \right )^{1/2}, \ \ \ 
i = 1, \cdots, n.
$$
It is known that if $T \sim$ BS$(\alpha, \beta)$, then $\ds \left \{ \left ( T/\beta \right )^{1/2} 
- \left (\beta/T \right )^{1/2} \right \}
\sim $ N$(0, \alpha^2)$, and using this fact, we may propose an estimate of $\alpha$ as
$$
\widehat{\alpha}_{\hbox{new}} = \sqrt{\frac{1}{n} \sum_{i=1}^n u_i^2}.
$$
Its properties and relative performance may need further investigation. 

\subsection{\sc Robust Estimators}

Dupuis and Mills \cite{DM:1998} proposed a robust estimation procedure for $\alpha$ and $\beta$ by employing 
the influence function (IF) approach of Hampel \cite{Hampel:1974}.  It has been observed by Hampel \cite{Hampel:1974} 
that the robustness of an estimator can be measured by means of its IF.  It is well known that [see Hampel \cite{Hampel:1974}]
the IF of an ML estimator is proportional to its score function, which for the BS model is given by 
\be
\underline{s}(t; \alpha, \beta) = \left [ \matrix{\frac{\partial \ln (f_T(t; \alpha, \beta))}{\partial \alpha} \cr  \cr
\frac{\partial \ln (f_T(t; \alpha, \beta))}{\partial \beta} \cr} \right ],
\ee
where 
\bea
\frac{\partial \ln (f_T(t; \alpha, \beta))}{\partial \alpha} & = & 
\frac{-\alpha^2 \beta t + t^2 - 2 \beta t + \beta^2}{\alpha^3 \beta t},   \label{score-1}  \\
\frac{\partial \ln (f_T(t; \alpha, \beta))}{\partial \beta} & = & 
\frac{-\beta \alpha^2 t^2 + \beta^2 \alpha^2 t + t^3 - \beta^2 t + \beta t^2 - \beta^3}{2 \beta^2 \alpha^2 t
(t+\beta)}.   \label{score-2}
\eea
Clearly, the score functions are unbounded in $t$, which implies that the corresponding IF is also unbounded
in $t$.   For this reason, the ML estimators become biased and inefficient when the model is incorrect.

Dupuis and Mills \cite{DM:1998}, therefore,  proposed the following optimal biased robust estimators (OBREs) 
of $\alpha$ and $\beta$.  The OBREs of $\alpha$ and $\beta$, say 
$\widehat{\alpha}_{R}$ and $\widehat{\beta}_{R}$, respectively,  are the 
solutions of 
\be
\sum_{i=1}^n \underline{\Psi}(t_i; \alpha, \beta) = 0
\ee
for some function $\underline{\Psi}: \mr_+^3  \rightarrow \mr_+^2$.  Moreover, the IF function of an $M$-estimator
$\underline{\Psi}$, at $F_T(t; \alpha, \beta)$, is given by
\beanno
IF(t; \underline{\Psi}, F_T) & = & [{\ve M}(\underline{\Psi}, F_T)]^{-1} \underline{\Psi}(t; \alpha, \beta), \\
{\ve M}(\underline{\Psi}, F_T) & = & - \int_0^{\infty} \underline{\Psi}(t; \alpha, \beta) [\underline{\Psi}(t; \alpha, \beta)]^{\top} \hbox{d}F_T(t; \alpha, \beta).
\eeanno
In any robust M-estimation procedure, $\underline{\Psi}(\cdot)$ needs to be chosen properly.  There are several versions
of OBREs that are available in the literature.  Dupuis and Mills \cite{DM:1998} proposed the standardized OBREs.  They can be 
defined as follows for a given bound $c$ on the IF.  The function $\underline{\Psi}(\cdot)$
is defined implicitly by
$$
\sum_{i=1}^n \underline{\Psi}(t_i; \alpha, \beta) = \sum_{i=1}^n \left \{\underline{s}(t_i; \alpha, \beta) - 
\underline{a}(\alpha, \beta) W_c(t_i; \alpha, \beta) \right \},
$$
where
$$
W_c(t; \theta) = \min \left [ 1, \frac{c}{||{\ve A}(\alpha, \beta) [\underline{s}(t; \alpha, \beta) - \underline{a} (\alpha, \beta)]||}
\right ]
$$
with $|| \cdot ||$ being the Euclidean norm.  The $2 \times 2$ matrix ${\ve A}(\alpha, \beta)$ and the 
$2 \times 1$ vector $\underline{a}(\alpha, \beta)$ are defined implicitly as
\beanno
\hbox{E}\left [ \underline{\Psi}(t; \alpha, \beta) [ \underline{\Psi}(t; \alpha, \beta) ]^{\top} \right ] & = & 
\{{\ve A}(\alpha, \beta)^{\top} {\ve A}(\alpha, \beta) \}^{-1},  \\
E\left [ \underline{\Psi}(t; \alpha, \beta) \right ]
& = & {\underline 0}.
\eeanno
The following algorithm can be used to compute the OBREs, as suggested by Dupuis and Mills \cite{DM:1998}:

%\vspace{1.00 in}

\noindent {\sc Algorithm}

\noindent Step 1: Fix an initial value of $(\alpha,\beta)$;

\noindent Step 2: Fix $\eta$, $c$, $\underline{a} = \underline{0}$, and ${\ve A} = 
({\ve I}(\alpha,\beta))^{-1/2}$.  Here, the matrix ${\ve I}(\alpha,\beta)$ is the Fisher information matrix
 given in (\ref{fish-info});

\noindent Step 3:  Solve the following two equations for ${\underline a}$ and ${\ve A}$:
\beanno
{\ve A}^{\top}{\ve A} & = & {\ve M}_2^{-1}, \\
{\underline a} & = & \frac{\int_0^{\infty} \underline{s}(t;\alpha,\beta) W_c(t;\alpha,\beta)\hbox{d}F_T(t;\alpha,\beta))}
{\int_0^{\infty} W_c(t;\alpha,\beta) \hbox{d}F_T(t;\alpha,\beta) },
\eeanno
where for $k$ = 1 and 2,
$$
{\ve M}_k = \int_0^{\infty} \{(s(t;\alpha,\beta) - {\underline a})(\underline{s}(t;\alpha,\beta) - {\underline a})^{\top}\}
(W_c(t;\alpha,\beta))^k \hbox{d}F_T(t;\alpha,\beta).
$$
The current values of $\alpha$, $\beta$, ${\underline a}$ and ${\ve A}$ are used as starting 
values for solving the equations in this step;

\noindent Step 4:  Compute ${\ve M}_1$, using ${\underline a}$ and ${\ve A}$ from Step 3, and
$$
\Delta(\alpha,\beta) = {\ve M}_1^{-1} \left [ \frac{1}{n} \sum_{i=1}^n 
(\underline{s}(t_i;\alpha,\beta) - {\underline a})(W_c(t_i;\alpha,\beta))  \right ];
$$

\noindent Step 5: If
$$
\max \left \{ \left | \frac{\Delta \alpha}{\alpha} \right |, \left | 
\frac{\Delta \beta}{\beta} \right | \right \} > \eta,
$$
then a new $(\alpha,\beta)$ can be obtained as previous $(\alpha,\beta) + \Delta(\alpha,\beta)$,
and then return to Step 3.  Otherwise, we  stop the iteration.

\section{\sc Interval Estimation}  \label{section-interval-estimation}

In the last section, we have discussed several point estimators of the parameters $\alpha$ and $\beta$.  In this section, we 
discuss different interval estimators of $\alpha$ and $\beta$ based on a random sample
$\{t_1, \cdots, t_n\}$ of size $n$ from BS$(\alpha, \beta)$.

\subsection{\sc Ng-Kundu-Balakrishnan Estimators}

Ng et al. \cite{NKB:2003} proposed interval estimation of the parameters based on the 
asymptotic distribution of the ML and MM estimators, as provided in (\ref{asymp-dist-mle}) and 
(\ref{asymp-dist-mme}), respectively.  Simulation results revealed that the ML and MM estimators are 
highly biased, and that they are of the order given in (\ref{bias-mle-mme}).  Almost unbiased bias-corrected 
ML (UML) and MM (UMM) estimators have been proposed by these authors as follows:
\beanno
\widehat{\alpha}^* & = & \left ( \frac{n}{n-1} \right ) \widehat{\alpha}, \ \ \ \ \ \ 
\widehat{\beta}^* = \left ( 1 + \frac{\widehat{\alpha}^{*^2}}{4n} \right )^{-1} \widehat{\beta},  \\
\widetilde{\alpha}^* & = & \left ( \frac{n}{n-1} \right ) \widetilde{\alpha}, \ \ \ \ \ \ 
\widetilde{\beta}^* = \left ( 1 + \frac{\widetilde{\alpha}^{*^2}}{4n} \right )^{-1} \widetilde{\beta}.
\eeanno
Here, $\widehat{\alpha}$ and $\widehat{\beta}$ are the ML estimators, while $\widetilde{\alpha}$ and 
$\widetilde{\beta}$ are the MM estimators of $\alpha$ and $\beta$, respectively.  Based on the UML and UMM estimators,  
100(1-$\gamma$)\% confidence intervals of $\alpha$ and $\beta$ can be obtained as follows:
\beanno
\left [ \widehat{\alpha}^* \left ( \sqrt{\frac{n}{2}} \frac{z_{\gamma/2}}{(n-1)} + 1 \right )^{-1},
\widehat{\alpha}^* \left ( \sqrt{\frac{n}{2}} \frac{z_{1-\gamma/2}}{(n-1)} + 1 \right )^{-1} \right ],
\\
\left [ \widehat{\beta}^* \left ( \frac{n}{h_1(\widehat{\alpha}^*)} \frac{4 z_{\gamma/2}}{(4 n + 
\widehat{\alpha}^{*^2})} + 1 \right )^{-1}, 
\widehat{\beta}^* \left ( \frac{n}{h_1(\widehat{\alpha}^*)} \frac{4 z_{1-\gamma/2}}{(4 n + 
\widehat{\alpha}^{*^2})} + 1 \right )^{-1} \right ],
\eeanno
and
\beanno
\left [ \widetilde{\alpha}^* \left ( \sqrt{\frac{n}{2}} \frac{z_{\gamma/2}}{(n-1)} + 1 \right )^{-1},
\widetilde{\alpha}^* \left ( \sqrt{\frac{n}{2}} \frac{z_{1-\gamma/2}}{(n-1)} + 1 \right )^{-1} \right ],
\\
\left [ \widetilde{\beta}^* \left ( \frac{n}{h_2(\widetilde{\alpha}^*)} \frac{4 z_{\gamma/2}}{(4 n + 
\widetilde{\alpha}^{*^2})} + 1 \right )^{-1}, 
\widetilde{\beta}^* \left ( \frac{n}{h_2(\widetilde{\alpha}^*)} \frac{4 z_{1-\gamma/2}}{(4 n + 
\widetilde{\alpha}^{*^2})} + 1 \right )^{-1} \right ],
\eeanno
respectively; see also Wu and Wong \cite{WW:2004}.  Here, $\ds h_1(x) = 0.25 + x^{-2} + I(x)$, $\ds h_2(x) = 
(1+3/4x^2)/(1+x^2/2)^2$, with $I(\cdot)$ being as defined in (\ref{i-func}), and $z_p$ is the $p$-th 
percentile point of a standard normal distribution.  Extensive simulation results suggest 
that the performance of the bias-corrected technique is quite good if the sample size is at least 
20.  If the sample size is less than 20, the coverage percentages are slightly lower than the 
corresponding nominal levels.

\subsection{\sc Wu and Wong Estimators}

Since the ML, UML, MM and UMM estimators do not work satisfactorily in case of small sample sizes, Wu and 
Wong \cite{WW:2004} used the higher-order likelihood-based method, as proposed by Barndorff-Nielsen 
\cite{BN:1986, BN:1991}, for constructing confidence intervals for the parameters.  This is known as the modified signed log-likelihood 
ratio statistic, and is known 
to have a higher-order accuracy, and it performs quite well even when the sample is quite small.  To be specific, let 
$$
r^*(\beta) = r(\beta) + r(\beta)^{-1} \ln \left \{ \frac{q(\beta)}{r(\beta)} \right \},
$$ 
where $r(\beta)$ is the signed log-likelihood ratio statistic defined by
$$
r(\beta) = \hbox{sgn}(\widehat{\beta} - \beta) \{2[l(\widehat{\theta}) - l(\widehat{\theta}_{\beta})]\}^{1/2},
$$
with $\underline{\theta} = (\alpha,\beta)$ being the unknown set of parameters, $l(\cdot)$ being the log-likelihood 
function as defined in (\ref{ll-comp}),  
$\underline{\widehat{\theta}} = (\widehat{\alpha},\widehat{\beta})$ being the overall ML estimator of $\underline{\theta}$, and 
$\underline{\widehat{\theta}}_{\beta} = (\widehat{\alpha}_{\beta},\beta)$ being the constrained ML estimator of $\underline{\theta}$ 
for a given $\beta$, $i.e.$,
$$
\widehat{\alpha}_{\beta} = \left [ \frac{s}{\beta} + \frac{\beta}{r} - 2 \right ]^{1/2}.
$$
Moreover, a general form of $q(\beta)$ takes on the following form [see Fraser et al. \cite{FRW:1999}]:
$$
q(\beta) = \frac{|l_{;V}(\underline{\widehat{\theta}}) - l_{;V}(\underline{\widehat{\theta}}_{\beta}) 
l_{\alpha;V}(\underline{\widehat{\theta}}_{\beta})|}{|l_{\theta;V}(\underline{\widehat{\theta}})|} \left \{ \frac{|j_{\underline{\theta} \ \underline{\theta}}
(\underline{\widehat{\theta}})|}{|j_{\alpha \alpha}(\underline{\widehat{\theta}}_{\beta})|} \right \}^{1/2},
$$
where $j_{\alpha \alpha}(\underline{\theta})$ is the observed Fisher information matrix and $j_{\alpha \alpha}(\underline{\theta})$ 
is the observed nuisance information matrix.  The quantity $l_{;V}(\underline{\theta})$ is known as the likelihood 
gradient, and  is defined as
$$
\underline{V} = (v_1,v_2) = \left . - \left ( \frac{\partial z(t;\underline{\theta})}{\partial t} \right )^{-1}
\left ( \frac{\partial z(t; \underline{\theta})}{\partial \underline{\theta}} \right ) \right |_{\underline{\widehat{\theta}}};
$$
it is a vector array with a vector pivotal quantity $\underline{z}(t;\underline{\theta})$ = 
$(z_1(t;\underline{\theta}), \cdots, z_n(t;\underline{\theta}))$.  The likelihood gradient becomes
$$
l_{;V} = \left \{ \frac{d}{d v_1} l(\underline{\theta};t), \frac{d}{d v_2} l(\underline{\theta};t) \right \}^{\top},
$$
where
$$
\frac{d}{dv_k} l(\underline{\theta};t) = \sum_{i=1}^n l_{t_i}(\underline{\theta};t) v_{ki}; \ \ \ k = 1,2,
$$
is the directional derivative of the log-likelihood function taken in the direction $\underline{v}_k = (v_{k1}, 
\cdots, v_{kn})$ on the data space with gradient $\ds l_{t_i}(\underline{\theta};t) = 
(\partial/\partial t_i)l(\underline{\theta};t)$, $i = 1, \cdots, n$.  Moreover,
$$
l_{\underline{\theta};V}(\underline{\theta}) = \frac{\partial l_{;V}(\underline{\theta})}{\partial \underline{\theta}}.
$$
It has been shown by Fraser et al. \cite{FRW:1999} that $r^*(\beta)$ is asymptotically distributed 
as $N(0,1)$, and with order of accuracy $O(n^{-3/2})$, and a 100(1-$\gamma$)\% confidence interval 
for $\beta$ based on $r^*(\beta)$ is then
$$
\{\beta: |r^*(\beta)| \le z_{\gamma/2}\}.
$$

Based on a complete data from the BS distribution, Wu and Wong \cite{WW:2004} constructed confidence intervals of $\alpha$ and $\beta$
by considering $z = (z_1, \cdots, z_n)$ as the pivotal quantity, since
$$
z_i = z_i(t_i;\underline{\theta}) = \frac{1}{\alpha} \left [ \left ( \frac{t_i}{\beta} \right )^{1/2} - 
\left ( \frac{t_i}{\beta} \right )^{-1/2} \right ], \ \ \ i = 1, \cdots, n,
$$
are distributed as N$(0,1)$.  These authors then performed an extensive Monte Carlo simulation study mainly for small sample 
sizes and from it, they observed that even for a sample of size 5, the coverage probabilities are quite close to the 
corresponding nominal values.

\section{\sc Bayes Estimations}

In the last two sections, we have discussed point and interval estimation of the parameters based on the frequentist 
approach.  In this section, we take on the Bayesian approach.

\subsection{\sc Achcar Estimators}

Achcar \cite{Achcar:1993} first considered the Bayesian inference of the parameters of a BS  
distribution.  He considered different non-informative priors on $\alpha$ and $\beta$ for developing the 
Bayesian inference.  The Jeffreys prior density for $\alpha$ and $\beta$ is given by
$$
\pi(\alpha,\beta) \propto \{\hbox{det } {\ve I}(\alpha,\beta)\}^{1/2},
$$
where ${\ve I}(\alpha,\beta)$ is the Fisher information matrix given in (\ref{fish-info}).  Considering 
Laplace approximation $E(T+\beta)^{-2} \approx 1/(4\beta)^2$, Jeffreys' non-informative prior takes on
the form
\be
\pi(\alpha,\beta) \propto \frac{1}{\alpha \beta} \left ( \frac{1}{\alpha^2} + \frac{1}{4} \right )^{1/2},
\ \ \ \alpha > 0, \beta > 0.     \label{prior-1} 
\ee
Another non-informative prior considered by Achcar \cite{Achcar:1993} is as follows:
\be
\pi(\alpha,\beta) \propto \frac{1}{\alpha \beta},  \ \ \ \alpha > 0, \beta > 0.    \label{prior-2}
\ee
Note that the prior in (\ref{prior-2}) can be obtained from the prior in (\ref{prior-1}) by assuming $H(\alpha^2) = 1$,
where 
\be
H(\alpha^2) = \left ( \frac{1}{\alpha^2} + \frac{1}{4} \right )^{-1/2}.   \label{h-func}
\ee
Based on the prior in (\ref{prior-1}), the posterior density of $\alpha$ and $\beta$ becomes
\be
\pi(\alpha,\beta|\hbox{data}) \propto \frac{\prod_{i=1}^n (\beta + t_i) exp\{-A(\beta)/\alpha^2\}}{\alpha^{n+1}
\beta^{(n/2)+1} H(\alpha^2)},   \label{posterior}
\ee
where $H(\alpha^2)$ is as defined in (\ref{h-func}) and 
$$
A(\beta) = \frac{n s}{2 \beta} + \frac{n \beta}{2 r} - n.
$$
Based on the prior in (\ref{prior-2}), the posterior density $\alpha$ and $\beta$ becomes the same as in (\ref{posterior}), 
with $H(\alpha^2)$ = 1.  Using Laplace approximation, Achcar \cite{Achcar:1993} then provided  approximate 
posterior density functions of $\alpha$, based on the priors in (\ref{prior-1}) and (\ref{prior-2}), to be 
$$
\pi(\alpha|\hbox{data}) \propto \alpha^{-(n+1)} (4+\alpha^2)^{1/2} \exp \left \{ - \frac{n}{\alpha^2} 
(\sqrt{s/r} - 1 ) \right \}
$$
and
$$
\pi(\alpha| \hbox{data}) \propto \alpha^{-n} \exp \left \{ - \frac{n}{\alpha^2} 
(\sqrt{s/r} - 1 ) \right \},
$$
respectively.  Similarly, the approximate posterior density functions of $\beta$, based on the priors in  
(\ref{prior-1})  and (\ref{prior-2}), are obtained
as 
$$
\pi(\beta| \hbox{data}) \propto \frac{\prod_{i=1}^n (\beta+t_i)\left \{4 + [2n/(n+2)][s/(2\beta) + 
\beta/(2r)-1]\right \}^{1/2}}{\beta^{n/2+1} \left \{s/(2\beta) + \beta/(2r) - 1\right \}^{(n+1)/2}}, 
\ \ \ \beta > 0,
$$
and
$$
\pi(\beta| \hbox{data}) \propto \frac{\beta^{-(n/2+1)} \prod_{i=1}^n(\beta+t_i)}{(s/(2\beta) + \beta/(2r)
-1)^{n/2}}, \ \ \  \beta > 0,
$$
respectively.  Evidently, the Bayes estimates of $\alpha$ and $\beta$ cannot be obtained in closed-form.
Achcar \cite{Achcar:1993} proposed to use the mode of the corresponding posterior density functions 
as Bayes estimates of the unknown parameters.  He noted that the posterior modes are quite close to 
the corresponding ML estimates of the parameters.

\subsection{\sc Xu and Tang Estimators}

More recently, Xu and Tang \cite{XT:2010} considered the Bayesian inference of the parameters based on
the prior in (\ref{prior-2}) and showed that the prior in (\ref{prior-2}) is the reference prior,
as introduced by Bernardo \cite{Bernado:1979} and further developed by Berger and Bernardo \cite{BB:1992}.  Based on Lindley's
\cite{Lindley:1980} method, the approximate Bayes estimates of $\alpha$ and $\beta$, with respect to squared error loss 
function, are obtained as
$$
\widehat{\alpha}_B = \widehat{\alpha} + \frac{1}{2} \left [ \left ( - \frac{\widehat{\alpha}}{2n}
+ \frac{3}{\widehat{\alpha} n^2} \sum_{i=1}^n \left ( \frac{t_i}{\widehat{\beta}} + 
\frac{\widehat{\beta}}{t_i} - 2 \right ) \right ) + \frac{\sum_{i=1}^n \widehat{\alpha} t_i}
{2 n^2 \widehat{\beta}(1 + \widehat{\alpha}(2 \pi)^{-1/2} H(\widehat{\alpha}))} \right ] - 
\frac{\widehat{\alpha}}{2n}
$$
and
\beanno
\widehat{\beta}_B & = & \widehat{\beta} + \frac{1}{2} \left ( - \frac{n}{\widehat{\beta}^3} + 
\sum_{i=1}^n \frac{2}{(t_i+\widehat{\beta})^3} + \sum_{i=1}^n \frac{3 t_i}{\widehat{\alpha}^2 
\widehat{\beta}^4} \right ) \left ( \frac{1 + \widehat{\alpha}(2 \pi)^{-1/2}H(\widehat{\alpha})}
{\widehat{\alpha}^2 \widehat{\beta}^2} \right )^2  \nonumber \\
& & + \frac{3 \sum_{i=1}^n \widehat{\beta}^2/t_i - \widehat{\beta}t_i}{4 n^2 \widehat{\beta}
(1 + \widehat{\alpha}(2 \pi)^{-1/2}H(\widehat{\alpha}))} - \frac{\widehat{\alpha}^3 \widehat{\beta}^2}
{2 n(1 + \widehat{\alpha}(2 \pi)^{-1/2}H(\widehat{\alpha}))},
\eeanno
respectively.  Here, $\widehat{\alpha}$ and $\widehat{\beta}$ are the ML estimators of $\alpha$ and $\beta$, 
respectively, and $h(\cdot)$ is 
as defined earlier in (\ref{h-func}).  Although Bayes estimates with respect to squared error loss function 
can be obtained using Lindley's approximation, the associated credible intervals cannot be obtained by the
same method.  So, Xu and Tang \cite{XT:2010} used Gibbs sampling procedure to generate posterior samples of
$\alpha$ and $\beta$ and based on those posterior samples, Bayes estimates and the associated credible 
intervals were obtained.  Simulation results showed that the Bayes estimates based on reference priors 
performed quite well even for small sample sizes.  Based on large scale simulations, it was also observed that 
the Bayes estimates based on Gibbs sampling technique performed better than the ML estimates as well as the approximate Bayes
estimates based on Lindley's approximation, and that the performance was satisfactory even for small sample
sizes.

%\section{\sc Estimation of the Change Point

\section{\sc Point and Interval Estimation Based on Censored Samples}

In this section, we discuss point and interval estimation of the parameters of a BS distribution when the 
data are Type-II and progressively Type-II censored.

\subsection{\sc Type-II Censoring}

Ng et al. \cite{NKB:2006} considered the point and interval estimation of the two-parameter BS 
distribution based on Type-II censored samples.  They mainly considered the ML estimators of the parameters 
and also used the observed Fisher information matrix to construct confidence intervals for the parameters.
Let $\{t_{1:n}, \dots, t_{r:n}\}$ be an ordered Type-II right censored
sample from $n$ units placed on a life-testing experiment.  Suppose the lifetime distribution 
of each unit follows the two-parameter BS distribution with PDF as in (\ref{bs-pdf}), and that the largest $(n-r)$
lifetimes have been censored.  Based on the observed Type-II right censored sample, the log-likelihood 
can be written as [see Balakrishnan and Cohen \cite{BC:1991}]
\bea
\ln (L) & = & \hbox{const.} + (n-r) \ln \left \{1 - \Phi \left [ \frac{1}{\alpha} \epsilon \left ( \frac{t_{r:n}}{\beta} 
\right ) \right ] \right \} - r \ln (\alpha) - r \ln (\beta)  \nonumber  \\
&  &  + \sum_{i=1}^r \epsilon' \left ( \frac{t_{i:n}}{\beta} \right ) - \frac{1}{2 \alpha^2} 
\sum_{i=1}^r \epsilon^2 \left ( \frac{t_{i:n}}{\beta} \right ),   \label{ll-censored-sample}
\eea
where $\epsilon(\cdot)$ and $\epsilon'(\cdot)$ are 
as defined earlier in (\ref{cubic-eq}).

The ML estimators of the parameters can then be obtained by differentiating the log-likelihood function in  
(\ref{ll-censored-sample}) with respect to the parameters $\alpha$ and $\beta$ and equating them to zero.  They 
cannot be obtained in explicit forms.  If we use the notation
$$
t^*_{i:n} = \frac{t_{i:n}}{\beta}, \ \ H(x) = \frac{\phi(x)}{\Phi(x)}, \ \ g(\alpha,\beta) = 
\frac{\alpha(n-r)}{r} H \left [ \frac{1}{\alpha} \epsilon(t_{r:n}^*) \right ], \ \ \ 
h_1(\beta) = \epsilon(t^*_{r:n}),  
$$
$$ 
h_2(\beta) = - \frac{1}{r} \sum_{i=1}^r \epsilon^2(t^*_{i:n}), \ \ 
h_3(\beta) = \left [ 1 + \sum_{i=1}^r \frac{t^*_{i:n} \epsilon''(t_{i:n}^*)}{\epsilon'(t_{i:n}^*)}
\right ]^{-1} t_{r:n}^* \epsilon'(t^*_{r:n}),
$$ 
$$
h_4(\beta) = \left [ 1 + \sum_{i=1}^r \frac{t^*_{i:n} \epsilon''(t_{i:n}^*)}{\epsilon'(t_{i:n}^*)}
\right ]^{-1} \left [ - \frac{1}{r} \sum_{i=1}^r t_{i:n}^* \epsilon(t^*_{i:n})t_{r:n}^* 
\epsilon'(t^*_{i:n}) \right ],
$$
$$
\psi^2(\beta) = \frac{h_2(\beta) h_3(\beta) - h_1(\beta) h_4(\beta)}{h_1(\beta) - h_3(\beta)}, \ \ \
u^* = \frac{1}{r} \sum_{i=1}^r t_{i:n}^*, \ \ \ v^* = \left [ \frac{1}{r} \sum_{i=1}^r (t_{i:n}^*)^{-1}
\right ]^{-1},
$$
$$
K^*(\beta) = \left [ \frac{1}{r} \sum_{i=1}^r (1+ t_{i:n}^*)^{-1} \right ]^{-1}, \ \ \ 
K^{'*}(\beta) = [K^*(\beta)]^2 \left [ \frac{1}{r} \sum_{i=1}^r (1+ t_{i:n}^*)^{-2} \right ],
$$
then the ML estimator of $\beta$ can be obtained as the unique solution of the non-linear 
equation
$$
Q(\beta) = \psi^2(\beta) \left [ \frac{1}{2} - \frac{1}{K^*(\beta)} \right ] - \frac{u^*}{2} 
+ \frac{1}{2 v^*} - \frac{\psi(\beta)(n-r)}{r} H \left [ \frac{\epsilon(t_{r:n}^*)}{\psi(\beta)}
\right ] t_{r:n}^* \epsilon'(t_{r:n}^*) = 0,
$$
and once $\widehat{\beta}$, the ML estimator of $\beta$, is obtained, the ML estimator of $\alpha$ can be obtained as
$$ 
\widehat{\alpha} = \psi(\widehat{\beta}); 
$$  
see Ng et al. \cite{NKB:2006} for pertinent details. Balakrishnan and Zhu \cite{BZ:2013} showed that for $r$ = 1,
the ML estimators of $\alpha$ and $\beta$ do not exist.  For $n > r > 1$, the ML estimators of $\alpha$ and $\beta$ may not
always exist, and that if they do exist, they are unique.  They provided some sufficient conditions for the existence of 
ML estimators of the parameters in this case.

For the construction of confidence intervals, Ng et al. \cite{NKB:2006} used the observed Fisher information matrix 
and the standard asymptotic properties of the ML estimators.  They also performed extensive Monte Carlo simulations to check
the performance of their  method.  They observed that the ML estimator of $\alpha$ is especially biased, and 
if the sample size is small, it is of the form
\be
\hbox{Bias}(\widehat{\alpha}) \approx - \frac{\alpha}{n} \left [ 1 + 2.5 \left ( 1 - \frac{r}{n} \right )
\right ].
\ee
Bias-correction method, as proposed in Section \ref{section-interval-estimation}, has been 
suggested, and it has been observed that the biased-corrected ML estimators, as well as the confidence intervals 
based on biased-corrected ML estimators, both  perform quite well even for small 
sample sizes.  Barreto et al. \cite{BCC:2013} provided improved Birnbaum-Saunders estimators under Type-II censoring.
Similar methods may be adopted for Type-I and hybrid censoring schemes as well.  Recently, 
Balakrishnan and Zhu \cite{BZ:2014b} discussed the existence and uniqueness of the ML estimators of the parameters $\alpha$ and $\beta$ of 
BS distribution in the case of Type-I, Type-II and hybrid censored samples.

\subsection{\sc Progressive Censoring}

Although Type-I and Type-II censoring schemes are the most popular censoring schemes, progressive 
censoring scheme has received considerable attention in the past two decades.  Progressive censoring scheme was first 
introduced by Cohen \cite{Cohen:1963}, but it became very popular since the publication of the book 
by Balakrishnan and Aggarwala \cite{BA:2000}.  A progressive censoring scheme 
can be briefly described as follows.  For a given $n$ and $m$, choose $m$ non-negative integers, 
$R_1, \dots, R_m$, such that 
$$
R_1 + \dots + R_m = n - m.
$$ 
Now, consider the following experiment.  Suppose $n$ identical units are placed on a life-test.  At the time of the 
first failure, say $t_{1:m:n}$, $R_1$ units from the $(n-1)$ surviving units are chosen at random and 
removed.  Similarly, at the time of the second failure, say $t_{2:m:n}$, $R_2$ units from the  
$n-R_1-1$ surviving units are removed, and so on.  Finally, at the time of the $m$-th failure, say $t_{m:m:n}$, 
all remaining  $R_m$ surviving units are removed, and the experiment terminates at $t_{m:m:n}$.  For more elaborate 
details on progressive censoring, the readers are referred to the discussion article by Balakrishnan \cite{Bala:2007} and the recent 
book by Balakrishnan and Cramer \cite{BC:2014}.  

Pradhan and Kundu \cite{PK:2013} considered the inference for the parameters $\alpha$ and $\beta$ of the BS  
distribution when the data are Type-II progressively censored.  
In a Type-II 
progressive censoring experiment, the data are of the form $\ds \{(t_{1:m:n},R_1), \dots, (t_{m:m:n},R_m)\}$.
Based on such a progressively censored data, without the additive constant, the log-likelihood function 
of the observed data can be written as
$$
l(\alpha,\beta|\hbox{data}) = \sum_{i=1}^m 
\{ \ln (f_T(t_{i:m:n}; \alpha,\beta)) + R_i \ln (\Phi(-g(t_{i:m:n};\alpha,\beta)))\},
$$
where $f_T(\cdot)$ is the PDF of the BS distribution given in (\ref{bs-pdf}) and
$$
g(t;\alpha,\beta) = \frac{1}{\alpha} \left \{ \left ( \frac{t}{\beta} \right )^{1/2} - \left ( 
\frac{\beta}{t} \right )^{1/2} \right \}. 
$$
As expected, the ML estimators of the parameters cannot be obtained in closed-form.  Using the property that a
BS distribution can be written as a mixture of an inverse Gaussian distribution and its reciprocal
as mentioned earlier in Section 3.3, Pradhan and Kundu \cite{PK:2013} provided an EM algorithm which can be used quite 
effectively to compute the ML estimates of the parameters.  It is observed that in each `E-step' of the algorithm,
the corresponding `M-step' can be obtained in an explicit form.  Moreover, by using the method of Louis \cite{Louis:1982}, 
the confidence intervals of the parameters can also be obtained conveniently.

\section{\sc Other Univariate Issues}

In this section we will discuss some of the other issues related to the univariate BS distribution.  First, we will consider 
the stochastic orderings of BS classes of distribution functions under different conditions.  Then, we will discuss 
mixture of two BS distributions, its properties and different inferential aspects.

\subsection{\sc Stochastic Orderings}

Stochastic orderings have been studied rather extensively for various lifetime distributions; see, for example, the book
by Shaked and Shantikumar \cite{SS:2007}.  However, very little work seems to have been done for the BS model in this regard with the first
work possibly by Fang et al. \cite{FZB:2016}.  By considering two independent and non-identically distributed BS random variables,
they established stochastic ordering results for parallel and series systems.  Specifically, let $X_i \sim$ BS$(\alpha_i, \beta_i)$,
independently, for $i = 1, \ldots, n$.  Further, let $X_i^* \sim$ BS$(\alpha_i^*, \beta_i^*)$, independently, for $i = 1, \ldots, n$. 
Then, Fang et al. \cite{FZB:2016} established that when $\alpha_1 = \ldots = \alpha_n = \alpha_1^* = \ldots = \alpha_n^*$, 
$(\beta_1, \ldots, \beta_n) \ge_{m} (\beta_1^*, \ldots, \beta_n^*)$ implies $X_{n:n} \ge_{\hbox{st}} X_{n:n}^*$.  Here, $X \ge_{\hbox{st}} Y$ denotes 
that $Y$ is smaller than $X$ in the usual stochastic order meaning $S_X(x) \ge S_Y(x)$ for all $x$, where $S_X(\cdot)$ and 
$S_Y(\cdot)$ denote the survival functions of $X$ and $Y$, respectively.  Also, if $(\lambda_1, \ldots, \lambda_n)$ and 
$(\lambda_1^*, \ldots, \lambda_n^*)$ are two real vectors and $\lambda_{[1]} \ge \ldots \ge \lambda_{[n]}$ and 
$\lambda_{[1]}^* \ge \ldots \ge \lambda_{[n]}^*$ denote their ordered components, then $(\lambda_1, \ldots, \lambda_n) \ge_m 
(\lambda_1^*, \ldots, \lambda_n^*)$ denotes that the second vector is majorized by the first vector meaning 
$\ds \sum_{i=1}^k \lambda_{[i]} \ge \sum_{i=1}^k \lambda_{[i]}^*$ for $k = 1, \ldots, n-1$ and $\ds \sum_{i=1}^n \lambda_{[i]} = 
\sum_{i=1}^n \lambda_{[i]}^*$.  Similarly, Fang et al. \cite{FZB:2016} established that when $\alpha_1 = \ldots = \alpha_n = 
\alpha_1^* = \ldots = \alpha_n^*$, $\ds \left ( 1/\beta_1, \ldots, 1/\beta_n \right ) 
\ge_m \left ( 1/\beta_1^*, \ldots, 1/\beta_n^* \right )$ implies $X_{1:n}^* \ge_{\hbox{st}} X_{1:n}$.  They also further proved
that when $\beta_1 = \ldots = \beta_n = \beta_1^* = \ldots = \beta_n^*$, 
$\ds \left ( 1/\alpha_1, \ldots, 1/\alpha_n \right ) 
\ge_m \left ( 1/\alpha_1^*, \ldots, 1/\alpha_n^* \right )$ implies $X_{n:n} \ge_{\hbox{st}} X_{n:n}^*$ and $X_{1:n}^* 
\ge_{\hbox{st}} X_{1:n}$.

There are several open problems in this direction; for example, are there such results for other order statistics, and for other 
stochastic orderings like hazard rate ordering and likelihood-ratio ordering?

\subsection{\sc Mixture of Two BS Distributions} 

Balakrishnan et al. \cite{BGKLS:2010} studied the properties of a mixture of two BS  
distributions and then discussed inferential issues for the parameters of such a mixture model.  The mixture of two 
BS distributions has its PDF as
\be
f_Y(t; \alpha_1,\beta_1,\alpha_2,\beta_2) = p f_{T_1}(t; \alpha_1,\beta_1) 
+ (1-p) f_{T_2}(t; \alpha_2,\beta_2), \ \ \ \  t > 0,   \label{mtbs-pdf}
\ee
where $T_i \sim$ BS$(\alpha_i, \beta_i)$, $f_{T_i}(\cdot)$ denotes the PDF of $T_i$, for $i$ = 1, 2, 
and $0 \le p \le 1$.
The distribution of the random variable $Y$, with PDF in (\ref{mtbs-pdf}), is known as the mixture 
distribution of two BS models (MTBS), and is denoted by MTBS$(\alpha_1,\beta_1,\alpha_2,
\beta_1,p)$.  Although  it has not been formally proved, it has been observed graphically that the PDF and HF of 
$Y$ can be either unimodal or bimodal depending on the parameter values.   Note that the first two raw moments of $Y$ can be easily obtained as 
\beanno
\hbox{E}(Y) & = & p \beta_1 \left [ 1 + \frac{1}{2} \alpha_1^2 \right ] + (1-p) \beta_2 \left [ 1 + 
\frac{1}{2} \alpha_2^2 \right ],  \\
\hbox{E}(Y^2) & = & \frac{3}{2} p \beta_1^2 \left [ \alpha_1^4 + 2 \alpha_1^2 + \frac{2}{3} \right ] + 
\frac{3}{2} (1-p) \beta_2^2 \left [ \alpha_2^4 + 2 \alpha_2^2 + \frac{2}{3} \right ].
\eeanno
The higher-order moments also can be obtained involving some infinite series expressions.
Moreover, if $Y \sim$ MTBS$(\alpha_1,\beta_1,\alpha_2,\beta_2,p)$, then $1/Y \sim$ MTBS$(\alpha_1,1/\beta_1,\alpha_2,1/\beta_2,p)$.  
Therefore, the inverse moments of $Y$ also can be obtained as done earlier.  

We shall now discuss the estimation of the parameters of MTBS$(\alpha_1,\beta_1,\alpha_2,\beta_2,p)$
based on a random sample $\{y_1,\cdots, y_n\}$ from $Y$.  The ML estimators of the parameters can be 
obtained by solving a five-dimensional optimization problem.  To avoid that, Balakrishnan et al. 
\cite{BGKLS:2010} suggested the use of the EM algorithm, which can be implemented rather easily as follows.  Let us rewrite the PDF of $Y$ 
in (\ref{mtbs-pdf}) as
$$
f_Y(y;\mu_1,\lambda_1,\mu_2,\lambda_2,p_1,p_2) = \frac{1}{2} \sum_{j=1}^2 p_j 
f_{X_1}(y;\mu_j,\lambda_j) + \frac{1}{2} \sum_{j=1}^2 p_j 
f_{X_2}(y;\mu_j,\lambda_j),
$$ 
where $\ds f_{X_1}(y;\mu_j,\lambda_j)$ is defined in (\ref{pdf-ig}), 
$\ds f_{X_2}(y;\mu_j,\lambda_j) = y f_{X_1}(y;\mu_j,\lambda_j)/\mu_j$ as defined in Section \ref{section-ig}, $\mu_j = \beta_j$, $\lambda_j = \beta_j/\alpha_j^2$, for $j$ = 1, 2, $p_1 = p$ and $p_2 = 1-p$.  To 
implement the EM algorithm, we may treat this as a missing value problem. For the random
variable $Y$, define an associated random vector $W = (U_1,V_1,U_2,V_2)$ as follows.  Here, each $U_j$
and $V_j$ can take on values 0 or 1, with $\ds \sum_{j=1}^2 (U_j+V_j)$ = 1, where $P(U_j = 1) = 
P(V_j = 1) = p_j/2$, for $j = 1, 2$.  Further, $Y|(U_j=1)$ and $Y|(V_j=1)$ have densities 
$f_{X_1}(\cdot; \mu_j,\lambda_j)$ and $f_{X_2}(\cdot;\mu_j\lambda_j)$, respectively, for $j = 1,2$.  From 
these specifications, it is possible to get uniquely the joint distribution of $(Y,W)$.  Now, suppose 
we have the complete observations $(y_i,w_i)$, where $w_i = (u_{i1},v_{i1},u_{i2},v_{i2}),$ for $i = 1,
\cdots, n$.  Then, with ${\underline{\theta}} = (\mu_1,\lambda_1,p_1,\mu_2,\lambda_2,p_2)$ being the parameter vector,
the complete data log-likelihood function is given by
\bea
l^{(c)}(\ul \theta|y,w) = c & + & \sum_{j=1}^2(u_{.j} + v_{.j})\log(p_j) + \sum_{i=1}^n \sum_{j=1}^2 u_{ij}
\log(f_{X_1}(y_i;\mu_j,\lambda_j)) \nonumber  \\  & + & \sum_{i=1}^n \sum_{j=1}^2 v_{ij}
\log(f_{X_2}(y_i;\mu_j,\lambda_j)),   \label{com-ll}
\eea
where $u_{.j} = \sum_{i=1}^n u_{ij}$, $v_{.j} = \sum_{i=1}^n v_{ij}$, and $c$ is a constant independent
of the unknown parameters.  Based on the complete data log-likelihood function in (\ref{com-ll}), it 
can be shown that the ML estimators of the model parameters are as follows:
$$ 
\widehat{p}_j = \frac{u_{.j} + v_{.j}}{n}, \ \ \ \ \ j = 1,2,
$$
and
\beanno
\widehat{\mu}_j & = & \frac{B_j(A_j-B_j) + \sqrt{(B_j(A_j-B_j))^2 + A_jC_jD_j(2B_j-A_j)}}{D_jA_j},  \\
\widehat{\lambda}_j & = & \frac{[u_{.j} + v_{.j}] \widehat{\mu}_j^2}{\sum_{i=1}^n (u_{ij}+v_{ij})(y_i-
\mu_j)^2/y_i}.
\eeanno
Here, $A_j = v_{.j}$, $B_j = (u_{.j}+v_{.j})/2$, $\ds C_j = \sum_{i=1}^n y_i(u_{ij}+v_{ij})y_i/2$ and
$\ds D_j = \sum_{i=1}^n y_i(u_{ij}+v_{ij})/(2y_i)$.  Thus, in the case of complete data, the ML estimators can be 
expressed in explicit forms.  The existence of the explicit expressions of the ML estimators simplifies the 
implementation of the EM algorithm, and also makes it computationally efficient.  Now, the EM algorithm
can be described very easily.  Suppose, at the $m$-th stage of the EM algorithm, the estimate of the 
unknown parameter ${\ul \theta}$ is ${\ul \theta}^{(m)} = (\mu_1^{(m)},\lambda_1^{(m)},p_1^{(m)},
\mu_2^{(m)},\lambda_2^{(m)},p_2^{(m)})$.  Before proceeding further, we need to introduce the following notation:
$$
a_{ij}^{(m)} = \hbox{E}(U_{ij}|y,{\ul \theta}^{(m)}) \ \ \ \ \hbox{and} \ \ \  
b_{ij}^{(m)} = \hbox{E}(V_{ij}|y,{\ul \theta}^{(m)}).
$$
Note here that $a_{ij}^{(m)}$ and $b_{ij}^{(m)}$ are the usual posterior probabilities associated with 
the $i$-th observation, which are given by
\beanno
a_{ij}^{(m)} & = & \frac{p_j^{(m)} f_{X_1}(y_i;\mu_j^{(m)},\lambda_j^{(m)})}
{\sum_{l=1}^2 p_l^{(m)} f_{X_1}(y_i;\mu_l^{(m)},\lambda_l^{(m)}) + 
\sum_{l=1}^2 p_l^{(m)} f_{X_2}(y_i;\mu_l^{(m)},\lambda_l^{(m)})},   \\
b_{ij}^{(m)} & = & \frac{p_j^{(m)} f_{X_2}(y_i;\mu_j^{(m)},\lambda_j^{(m)})}
{\sum_{l=1}^2 p_l^{(m)} f_{X_1}(y_i;\mu_l^{(m)},\lambda_l^{(m)}) + 
\sum_{l=1}^2 p_l^{(m)} f_{X_2}(y_i;\mu_l^{(m)},\lambda_l^{(m)})}.   
\eeanno
Therefore, at the $m$-th iteration, the $E$-step of the EM algorithm can be obtained by computing  
the pseudo log-likelihood function as follows:
\be
l^{(ps)}({\ul \theta}|y,{\ul \theta}^{(m)})  =  \sum_{i=1}^n \sum_{j=1}^2 a_{ij}^{(m)} \log \left ( 
\frac{p_j}{2} f_{X_1}(y_i;\mu_j,\lambda_j) \right ) + 
\sum_{i=1}^n \sum_{j=1}^2 b_{ij}^{(m)} \log \left ( 
\frac{p_j}{2} f_{X_2}(y_i;\mu_j,\lambda_j) \right ).   \label{pseudo-ll}
\ee
The $M$-step of the EM algorithm involves the maximization of (\ref{pseudo-ll}) with respect to the unknown 
parameters, and for $j$ = 1 and 2, they are obtained as follows:
\beanno
\widehat{p}_j^{(m+1)} & = & \frac{a_{.j}^{(m)} + b_{.j}^{(m)}}{n},     \\
\widehat{\mu}_j^{(m+1)} & = & \frac{B_j^{(m)}(A_j^{(m)}-B_j^{(m)}) + 
\sqrt{(B_j^{(m)}(A_j^{(m)}-B_j^{(m)}))^2 + A_j^{(m)}C_j^{(m)}
D_j^{(m)}(2B_j^{(m)}-A_j^{(m)})}}{D_j^{(m)}A_j^{(m)}},  \\
\widehat{\lambda}_j^{(m+1)} & = & \frac{[a_{.j}^{(m)} + v_{.j}^{(m)}] [\widehat{\mu}_j^{(m+1)}]^2}
{\sum_{i=1}^n (a_{ij}^{(m)}+b_{ij}^{(m)})(y_i-
\mu_j^{(m+1)})^2/y_i}.
\eeanno
Here, $\ds a_{.j}^{(m)} = \sum_{i=1}^n a_{ij}^{(m)}$, $\ds b_{.j}^{(m)} = \sum_{i=1}^n b_{ij}^{(m)}$, $A_j^{(m)} 
= b_{.j}^{(m)}$, $B_j^{(m)} = (a_{.j}^{(m)}+b_{.j}^{(m)})/2$, $\ds C_j^{(m)} = \sum_{i=1}^n y_i(a_{ij}^{(m)}+
b_{ij}^{(m)})y_i/2$, and
$\ds D_j^{(m)} = \sum_{i=1}^n y_i(a_{ij}^{(m)}+b_{ij}^{(m)})/(2y_i)$.

\section{\sc Bivariate BS Distribution}

Until now, we have focused only on univariate BS distribution.  In this section, we introduce the bivariate BS (BVBS) distribution, discuss
several properties and address associated inferential issues.

\subsection{\sc PDF and CDF} 

Kundu et al. \cite{KBJ:2010} introduced the BVBS  distribution by using the same
idea as in the construction of the univariate BS distribution.  The bivariate random vector $(T_1, T_2)$ is said to 
have a BVBS, with parameters $\alpha_1 > 0$, $\beta_1 > 0$, $\alpha_2 > 0$, 
$\beta_2 > 0$, $-1 < \rho < 1$, if the joint cumulative distribution function of $T_1$ and $T_2$ can 
be expressed as
$$
F_{T_1, T_2}(t_1, t_2) = P(T_1 \le t_1, T_2 \le t_2) = \Phi_2 \left [ \frac{1}{\alpha_1} \left ( \sqrt{\frac{t_1}{\beta_1}} - 
\sqrt{\frac{\beta_1}{t_1}} \right ),
 \frac{1}{\alpha_2} \left ( \sqrt{\frac{t_2}{\beta_2}} - \sqrt{\frac{\beta_2}{t_2}} \right ); \rho \right ]
      %\label{joint-cdf}
$$
for $t_1 > 0$, $t_2 > 0$, where $\Phi_2(u, v; \rho)$ is the cumulative distribution function of a standard 
bivariate normal vector $(Z_1, Z_2)$ with correlation coefficient $\rho$.  The corresponding joint 
PDF is 
\beanno
f_{T_1, T_2}(t_1, t_2) & = & \phi_2 \left ( \frac{1}{\alpha_1} \left ( \sqrt{\frac{t_1}{\beta_1}} - 
\sqrt{\frac{\beta_1}{t_1}} \right ), \frac{1}{\alpha_2} \left ( \sqrt{\frac{t_2}{\beta_2}} - 
\sqrt{\frac{\beta_2}{t_2}} \right ); \rho \right )   \nonumber  \\
&  & \ \ \ \times \frac{1}{2 \alpha_1 \beta_1} \left \{ \left ( \frac{\beta_1}{t_1} \right )^{1/2} + 
\left ( \frac{\beta_1}{t_1} \right )^{3/2} \right \} \times \frac{1}{2 \alpha_2 \beta_2} 
\left \{ \left ( \frac{\beta_2}{t_2} \right )^{1/2} + 
\left ( \frac{\beta_2}{t_2} \right )^{3/2} \right \},    %\label{joint-pdf}
\eeanno
where $\phi_2(u, v; \rho)$ denotes the joint PDF of $Z_1$ and $Z_2$ given by
\be
\phi_2(u, v, \rho) = \frac{1}{2 \pi \sqrt{1 - \rho^2}} \exp \left \{- \frac{1}{2(1-\rho^2)}(u^2 + v^2 - 
2 \rho u v) \right \}.  \label{bv-nor-pdf}
\ee
From here on, we shall denote this distribution by BS$_2(\alpha_1, \beta_1, \alpha_2, \beta_2; \rho)$.  As expected, the 
joint PDF can take on different shapes depending on the shape parameters $\alpha_1$ and $\alpha_2$
and the correlation parameter $\rho$.  The surface plots of $\ds f_{T_1, T_2}(t_1, t_2)$, for different parameter
values, can be found in Kundu et al. \cite{KBJ:2010}.  It has been observed that the joint PDF of the BS distribution is 
unimodal.  It is an absolutely continuous distribution and due to its flexibility, it can be used quite  
effectively to analyze bivariate data provided there are no ties.  

\subsection{\sc Copula Representation}  

Recently, Kundu and Gupta \cite{KG:2016} observed that the CDF of BS$_2(\alpha_1, \beta_1, \alpha_2, \beta_2; \rho)$ has the 
following copula representation:
$$
F_{T_1, T_2}(t_1, t_2) = C_G(F_{T_1}(t_1; \alpha_1, \beta_1), F_{T_2}(t_2; \alpha_2, \beta_2); \rho).
$$
Here, $F_{T_1}(t_1)$ and $F_{T_2}(t_2)$ denote the CDFs of BS$(\alpha_1, \lambda_1)$ and BS$(\alpha_2, \lambda_2)$, respectively, 
and $C_G(u, v; \rho)$ denotes the Gaussian copula defined by, for $0 \le u, v \le 1$, 
\be
C_G(u, v; \rho) = \int_{-\infty}^{\Phi^{-1}(u)} \int_{-\infty}^{\Phi^{-1}(v)} \phi_2(x,y; \rho) \hbox{d}x \hbox{d}y = \Phi_2(\Phi^{-1}(u), \Phi^{-1}(v); \rho).
\label{bs-copula}
\ee
Using the above copula structure in (\ref{bs-copula}), it has been shown that a BVBS distribution has a total positivity of order two 
(TP$_2$) property if $\rho > 0$ and reverse rule of order two (RR$_2$) property for $\rho < 0$ for all values of $\alpha_1, \beta_1, 
\alpha_2$ and $\beta_2$.  It has also been shown, using the above copula structure, that for a 
BS$_2(\alpha_1, \beta_1, \alpha_2, \beta_2; \rho)$, for all values of $\alpha_1, \beta_1, \alpha_2, \beta_2$, the Blomqvist's beta 
($\beta$), Kendall's tau ($\tau$) and Spearman's rho ($\rho_S$) become
$$
\beta = \frac{2}{\pi} \hbox{arcsin}(\rho), \ \ \tau = \frac{2}{\pi} \hbox{arcsin}(\rho), \ \ \
\rho_S = \frac{6}{\pi} \hbox{arcsin} \left (\frac{\rho}{2} \right ),
$$
respectively.

\subsection{\sc Generation and Other Properties}  \label{generation-bs}

It is very simple to generate bivariate BS distribution using univariate normal random number 
generators.  The following algorithm has been 
suggested by Kundu et al. \cite{KBJ:2010} to generate bivariate BS distribution.

\noindent  Algorithm:

\noindent Step 1: Generate independent $U_1$ and $U_2$ from $N(0, 1)$;  

\noindent Step 2: Compute
$$
Z_1 = \frac{\sqrt{1 + \rho} + \sqrt{1 - \rho}}{2} U_1 + \frac{\sqrt{1 + \rho} - \sqrt{1 - \rho}}{2} U_2,
$$
$$
Z_2 = \frac{\sqrt{1 + \rho} - \sqrt{1 - \rho}}{2} U_1 + \frac{\sqrt{1 + \rho} + \sqrt{1 - \rho}}{2} U_2;
$$

\noindent Step 3: Set
$$
T_i = \beta_i \left [ \frac{1}{2} \alpha_i Z_i + \sqrt{\left ( \frac{1}{2} \alpha_i Z_i \right )^2 + 1}
\right ]^2, 
\ \ \ \hbox{for} \ \ \ i = 1, 2.
$$

Several properties of the bivariate BS distribution have been established by these authors.
It has been observed that if $(T_1, T_2) \sim$ BS$_2(\alpha_1, \beta_1, \alpha_2, \beta_2; \rho)$,
then $T_1 \sim$ BS($\alpha_1, \beta_1$) and $T_2 \sim$ BS($\alpha_2, \beta_2$).  $T_1$ and $T_2$ are 
independent if and only if $\rho$ = 0.  The correlation between $T_1$ and $T_2$ ranges from -1 to 1, depending
on the value of $\rho$.  Using the joint and marginal PDFs, the conditional PDF of $T_1$, given $T_2$, can be 
easily obtained.  Different moments and product moments can be obtained involving infinite series expressions.  
It has also been observed that for $\rho > 0$ and for all values of $\alpha_1, \beta_1, \alpha_2, \beta_2$, $T_1$ ($T_2$) 
is stochastically increasing in $T_2$ ($T_1$).  If $(h_1(t_1, t_2), h_2(t_1, t_2))$ denotes the bivariate hazard rate of Johnson 
and Kotz \cite{JK:1975}, then for $\rho > 0$ and for fixed $t_2$ ($t_1$), $h_1(t_1, t_2) (h_2(t_1, t_2))$ is an unimodal function of 
$t_1$ ($t_2$).  It may also be noted that for all values of $\alpha_1, \alpha_2$ and $\rho$, if $\beta_1 = \beta_2$, then the stress-strength
parameter $R = P(T_1 < T_2) = 1/2$.  Different reliability properties of a BVBS distribution can be found in Gupta \cite{Gupta:2015}.

\subsection{\sc ML and MM Estimators}

We shall now discuss the estimation of the parameters $\alpha_1, \beta_1, \alpha_2, \beta_2$ and $\rho$ 
based on a random sample, say 
$\{(t_{1i}, t_{2i}), i = 1, \cdots, n\}$, of size $n$ from BS$_2(\alpha_1, \beta_1, \alpha_2, \beta_2; \rho)$.  
Based on the random sample, the log-likelihood function is given by 
\bea
l(\underline{\theta}) & = & - n \ln (\alpha_1) - n \ln (\beta_1) - n \ln (\alpha_2) - n \ln (\beta_2) - \frac{n}{2} \ln (1 - \rho^2)
+ \sum_{i=1}^n \ln \left \{ \left ( \frac{\beta_1}{t_{1i}} \right )^{1/2} + 
 \left ( \frac{\beta_1}{t_{1i}} \right )^{3/2} \right \}  \nonumber  \\
&  & + \sum_{i=1}^n \ln \left \{ \left ( \frac{\beta_2}{t_{2i}} \right )^{1/2} + 
 \left ( \frac{\beta_2}{t_{2i}} \right )^{3/2} \right \} - \frac{1}{2(1-\rho^2)} \left \{ 
\sum_{i=1}^n \frac{1}{\alpha_1^2} \left ( \left ( \frac{t_{1i}}{\beta_1} \right )^{1/2} - 
\left ( \frac{\beta_1}{t_{1i}} \right )^{1/2} \right )^2  \right .  \nonumber \\
&  & \left . + \sum_{i=1}^n \frac{1}{\alpha_2^2} \left ( \left ( \frac{t_{2i}}{\beta_2} \right )^{1/2} - 
\left ( \frac{\beta_2}{t_{2i}} \right )^{1/2} \right )^2 - 
\frac{2 \rho}{\alpha_1 \alpha_2} \sum_{i=1}^n  \left ( \frac{t_{2i}}{\beta_2} \right ) 
\left ( \frac{\beta_2}{t_{2i}} \right )  \right \}.   \label{biv-ll}
\eea
The ML estimators of the parameters can be obtained by maximizing (\ref{biv-ll}) with respect to $\alpha_1$, $\alpha_2$, $\beta_1$, 
$\beta_2$ and $\rho$.  Evidently, the ML estimators cannot be obtained in explicit form.  From the observation that
$$
\left \{ \left ( \sqrt{\frac{T_1}{\beta_1}} - \sqrt{\frac{\beta_1}{T_1}} \right ), 
\left ( \sqrt{\frac{T_2}{\beta_2}} - \sqrt{\frac{\beta_2}{T_2}} \right ) \right \} \sim 
N_2 \left ( \left ( \matrix{0 \cr \cr 0} \right ), \left ( \matrix{\alpha_1^2 & \alpha_1 \alpha_2 \rho \cr
&  \cr
\alpha_1 \alpha_2 \rho & \alpha_2^2 \cr} \right ) \right ),
$$
it is clear that for fixed $\beta_1$ and $\beta_2$, the ML estimators of $\alpha_1$, $\alpha_2$ and $\rho$ are
$$
\widehat{\alpha}_i = \left ( \frac{s_i}{\beta_i} + \frac{\beta_i}{r_i} - 2 \right )^{1/2}, \ \ \ i = 1, 2,
$$ 
and
$$
\widehat{\rho}(\beta_1, \beta_2) = \frac{\sum_{i=1}^n  \left ( \left ( \frac{t_{1i}}{\beta_1} \right )^{1/2} - 
\left ( \frac{\beta_1}{t_{1i}} \right )^{1/2} \right )
\left ( \left ( \frac{t_{2i}}{\beta_2} \right )^{1/2} - 
\left ( \frac{\beta_2}{t_{2i}} \right )^{1/2} \right )  }
{\sum_{i=1}^n  \left ( \left ( \frac{t_{1i}}{\beta_1} \right )^{1/2} - 
\left ( \frac{\beta_2}{t_{2i}} \right )^{1/2} \right )^2
\left ( \left ( \frac{t_{1i}}{\beta_1} \right )^{1/2} - 
\left ( \frac{\beta_2}{t_{2i}} \right )^{1/2} \right )^2},
$$   
where 
$$
s_i = \frac{1}{n} \sum_{k=1}^n t_{ik} \ \ \ \hbox{and} \ \ \
r_i = \left [ \frac{1}{n} \sum_{k=1}^n t_{ik}^{-1} \right ]^{-1}, \ \ \ \hbox{for} \ \ \ 
i = 1, 2.
$$ 
It may be observed that $\widehat{\alpha}_1(\beta_1, \beta_2)$ is a function of $\beta_1$ only, while 
 $\widehat{\alpha}_2(\beta_1, \beta_2)$ is a function of $\beta_2$ only.  The ML estimators of $\beta_1$ and $\beta_2$
can then be obtained by maximizing the profile log-likelihood function of $\beta_1$ and $\beta_2$ given by
\beanno
l_{\hbox{profile}}(\beta_1, \beta_2) & = & l(\widehat{\alpha}_1(\beta_1), \beta_1, \widehat{\alpha}_2(\beta_2), 
\beta_2, \widehat{\rho}(\beta_1, \beta_2))    \\
& = & - n \ln (\widehat{\alpha}_1(\beta_1)) - n \ln (\beta_1) - 
n \ln (\widehat{\alpha}_2(\beta_2)) - n \ln (\beta_2) - \frac{n}{2} \ln (1 - \widehat{\rho}^2(\beta_1, \beta_2))
  \\
& & + 
\sum_{i=1}^n \ln \left \{ \left ( \frac{\beta_1}{t_{1i}} \right )^{1/2} + \left ( \frac{\beta_1}{t_{1i}}
\right )^{3/2} \right \} + 
\sum_{i=1}^n \ln \left \{ \left ( \frac{\beta_2}{t_{2i}} \right )^{1/2} + \left ( \frac{\beta_2}{t_{2i}}
\right )^{3/2} \right \}.  
\eeanno
Clearly, no explicit solutions exist, and so some numerical techniques like Newton-Raphson algorithm or some of its
variants need to be used for computing the ML estimators of $\beta_1$ and $\beta_2$.  Once the ML estimators of $\beta_1$ and $\beta_2$
are obtained, the ML estimators of $\alpha_1$, $\alpha_2$ and $\rho$ can be easily obtained.  The explicit expression 
of the Fisher information matrix, although complicated, has been provided by Kundu et al. \cite{KBJ:2010},
which can be used to obtain the asymptotic variance-covariance matrix of the ML estimators.

Since the ML estimators do not have explicit forms, Kundu et al. \cite{KBJ:2010} proposed modified moment 
estimators which do have explicit expressions.  Using the same idea as in the case of univariate 
modified moment estimators, in the bivariate case, the modified moment estimators take on the following forms:
$$ 
\widetilde{\alpha}_i = \left \{ 2 \left [ \left ( \frac{s_i}{r_i} \right )^{1/2} - 1 \right ] \right \}^{1/2}
\ \ \ \ \hbox{and} \ \ \ \widetilde{\beta}_i = (s_i r_i)^{1/2}, \ \ \ \hbox{for} \ \ \ i = 1,2,
$$
and
$$ 
\widetilde{\rho} = \frac{\sum_{i=1}^n  \left ( \left ( \frac{t_{1i}}{\widetilde{\beta}_1} 
\right )^{1/2} - 
\left ( \frac{\widetilde{\beta}_1}{t_{1i}} \right )^{1/2} \right )
\left ( \left ( \frac{t_{2i}}{\widetilde{\beta}_2} \right )^{1/2} - 
\left ( \frac{\widetilde{\beta}_2}{t_{2i}} \right )^{1/2} \right )  }
{\sum_{i=1}^n  \left ( \left ( \frac{t_{1i}}{\widetilde{\beta}_1} \right )^{1/2} - 
\left ( \frac{\widetilde{\beta}_2}{t_{2i}} \right )^{1/2} \right )^2
\left ( \left ( \frac{t_{1i}}{\widetilde{\beta}_1} \right )^{1/2} - 
\left ( \frac{\widetilde{\beta}_2}{t_{2i}} \right )^{1/2} \right )^2}.
$$   
Since the modified moment estimators have explicit forms, they can be used as initial values in the numerical computation of  
the ML estimators.  Kundu et al. \cite{KBJ:2010} performed some simulations to investigate the performance of 
the ML estimators for the bivariate BS distribution.  It has been observed by these authors that even for  
small sample sizes, say 10, the ML estimators of all the parameters are almost unbiased.  Interestingly, the bias
and the mean squared errors of the ML estimators do not depend on the true value of $\rho$.

Kundu et al. \cite{KBJ:2010} also constructed confidence intervals based on pivotal 
quantities associated with all these estimators obtained from the empirical Fisher information matrix.
Extensive Monte Carlo simulations have been performed by these authors to examine the effectiveness of the proposed method. It has been 
observed that the asymptotic normality does not work satisfactorily in the case of small sample sizes.  The coverage 
probabilities turn out to be quite low if the sample size is small.  Parametric or non-parametric bootstrap confidence intervals 
are recommended in this case.  Recently, Kundu and Gupta \cite{KG:2016} provided a two-step estimation procedure using the copula 
structure, mainly based on the idea of Joe \cite{Joe:2005}, and provided the asymptotic distribution of the two-step estimators.  
It has been observed, based 
on extensive simulations carried out by Kundu and Gupta \cite{KG:2016}, that the two-step estimators behave almost as efficiently as the ML 
estimators.  Hence,
they may be used effectively in place of the ML estimators to avoid solving a high-dimensional optimization problem.

\section{\sc Multivariate BS Distribution}

In the last section, we have introduced the bivariate BS distribution and discussed its properties.  In this section, we
introduce the multivariate BS distribution along the same lines.  We then present several properties of this multivariate 
BS distribution
and also discuss the ML estimation of the model parameters.

\subsection{\sc PDF and CDF}

The multivariate BS distribution has been introduced by Kundu et al. \cite{KBJ:2013} along the same lines as the bivariate BS distribution, discussed in the last section.

\noindent {\sc Definition 1:} Let $\underline{\alpha},  \underline{\beta} \in \mr^p$, where
 $\underline{\alpha} = (\alpha_1, \cdots, \alpha_p)^{\top}$ and $\underline{\beta} = (\beta_1, \cdots, \beta_p)^{\top}$, with
$\alpha_i > 0, \beta_i > 0$ for $i = 1, \cdots, p$.  Let $\ve{\Gamma}$ be a $p \times p$
positive-definite correlation matrix.  Then, the random vector ${\underline T} = (T_1, \cdots, T_p)^{\top}$ is said to
have a $p$-variate BS distribution with parameters $(\ul{\alpha}, \ul{\beta}, \ve{\Gamma})$ if
it has the joint CDF as
\bea
P(\ul{T} \le \ul{t}) &=& P(T_1 \le t_1, \cdots, T_p \le t_p) \nonumber\\
&=&
\Phi_p \left [ \frac{1}{\alpha_1}
\left ( \sqrt{\frac{t_1}{\beta_1}} - \sqrt{\frac{\beta_1}{t_1}}
\right ), \cdots, \frac{1}{\alpha_p} \left ( \sqrt{\frac{t_p}{\beta_p}} -
\sqrt{\frac{\beta_p}{t_p}} \right ); \ve{\Gamma} \right],
\label{3.1}
\eea
for  $t_1 > 0, \cdots, t_p > 0$.  Here, for $\ul{u} = (u_1, \cdots, u_p)^{\top}$, $ \Phi_p(\ul{u}; \ve{\Gamma})$ denotes the joint CDF of a standard normal
vector $\ul{Z} = (Z_1, \cdots, Z_p)^{\top}$ with correlation matrix $\ve{\Gamma}$.

The joint PDF of ${\ul T} = (T_1, \cdots, T_p)^{\top}$ can be obtained readily from (\ref{3.1}) as
\bea
f_{\ul{T}}(\ul{t}; \ul{\alpha}, \ul{\beta}, \ve{\Gamma})
& = & \phi_p \left ( \frac{1}{\alpha_1} \left (
\sqrt{\frac{t_1}{\beta_1}} - \sqrt{\frac{\beta_1}{t_1}} \right ), \cdots,
\frac{1}{\alpha_p} \left ( \sqrt{\frac{t_p}{\beta_p}} -
\sqrt{\frac{\beta_p}{t_p}} \right ); \ve{\Gamma}
\right )   \nonumber  \\
&  & \ \ \ \  \times \prod_{i=1}^p  \frac{1}{2 \alpha_i \beta_i} \left \{ \left (
\frac{\beta_i}{t_i} \right )^\frac{1}{2} +\left (
\frac{\beta_i}{t_i} \right )^\frac{3}{2} \right \},
     \label{3.2}
\eea
for  $t_1 > 0, \cdots, t_p > 0$; here, for $\ul{u} = (u_1, \cdots, u_p)^{\top}$,
$$ 
\phi_p(u_1, \cdots, u_p; \ve{\Gamma}) = \frac{1}{(2 \pi)^{\frac{p}{2}} |\ve{\Gamma}|^{\frac{1}{2}}}
\exp \left \{-\frac{1}{2}\ul{u}^{\top} {\ve{\Gamma}}^{-1} \ul{u} \right \},
$$ 
is the PDF of the standard normal vector with correlation matrix ${\ve{\Gamma}}$.  Hereafter, the  $p$-variate BS distribution, with joint PDF as  in (\ref{3.2}), will be denoted by
BS$_p(\ul{\alpha}, \ul{\beta}, \ve{\Gamma})$.

\subsection{\sc Marginal and Conditional Distributions}\label{mcd}

The marginal and conditional
distributions of BS$_p(\ul{\alpha}, \ul{\beta}, \ve{\Gamma})$ are as follows; the proofs can be 
seen in the work of Kundu et al. \cite{KBJ:2013}.

\noindent {\sc Result:} Let $\ul{T} \sim$ BS$_p(\ul{\alpha}, \ul{\beta}, \ve{\Gamma})$, and let
$\ul{T}, \ul{\alpha}, \ul{\beta}, \ve{\Gamma}$ be partitioned as follows:
\be
\ul{T} = \left (\matrix{\ul{T}_1 \cr \ul{T}_2} \right ), \ \
 \ul{\alpha} = \left (\matrix{\ul{\alpha}_1 \cr \ul{\alpha}_2} \right ), \ \
\ul{\beta} = \left (\matrix{\ul{\beta}_1 \cr \ul{\beta}_2} \right ), \ \
\ve{\Gamma} = \left (\matrix{\ve{\Gamma}_{11} & \ve{\Gamma}_{12}  \cr
\ve{\Gamma}_{21} & \ve{\Gamma}_{22}} \right ), \ \
\ee
where $\ul{T}_1, \ul{\alpha}_1, \ul{\beta}_1$ are all $q \times 1$ vectors and $\ve{\Gamma}_{11}$ is a
$q \times q$ matrix.  The remaining elements are all defined suitably.
Then, we have:

\noindent (a)  $\ul{T}_1 \sim$ BS$_q(\ul{\alpha}_1, \ul{\beta}_1, \ve{\Gamma}_{11})$ \ \ \ \hbox{and}
\ \ \  $\ul{T}_2 \sim$ BS$_{p-q}(\ul{\alpha}_2, \ul{\beta}_2, \ve{\Gamma}_{22})$;

\noindent (b) The conditional CDF of $\ul{T}_1$, given $\ul{T}_2 = \ul{t}_2$, is
$$
P[\ul{T}_1 \le \ul{t}_1|\ul{T}_2 = \ul{t}_2] = \Phi_q(\ul{w}; \ve{\Gamma}_{11.2});
$$

\noindent (c) The conditional PDF of $\ul{T}_1$, given $\ul{T}_2 = \ul{t}_2$, is
\be
f_{\ul{T}_1 \mid \left(\ul{T}_2 = \ul{t}_2\right)}(\ul{t}_1)
 =  \phi_q ( \ul{w}; \ve{\Gamma}_{11.2})
\ \prod_{i=1}^q  \frac{1}{2 \alpha_i \beta_i} \left \{ \left (
\frac{\beta_i}{t_i} \right )^\frac{1}{2} +\left (
\frac{\beta_i}{t_i} \right )^\frac{3}{2} \right\},
     \label{3.2a}
\ee
where
$$
\ul{w} = \ul{v}_1 - \ve{\Gamma}_{12} \ve{\Gamma}_{22}^{-1} \ul{v}_2, \ \ \ \ \ \ul{v} = (v_1 \cdots, v_p)^{\top},
\ \ \ \ \
v_i = \frac{1}{\alpha_i} \left ( \sqrt{\frac{t_i}{\beta_i}} -
\sqrt{\frac{\beta_i}{t_i}} \right ) \mbox{ for } \ i = 1, \cdots, p,
$$
$$
\ve{\Gamma}_{11.2} = \ve{\Gamma}_{11} - \ve{\Gamma}_{12}\ve{\Gamma}_{22}^{-1}\ve{\Gamma}_{21}, \ \ \
\ul{v} = \left ( \matrix{\ul{v}_1 \cr \ul{v}_2} \right ),
$$
and $\ul{v}_1$, $\ul{v}_2$ are vectors of dimensions $q \times 1$ and $(p-q) \times 1$, respectively;

\noindent (d) $\ul{T}_1$ and $\ul{T}_2$ are independent if and only if $\ve{\Gamma}_{12} = \ve{0}$.

\subsection{\sc Generation and Other Properties}

The following algorithm can be adopted to generate $\ul{T} =
(T_1, \cdots, T_p)^{\top}$ from BS$_p(\ul{\alpha},
\ul{\beta}, \ve{\Gamma})$ in (\ref{3.2}).

\noindent {\sc Algorithm}

\noindent Step 1: Make a Cholesky decomposition of $\ds \ve{\Gamma} = \ve{A} \ve{A}^{\top}$ \ (say);

\noindent Step 2: Generate $p$ independent standard normal random numbers, say, $U_1, \cdots, U_p$;

\noindent Step 3: Compute $\ds \ul{Z} = (Z_1, \cdots, Z_p)^{\top} = \ve{A} \ (U_1, \cdots, U_p)^{\top}$;

\noindent Step 4: Make the transformation
$$
T_i = \beta_i \left [ \frac{1}{2} \alpha_i Z_i + \sqrt{\left ( \frac{1}{2} \alpha_i Z_i \right )^2 + 1}
\right ]^2 \mbox{ for } \ \  i = 1, \cdots, p.
$$
Then, $\ul{T} =
(T_1, \cdots, T_p)^{\top}$ has the required BS$_p(\ul{\alpha},
\ul{\beta}, \ve{\Gamma})$ distribution.   

Now, let use the notation $\ds 1/\ul{a} =\left ( 1/a_1, \cdots, 1/a_k \right )^{\top}$ for a vector
$\ds \ul{a} = (a_1, \cdots, a_k)^{\top} \in {\bf R}^k$, where $a_i \ne 0$, for $i = \cdots, k$.  Then, we 
have the following result for which a detailed proof can be found in Kundu et al. \cite{KBJ:2013}.

\noindent {\sc Result:} Let $\ul{T} \sim$ BS$_p(\ul{\alpha}, \ul{\beta}, \ve{\Gamma})$, and 
${\ul T}_1$, ${\ul T}_2$ be the same as defined in Section \ref{mcd}. Then:

\noindent (a)
$$
\left ( \matrix{\ul{T}_1 \cr \cr \frac{1}{\ul{T}_2}} \right ) \sim
\hbox{BS}_p \left (\left ( \matrix{\ul{\alpha}_1 \cr \cr \ul{\alpha}_2} \right ),
\left ( \matrix{\ul{\beta}_1 \cr \cr \frac{1}{\ul{\beta}_2}} \right ), \left ( \matrix{
\ve{\Gamma}_{11} & - \ve{\Gamma}_{12}  \cr  \cr - \ve{\Gamma}_{21} & \ \ \ve{\Gamma}_{22}}
\right )
\right ),
$$
\noindent (b)
$$
\left ( \matrix{\frac{1}{\ul{T}_1} \cr \cr \ul{T}_2} \right ) \sim
\hbox{BS}_p \left (\left ( \matrix{\ul{\alpha}_1 \cr \cr \ul{\alpha}_2} \right ),
\left ( \matrix{\frac{1}{\ul{\beta}_1} \cr \cr \ul{\beta}_2} \right ), \left ( \matrix{
\ve{\Gamma}_{11} & - \ve{\Gamma}_{12}  \cr  \cr - \ve{\Gamma}_{21} & \ \ \ve{\Gamma}_{22}}
\right )
\right ),
$$
\noindent (c)
$$
\left ( \matrix{\frac{1}{\ul{T}_1} \cr \cr \frac{1}{\ul{T}_2}} \right ) \sim
\hbox{BS}_p \left (\left ( \matrix{\ul{\alpha}_1 \cr \cr \ul{\alpha}_2} \right ),
\left ( \matrix{\frac{1}{\ul{\beta}_1} \cr \cr \frac{1}{\ul{\beta}_2}} \right ), \left ( \matrix{
\ve{\Gamma}_{11} &  \ve{\Gamma}_{12}  \cr  \cr \ve{\Gamma}_{21} & \ \ \ve{\Gamma}_{22}}
\right )
\right ).
$$

\subsection{\sc ML Estimation}

Now, we discuss the ML estimators of the model parameters
based on the data $\{(t_{i1}, \cdots, t_{ip})^{\top}; i = 1, \cdots, n\}$.  The log-likelihood
function, without the additive constant, is given by
\bea
l(\ul{\alpha}, \ul{\beta}, \ve{\Gamma}|\hbox{data}) & = & -\frac{n}{2} \ln (|\ve{\Gamma}|) - \frac{1}{2}
\sum_{i=1}^n \ul{v}_i^{\top} \ve{\Gamma}^{-1} \ul{v}_i - n \sum_{j=1}^p \ln (\alpha_j) - n \sum_{j=1}^p \ln (\beta_j)
\nonumber  \\
&  & \ \ \ \ \ +
\sum_{i=1}^n \sum_{j=1}^p \ln \left \{ \left ( \frac{\beta_{ij}}{t_{ij}} \right )^{\frac{1}{2}}
+ \left ( \frac{\beta_{ij}}{t_{ij}} \right )^{\frac{3}{2}}
\right \},    \label{4.1}
\eea
where
$$
\ul{v}_i^{\top} = \left [\frac{1}{\alpha_1} \left (
\sqrt{\frac{t_{i1}}{\beta_1}} - \sqrt{\frac{\beta_1}{t_{i1}}} \right ), \cdots,
\frac{1}{\alpha_p} \left ( \sqrt{\frac{t_{ip}}{\beta_p}} -
\sqrt{\frac{\beta_p}{t_{ip}}} \right )
\right ].
$$
Then, the ML estimators of the unknown parameters can be obtained by maximizing (\ref{4.1}) with respect
to the parameters $\ul{\alpha}, \ul{\beta}$ and $\ve{\Gamma}$, which would require a $\ds 2p +
{p \choose 2}$ dimensional optimization process.  For this purpose, the following procedure can be 
adopted for  reducing the computational effort significantly. Observe that
\be
\left [ \left (
\sqrt{\frac{T_1}{\beta_1}} - \sqrt{\frac{\beta_1}{T_1}} \right ), \cdots,
\left ( \sqrt{\frac{T_p}{\beta_p}} -
\sqrt{\frac{\beta_p}{T_p}} \right )
\right ]^{\top} \sim N_p \left ( \ul{0}, \ve{D} \ve{\Gamma} \ve{D}^{\top} \right ),    \label{beta-known}
\ee
where $\ve{D}$ is a diagonal matrix given by $\ve{D} = \hbox{diag} \{\alpha_1, \cdots, \alpha_p\}$.
Therefore, for given $\ul{\beta}$, the ML estimators of $\ul{\alpha}$ and $\ve{\Gamma}$ become
\bea
\widehat{\alpha}_j(\ul{\beta}) = \left ( \frac{1}{n} \sum_{i=1}^n
\left (\sqrt{\frac{t_{ij}}{\beta_j}} - \sqrt{\frac{\beta_j}{t_{ij}}} \right )^2 \right )^{\frac{1}{2}} & = & 
\left ( \frac{1}{\beta_j} \left \{ \frac{1}{n} \sum_{i=1}^n t_{ij} \right \}
 + \beta_j \left \{ \frac{1}{n} \sum_{i=1}^n \frac{1}{t_{ij}} \right \} - 2 \right )^{\frac{1}{2}},  \nonumber \\
&  & j = 1, \cdots, p,    \label{mle-alpha}
\eea
and
\be
\widehat{\ve{\Gamma}}(\ul{\beta}) = \ve{P}(\ul{\beta}) \ve{Q}(\ul{\beta}) \ve{P}^{\top}(\ul{\beta});
   \label{mle-gamma}
\ee
here, $\ds \ve{P}(\ul{\beta})$ is a diagonal matrix given by $\ds \ve{P}(\ul{\beta}) = \hbox{diag}
\{1/\widehat{\alpha}_1(\ul{\beta}), \cdots, 1/\widehat{\alpha}_p(\ul{\beta})\}$, and  the
elements $q_{jk}(\ul{\beta})$ of the matrix $\ve{Q}(\ul{\beta})$ are given by
\be
q_{jk}(\ul{\beta}) = \frac{1}{n} \sum_{i=1}^n
\left (\sqrt{\frac{t_{ij}}{\beta_j}} - \sqrt{\frac{\beta_j}{t_{ij}}} \right )
\left (\sqrt{\frac{t_{ik}}{\beta_k}} - \sqrt{\frac{\beta_k}{t_{ik}}} \right ), \ \ \
\mbox{ for} \ \ \ j,k = 1, \cdots, p.   \label{q-ele}
\ee
Thus, we obtain the $p$-dimensional profile log-likelihood function
$\ds l(\widehat{\ul{\alpha}}(\ul{\beta}), \ul{\beta}, \widehat{\ve{\Gamma}}(\ul{\beta})|data)$.  The
ML estimator of $\ul{\beta}$ can then be obtained by maximizing the $p$-dimensional profile log-likelihood
function, and  once we get the ML estimator of $\ul{\beta}$, say $\widehat{\ul{\beta}}$, the ML estimators of $\ul{\alpha}$ and
$\ve{\Gamma}$ can be obtained readily by substituting $\widehat{\ul{\beta}}$ in place of $\ul{\beta}$ in Eqs.
(\ref{mle-alpha}) and (\ref{mle-gamma}), respectively.

However, for computing the ML estimators of the parameters, we need to maximize the profile log-likelihood
function of $\ul{\beta}$ and we may use the Newton-Raphson iterative process for this purpose.  Finding a proper $p$-dimensional initial guess for $\ul{\beta}$ becomes quite
 important in this case.  Modified moment estimators, similar to those proposed by Ng et al. \cite{NKB:2003} and described earlier,
can be used effectively  as initial guess, and they are as follows:
$$ 
\beta_j^{(0)} = \left ( \frac{1}{n} \sum_{i=1}^n t_{ij} \left / \frac{1}{n} \sum_{i=1}^n \frac{1}{t_{ij}}
\right .
\right )^{\frac{1}{2}}, \ \ \ \ j = 1, \cdots, p.
$$ 
Note that if $\ul{\beta}$ is known, then the ML estimators of $\ul{\alpha}$ and $\ve{\Gamma}$ can be
obtained explicitly.

If $\ul{\alpha}$ and $\ul{\beta}$ are known, then the ML estimator of $\ve{\Gamma}$ is $\widehat{\ve{\Gamma}} =  \ve{D}^{-1}
\ve{Q} (\ul{\beta}) \ve{D}^{-1}$, where the elements of the matrix $\ve{Q}(\ul{\beta})$ are  as
in (\ref{q-ele}) and the matrix $\ve{D}$ is  as defined earlier.  From
(\ref{beta-known}), it immediately follows in this case that $\hat{\ve{\Gamma}}$ has
a Wishart distribution with parameters $p$ and $\ve{\Gamma}$.  Furthermore, if only $\ul{\beta}$ is known,
it is clear that $\widehat{\alpha}_j^2(\ul{\beta})$,  defined in (\ref{mle-alpha}), is
distributed as $\chi^2_1$, for $j = 1, \cdots, p$.

%\section{\sc Log-BS Distribution}

\section{\sc Some Related Distributions}

In this section, we describe various distributions related to the BS distribution that are derived from the univariate and bivariate BS distributions.  We discuss their  
properties and also address inferential issues associated with them.  We focus here mainly on sinh-normal or log-BS, 
bivariate sinh-normal, length-biased BS and epsilon-BS distributions.

\subsection{\sc Sinh-Normal or log-BS Distribution}

Let $Y$ be a real-valued random variable with cumulative distribution function $F_Y(\cdot)$ given by 
\be
P(Y \le y) = F_Y(y; \alpha, \gamma, \sigma) = \Phi \left \{ \frac{2}{\alpha} \sinh \left (\frac{y-\gamma}{\sigma} \right ) \right \}, \ \ \ 
\hbox{for} \ \ \ -\infty < y < \infty.   \label{sinh-cdf}
\ee
Here, $\alpha > 0$, $\sigma > 0$, $-\infty < \gamma < \infty$, and $\sinh(x)$ is the hyperbolic sine
function of $x$, defined as $\ds \sinh(x) = (e^x - e^{-x})/2$.  In this case, $Y$ is said
to have a sinh-normal distribution, and  is denoted by SN$(\alpha, \gamma, \sigma)$.  The 
PDF of sinh-normal distribution is given by
$$
f_Y(y; \alpha, \gamma, \sigma) = \frac{2}{\alpha \sigma \sqrt{2 \pi}} \times \cosh 
\left ( \frac{y-\gamma}{\sigma} \right ) 
\times \exp \left [ \left ( - \frac{2}{\alpha^2} \sinh^2 \left (\frac{y-\gamma}{\sigma} \right ) \right )
\right ].   %\label{sinh-pdf}
$$
Here, $\cosh(x)$ is the hyperbolic cosine function of $x$, defined as $\ds \cosh(x) = 
(e^x+ e^{-x})/2$.  In this case, $\alpha$ is the shape parameter, $\sigma$ is the scale parameter and
$\gamma$ is the location parameter.  Note that if $T \sim$ BS$(\alpha, \beta)$, $\ln (T) \sim$ SN$(\alpha, 
\ln (\beta), 2)$.  For this reason, this distribution is also known as log BS distribution.

\noindent It is clear from (\ref{sinh-cdf}) that if $Y \sim$ SN$(\alpha, \gamma, \sigma)$, then 
\be
Z = \frac{2}{\alpha} \sinh \left ( \frac{Y-\gamma}{\sigma} \right ) \sim \hbox{N}(0, 1).   \label{normal-sinh}
\ee
From (\ref{normal-sinh}), it follows that if $Z \sim$ N$(0, 1)$, then 
\be
Y = \sigma \ \hbox{arcsinh} \left ( \frac{\alpha Z}{2} \right ) + \gamma \sim \hbox{SN}(\alpha, \gamma, \sigma).
    \label{rel-nor-sinh}
\ee
The representation in (\ref{rel-nor-sinh}) of the sinh-normal distribution can be used for generation 
purposes.  Using this representation,  connecting with the standard normal distribution, sinh-normal distribution 
can be easily generated, and consequently it can be used for the generation of BS distribution as well.

The random variable $Y$ is said to have a standard sinh-normal 
distribution when $\gamma$ = 0 and $\sigma$ = 1.  Therefore, if $Y \sim$ SN($\alpha, 0, 1$), then the 
corresponding PDF becomes
\be
f_Y(y; \alpha, 0, 1) =  \frac{2}{\alpha \sqrt{2 \pi}} \cosh (y) 
\exp \left [ \left ( - \frac{2}{\alpha^2} \sinh^2 (y) \right )
\right ].   \label{std-sinh-pdf}
\ee
It is immediate from (\ref{std-sinh-pdf}) that the SN$(\alpha, 0, 1)$ is symmetric about 0.
The PDF in (\ref{std-sinh-pdf}) of $Y$, for different values of $\alpha$ when $\gamma$ = 0,
and $\sigma$ = 1, has been presented by Rieck \cite{Rieck:1989}.  It has been observed by Rieck \cite{Rieck:1989}
that for $\alpha \le 2$, the PDF is strongly unimodal and for $\alpha > 2$, it is bimodal.  Furthermore,
if $Y \sim$ SN$(\alpha, \gamma, \sigma)$, then $\ds U = 2 \alpha^{-1} (Y - \gamma)/\sigma$ converges to the 
standard normal distribution as $\alpha \rightarrow 0$

If $Y \sim$ SN$(\alpha, \gamma, \sigma)$, then the MGF of $Y$ can be obtained as 
[see Rieck  \cite{Rieck:1989} or Leiva et al. \cite{LBPG:2007}]
\be
M(s) = E(e^{sY}) = e^{\mu s} 
\left [ \frac{K_a(\delta^{-2}) + K_b(\delta^{-1})}{2 K_{1/2}(\delta^{-2})} \right ],  \label{s-10-mgf}
\ee
where $a = (\sigma s + 1)/2$, $b = (\sigma s - 1)/2$, and $K_{\lambda}(\cdot)$ is the modified Bessel function 
of the third kind, given by 
$$
K_{\lambda}(w) = \frac{1}{2} \left (\frac{w}{2} \right )^{\lambda} \int_0^{\infty} y^{-\lambda - 1} 
e^{-y - (w^2/4y)} \hbox{d}y;
$$
see Gradshteyn and Randzhik (\cite{GR:2000}, p. 907).  Thus, by differentiating the moment generating function 
in (\ref{s-10-mgf}), moments of $Y$ can be obtained readily.  It is clear that the variance or the fourth moment cannot be expressed in
closed-form.  However, it has been observed that as $\alpha$ increases, the kurtosis of $Y$ increases.
Moreover, for $\alpha \le 2$, the kurtosis of $Y$ is smaller than that of the normal.  For $\alpha > 2$, when 
$\alpha$ increases, the SN distribution begins to display bimodality, with modes that are more separated,
and the kurtosis is greater than that of the normal.

We shall now discuss the estimation of the model parameters based on a random sample of size $n$,
say $\{y_1, \cdots, y_n\}$, from SN$(\alpha, \gamma, \sigma)$.  The log-likelihood function of the observed
sample, without the additive constant, is given by 
$$
l(\alpha, \gamma, \sigma| \hbox{data}) = -n \ln (\alpha) - n \ln (\sigma) + \sum_{i=1}^n \cosh \left ( \frac{y_i - \gamma}
{\sigma} \right ) - \frac{2}{\alpha^2} \sum_{i=1}^n \sinh^2 \left ( \frac{y_i - \gamma}{\sigma} \right ).
$$
It is clear that when all the parameters are unknown, the ML estimators of the unknown parameters cannot be obtained 
in closed-form.  But, for a given $\gamma$ and $\sigma$, the ML estimator of $\alpha$, say $\ds \widehat{\alpha}
(\gamma, \sigma)$, can be obtained as
\be
\widehat{\alpha}(\gamma, \sigma) = \left [ \frac{4}{n} \sum_{i=1}^n \sinh^2 \left ( \frac{y_i-\gamma}{\sigma}
\right ) \right ]^{1/2}.   \label{sn-alpha}
\ee
Then, the ML estimators of $\gamma$ and $\sigma$ can be obtained by maximizing the profile log-likelihood function
of $\gamma$ and $\sigma$, namely, $\ds l(\widehat{\alpha}(\gamma, \sigma), \gamma, \sigma)$.  It may be 
noted that for $\sigma$ = 2, $\gamma = \ln (\beta)$ and $y_i = \ln x_i$, $\widehat{\alpha}$ obtained 
from (\ref{sn-alpha}) is the same as that obtained from (\ref{ne-alpha}) presented later in Section 13.1.

\subsection{\sc Bivariate Sinh-Normal distribution}

Proceeding in the same way as we did with the univariate sinh-normal distribution, we can introduce bivariate 
sinh-normal distribution as follows.  The random vector $(Y_1, Y_2)$ is said to have a bivariate 
sinh-normal distribution, with parameters $\alpha_1 > 0$, $-\infty < \gamma_1 < \infty$, $\sigma_1 > 0$, 
$\alpha_2 > 0$, $-\infty < \gamma_2 < \infty$, $\sigma_2 > 0$, $-1 < \rho < 1$, if the joint CDF of 
$(Y_1, Y_2)$ is
$$
P(Y_1 \le y_1, Y_2 \le y_2) = F_{Y_1, Y_2}(y_1, y_2; {\ul \theta}) = 
\Phi_2 \left \{ \frac{2}{\alpha_1} \sinh \left (\frac{y_1 - \gamma_1}{\sigma_1}
\right ), \frac{2}{\alpha_2} \sinh \left (\frac{y_2 - \gamma_2}{\sigma_2}
\right ); \rho \right \}    %\label{bvsn-cdf}
$$
for $- \infty < y_1, \ y_2 < \infty$.  Here, ${\ul \theta}$ = 
($\alpha_1, \gamma_1, \sigma_1, \alpha_2, \gamma_2, \sigma_2, \rho$)  and 
$\Phi_2(u, v; \rho)$ is the joint cumulative distribution function 
of a standard bivariate normal vector $(Z_1, Z_2)$ with correlation coefficient $\rho$.  The joint
PDF of $Y_1$ and $Y_2$ is
\bea
f_{Y_1, Y_2}(y_1, y_2; {\ul \theta}) & = & \frac{2}{\alpha_1 \alpha_2 \sigma_1 \sigma_2 \pi} 
\phi_2 \left \{ \frac{2}{\alpha_1} \sinh \left (\frac{y_1 - \gamma_1}{\sigma_1}
\right ), \frac{2}{\alpha_2} \sinh \left (\frac{y_2 - \gamma_2}{\sigma_2}
\right ); \rho \right \}     \nonumber  \\
& & \ \ \cosh \left ( \frac{y_1 - \gamma_1}{\sigma_1} \right ) 
\cosh \left ( \frac{y_2 - \gamma_2}{\sigma_2} \right ),  \label{bv-sinh-pdf}
\eea
where $\phi_2(u, v, \rho)$ is as in (\ref{bv-nor-pdf}).  The contour plots of (\ref{bv-sinh-pdf}), for
different values of $\alpha_1$ and $\alpha_2$, can be found in Kundu \cite{Kundu:2015b} from which it is evident 
that it can take on
different shapes.  It is symmetric along the axis $x = y$, but it can be both unimodal and bimodal, and so 
it can be used effectively to analyze bivariate data.  From here on, if a bivariate random vector  
$(Y_1, Y_2)$ has the joint PDF as in (\ref{bv-sinh-pdf}), we will denote it by BSN$(\alpha_1, \gamma_1, \sigma_1, 
\alpha_2, \gamma_2, \sigma_2, \rho)$.

Some simple properties can be easily observed for the bivariate sinh-normal distribution.  For example,
if $(Y_1, Y_2) \sim$ BSN$(\alpha_1, \gamma_1, \sigma_1, \alpha_2, \gamma_2, \sigma_2, \rho)$, then 
$Y_1 \sim$ SN$(\alpha_1, \gamma_1, \sigma_1)$ and $Y_2 \sim$ SN$(\alpha_2, \gamma_2, \sigma_2)$.  The 
conditional PDF of $Y_1$, given $Y_2$, can be obtained easily as well.  Note that by using an algorithm similar to the one 
described in Section \ref{generation-bs}, we can first generate $(Z_1, Z_2)$ and then make the 
transformation
$$
Y_i = \sigma_i \ \hbox{arcsinh} \left ( \frac{\alpha_i Z_i}{2} \right ) + \gamma_i, \ \ \ i = 1,2.
$$
Then, $(Y_1, Y_2) \sim$ BSN$(\alpha_1, \gamma_1, \sigma_1, \alpha_2, \gamma_2, \sigma_2, \rho)$, as required. 

It may be easily seen that if $(X_1, X_2) \sim$ BS$_2(\alpha_1, \beta_1, \alpha_2, \beta_2, \rho)$, then 
$(Y_1, Y_2) = $ \linebreak $(\ln (X_1), \ln (X_2)) \sim$ BSN$(\alpha_1, \ln (\beta_1), 2, \alpha_2, \ln (\beta_2), 2, \rho)$.
It will be interesting to develop other properties of this distribution along with  
inferential procedures.  Multivariate sinh distribution can also be defined in an analogous manner.  Much work remains to be 
done in this direction.

\subsection{\sc Length-Biased BS distribution}

The length-biased distribution plays an important role in various applications.  
Length-biased distribution is a special case of weighted distributions.  The length-biased distribution of 
a random variable $X$ can be described as follows.  If $Y$ is a positive random variable with  
PDF $f_Y(\cdot)$ and $E(Y) = \mu < \infty$, then the corresponding length-biased random variable $T$ has its PDF
$f_T(\cdot)$ as 
$$
f_T(t) = \frac{t f_X(t)}{\mu}, \ \ \  t > 0.    %\label{lb-pdf}
$$
Therefore, it is clear that the length-biased version of a distribution does not introduce any extra 
parameter and has the same number of parameters as the original distribution.

The length-biased version of different distributions have found applications in diverse fields, such as biometry,
ecology, reliability and survival analysis.  A review of different length-biased distributions
can be found in Gupta and Kirmani \cite{GK:1990}.  Patil \cite{Patil:2002} has provided several applications of 
length-biased distributions in environmental science.

Recently, Leiva et al. \cite{LSA:2009} studied the length-biased BS (LBS) distribution.
If $Y \sim$ BS$(\alpha, \beta)$, then the LBS of $Y$, say $T$, has its PDF as
$$
f_T(t) = \phi \left ( \frac{1}{\alpha} \left ( \left [ \frac{t}{\beta} \right ]^{1/2} - \left [ 
\frac{\beta}{y} \right ]^{1/2} \right ) \right ) \times 
\frac{1}{(\alpha^3 + 2 \alpha)\beta} \times \left ( \left [ \frac{t}{\beta} \right ]^{1/2} + \left [
\frac{\beta}{t} \right ]^{1/2} \right ),  \ \ \  t > 0.   %\label{lbbs-pdf}
$$
From here on, this distribution will be denoted by LBS$(\alpha, \beta)$.  Here also, $\beta$ plays the role of the scale 
parameter while $\alpha$ plays the role of a shape parameter.  The PDFs of LBS$(\alpha, 1)$, for different 
values of $\alpha$, indicate that the PDF is unimodal. The mode
can be obtained as a solution of the cubic equation
$$
t^3 + \beta(1-\alpha^2) t^2 - \beta^2(1+\alpha^2) t - \beta^2 = 0.
$$
Moreover, the HF of LBS is always unimodal just as in the case of the BS distribution.  The moments of $T$ depend
on the moments of $Y$ and they are related as follows:
\beanno
\hbox{E}(T^r) &  = & \frac{1}{\alpha^2+2} \left \{ \frac{2^{r+1} \Gamma(r + \frac{3}{2}) \alpha^{2r+2} \beta^r}{\sqrt{\pi}}
- \sum_{k=1}^{r+1} (-1)^k {{2r+1}\choose{k}} \frac{E(Y^{r-k+1})}{\beta^{1-k}} \right .   \nonumber  \\
&  & 
\left . - \sum_{k=r+2}^{2r+1} (-1)^k {{2r+2}\choose{k}} \frac{E(Y^{k-r-1})}{\beta^{k-2r-1}}  \right \}.
\eeanno
We shall now briefly discuss the estimation of the model parameters $\alpha$ and $\beta$ based on a sample of size $n$, say 
$\{t_1, \cdots, t_n\}$, from a 
LBS$(\alpha, \beta)$.  The log-likelihood function of the data, without the additive constant, is given by 
\be
l(\alpha, \beta|\hbox{data}) = - \frac{1}{2 \alpha^2} \sum_{i=1}^n \left ( \frac{t_i}{\beta} + \frac{\beta}{t_i}
- 2 \right ) - n \ln(2 \alpha + \alpha^2)- \frac{3}{2} n \ln (\beta) + \sum_{i=1}^n \ln (t_i + \beta).
\label{ll-lbs}
\ee
The ML estimators of $\alpha$ and $\beta$ can then be obtained by maximizing the log-likelihood function in (\ref{ll-lbs})
with respect to $\alpha$ and $\beta$.  Since they do not have explicit forms, they need to be obtained 
numerically.  For constructing confidence intervals of the parameters, since the exact distributions
of the ML estimators are not possible to obtain, the asymptotic distribution of the ML estimators can be used.  It has been 
shown by Leiva et al. \cite{LSA:2009} that $(\widehat{\alpha}, \widehat{\beta})$, the ML estimator of 
$(\alpha, \beta)$, is asymptotically bivariate normal with mean $(\alpha, \beta)$ and 
variance-covariance matrix $\widehat{\Sigma}$, where the elements of $\widehat{\Sigma}^{-1}$ are as 
follows:
$$
\widehat{\Sigma}^{-1} = - \left [ \matrix{\sigma^{11} & \sigma^{12} \cr \sigma^{21} & \sigma^{22} \cr} \right ], 
$$
with
\beanno
\sigma^{11} & = & \sum_{i=1}^n \frac{3\left ( 2 - \frac{t_i}{\beta} - \frac{\beta}{t_i} \right )}{\alpha^4} - 
\frac{n(4 + 3\alpha^4)}{(2 \alpha + \alpha^3)^2},    \\ 
\sigma^{12} & = & \sigma^{21} = \sum_{i=1}^n \frac{\left ( \frac{1}{t_i} - \frac{t_i}{\beta^2} \right )}
{\alpha^3},   \\
\sigma^{22} & = & \frac{3n}{2 \beta^2} - \sum_{i=1}^n \left \{ \frac{t_i}{\alpha^2 \beta^3} - 
\frac{1}{(t_i+\beta)^2} \right \}.
\eeanno
The asymptotic distribution of the ML estimators can be used for constructing approximate confidence intervals.  
Alternatively, bootstrap confidence intervals can also be constructed and this may be especially suitable for 
small sample sizes.  

\subsection{\sc Epsilon BS Distribution}

Mudholkar and Hutson \cite{MH:2000} introduced the epsilon skew-normal distribution defined on the entire real
line as follows.  A random variable $X$ is said to have an epsilon skew-normal distribution if it
has the PDF
$$
g(x; \epsilon) = \left \{ \matrix{\phi \left ( \frac{x}{(1+\epsilon)} \right ) & \hbox{if} & 
x < 0, \cr &  &  \cr \phi \left ( \frac{x}{(1-\epsilon)} \right ) & \hbox{if} & x \ge 0,} \right .
$$
for $- 1 \le \epsilon \le 1$.  It is clear that for $\epsilon$  = 0, the epsilon skew-normal distribution 
becomes the standard normal distribution.  For $\epsilon = 1$, it is a negative half normal distribution and 
for $\epsilon = -1$, it is a positive half normal distribution.

Arellano-Valle et al. \cite{AGQ:2005} generalized the epsilon skew-normal distribution to any symmetric
distribution as follows.  A random variable $X$ is said to have a epsilon skew-symmetric distribution if
it has the density function
$$
h(x; \epsilon) = \left \{ \matrix{f \left ( \frac{x}{(1+\epsilon)} \right ) & \hbox{if} & x < 0,  \cr
&  &   \cr 
f \left ( \frac{x}{(1-\epsilon)} \right ) & \hbox{if} & x \ge 0,  \cr}  \right .
$$
where $f(\cdot)$ is any symmetric density function  and $- 1 < \epsilon < 1$.  It will be denoted by
$X \sim \hbox{ES}f(\epsilon)$.

Using the same transformation as in (\ref{relt}), Castillo et al.  \cite{CGB:2011}, see also 
Vilca et al. \cite{VSLC:2010} in this aspect, 
 defined the 
epsilon generalized-BS distribution as follows.  A random variable $W$ is said to 
have an epsilon generalized-BS distribution, with parameters $\alpha > 0$, $\beta > 0$, 
$-1 < \epsilon < 1$, if 
$$
W = \frac{\beta}{4} \left [ \alpha X + \sqrt{\{\alpha X\}^2 + 4} \right ]^2, 
$$
where $\ds X \sim \hbox{ES}f(\epsilon)$.  This distribution will be denoted by EGBS$(\alpha, \beta, \epsilon)$.  If
$W \sim$ EGBS$(\alpha, \beta, \epsilon)$, then the PDF of $W$ is
$$
f_W(w| \alpha, \beta, \epsilon) = \frac{w^{-3/2} (w +\beta)}{2 \alpha \sqrt{\beta}} \times
\left \{ \matrix{f \left ( \frac{t(w)}{1+\epsilon} \right ) & \hbox{if} & w < \beta, \cr 
&  &  \cr
f \left ( \frac{t(w)}{1-\epsilon} \right ) & \hbox{if} & w \ge \beta, \cr }  \right .
$$
where
$$
t(w) = \frac{1}{\alpha} \left ( \sqrt{\frac{w}{\beta}} - \sqrt{\frac{\beta}{w}} \right ). 
$$
These authors then discussed different properties of the epsilon generalized-BS distribution.
For example, it can be easily seen that if $W \sim$ EGBS$(\alpha, \beta, \epsilon)$, then 
(a) $a W \sim$ EGBS$(\alpha, a \beta, \epsilon)$, for $a > 0$, and (b)
$W^{-1} \sim$ EGBS$(\alpha, \beta^{-1}, \epsilon)$.  Moreover, the moments of $W$,
for $k = 1, 2, \cdots$, can be expressed as follows :
$$
\hbox{E}(W^k) = \beta^k \sum_{r=0}^{2k} \left ( \frac{\alpha}{2} \right )^{2k - r} {{2k}\choose{r}} 
\hbox{E} \left ( X^{2k-r} \left ( (\alpha X/2)^2 + 1 \right )^{r/2} \right ).
$$
When $f(\cdot)  = \phi(\cdot)$, the standard normal PDF, the 
epsilon generalized-BS distribution  is known as the epsilon BS distribution.
Castillo et al. \cite{CGB:2011} derived the first and second moments explicitly in terms of
incomplete moments of the standard normal distribution.  These authors also discussed 
the estimation of the model parameters of epsilon BS distribution.  The ML estimators are obtained by solving three non-linear equations, 
and explicit expressions of the elements of the Fisher information matrix have also been provided.

\section{\sc Log-BS Regression}

In this section we introduce log-BS regression model which has been used quite extensively in practice in recent years.  
First, we provide the basic formulation of the model, and then we discuss about different inferential issues related 
to this model.  

\subsection{\sc Basic Formulation}

Rieck and Nedelman \cite{RN:1991} first introduced the log-BS regression model as 
follows.  Let $T_1, \cdots, T_n$ be $n$ independent random variables, and the distribution of $T_i$ 
be a BS distribution with shape parameter $\alpha_i$ and scale parameter $\beta_i$.  Let us 
further assume that the distribution of $T_i$ depends on a set of $p$ explanatory variables 
${\ul x}_i = (x_{i1}, \cdots, x_{ip})$ as follows:

\noindent 1. $\ds \beta_i = \exp({\ul x}^{\top}_i {\ul \theta})$, for $i = 1, \cdots, n$, ${\ul \theta} = 
(\theta_1, \cdots, \theta_p)$ is a set of $p$ unknown parameters to be estimated;   

\noindent 2.  The shape parameter is independent of the explanatory vector ${\ul x}_i$, and they are 
constant, i.e. $\alpha_1 = \cdots = \alpha_n = \alpha$.

Now, from the properties of the BS distribution, it immediately follows that
$\ds T_i = \exp({\ul x}^{\top}_i {\ul \theta}) U_i$, where $U_i$ is distributed as BS with scale parameter one 
and shape parameter $\alpha$.  Therefore, if we denote
$Y_i = \ln (T_i)$, for $i = 1, \cdots, n$, then 
\be
Y_i = {\ul x}^{\top}_i {\ul \theta} + \epsilon_i, \ \ \ \ i = 1, \cdots, n.    \label{ln-bs-reg}
\ee
Here, $\epsilon_i = \ln (U_i)$ has a log-BS or sinh distribution, $i.e.$, SN$(\alpha, 0, 2)$.
Because of the form, the model in (\ref{ln-bs-reg}) is known as the log-BS regression model.
Here also, the problem is same as the standard regression problem, $i.e.$, to estimate the unknown parameters,
namely ${\ul \theta}$ and $\alpha$, based on the sample $t_i$ and the associated covariates ${\ul x}_i$, for
$i = 1, \cdots, n$.

\subsection{\sc ML Estimators}

Based on independent observations $y_1, \cdots, y_n$, from the regression model in (\ref{ln-bs-reg}), the 
log-likelihood function, without the additive constant, is given by [see Rieck and Nedelman \cite{RN:1991}]
\be
l({\ul \theta}, \alpha| \hbox{data}) = \sum_{i=1}^n \ln (w_i) - \frac{1}{2} \sum_{i=1}^n z_i^2,
   \label{ll-bs-reg}
\ee
where
$$
w_i = \frac{2}{\alpha} \ \cosh \left (\frac{y_i - {\ul x}_i^{\top}{\ul \theta}}{2} \right ) \ \ \ \hbox{and} 
\ \ \ \
z_i = \frac{2}{\alpha} \ \sinh \left (\frac{y_i - {\ul x}_i^{\top}{\ul \theta}}{2} \right ).
$$
The ML estimators of $\alpha$ and ${\ul \theta}$ can be obtained by maximizing (\ref{ll-bs-reg}) with respect
to the parameters.  The normal equations are 
\bea
\frac{\partial l({\ul \theta}, \alpha| \hbox{data})}{\partial \theta_j} & = & 
\left ( \frac{1}{2} \right ) \sum_{i=1}^n x_{ij} \left (z_i w_i - \frac{z_i}{w_i} \right ) = 0, 
\label{ne-theta}  \\
\frac{\partial l({\ul \theta}, \alpha| \hbox{data})}{\partial \alpha} & = & 
-\frac{n}{\alpha} + \left ( \frac{1}{\alpha} \right )\sum_{i=1}^n z_i^2 = 0.  \label{ne-alpha}
\eea
From (\ref{ne-alpha}), it is clear that, for given ${\ul \theta}$, the ML estimator of $\alpha$ can be 
obtained as
$$
\widehat{\alpha}(\theta) = \left \{ \frac{4}{n} \sum_{i=1}^n \sinh^2 \left (\frac{y_i - {\ul x}_i^{\top} {\ul \theta}}
{2} \right ) \right \}^{1/2},
$$
and the ML estimators of $\theta_j$'s can then be obtained from (\ref{ne-theta}), or by maximizing
the profile log-likelihood function of ${\ul \theta}$, $i.e.$, 
$\ds  l({\ul \theta}, \widehat{\alpha}({\ul \theta})| \hbox{data})$, with respect to ${\ul \theta}$.  No explicit
solutions are available in this case, and so suitable numerical techniques like Newton-Raphson or the 
Fisher's scoring method may be applied to determine the ML estimates.   Rieck \cite{Rieck:1989} compared, by 
extensive Monte Carlo simulation
studies, the performances of the Newton-Raphson and the Fisher's scoring algorithm, and observed that the 
Newton-Raphson method performed better than the Fisher's scoring algorithm particularly for small sample sizes and
when the number of parameters is small.

It has been observed that the likelihood function may have more than one maxima mainly due to the fact 
that sinh-normal is bimodal for $\alpha > 2$.  It has been shown by Rieck \cite{Rieck:1989} that if $\alpha < 2$ is
known, then the ML estimator of ${\ul \theta}$ exists and is unique.  Under the assumptions that $|x_{ij}|$ are uniformly bounded  and 
as $n \rightarrow \infty$, 
$$
\frac{1}{n} \sum_{i=1}^n {\ul x}_i {\ul x}_i^{\top} = {\bf A} > 0,
$$
Rieck and Nedelman \cite{RN:1990} showed that there exists at least one solution of the normal equations
that is a consistent estimate of the true parameter vector $({\ul \theta}^{\top}, \alpha)$.  Moreover,
among all possible solutions of the normal equations, one and only one tends to $({\ul \theta}^{\top}, \alpha)$
with probability one.  They also proved that $\ds \sqrt{n}((\widehat{{\ul \theta}} - {\ul \theta})^{\top}, 
\widehat{\alpha} -\alpha)$ converges in distribution to a multivariate normal distribution with mean 
vector ${\ul 0}$, and the asymptotic variance-covariance matrix $\Sigma$, where 
$$
\Sigma = \left [ \matrix{\frac{4{\bf A}^{-1}}{C(\alpha)} & {\ul 0} \cr  \cr  {\ul 0}^{\top} & \frac{\alpha^2}{2}}
\right ],
$$
with 
$$
C(\alpha) = 2 + \frac{4}{\alpha^2} - \left (\frac{2\pi}{\alpha^2} \right )^{1/2} \left \{
1 - \hbox{erf}[(2/\alpha^2)^{1/2}] \right \}\exp(2/\alpha^2)
$$
and $\ds \hbox{erf}(x) = 2 \Phi(\sqrt{2} x) - 1$; see Rieck and Nedelman \cite{RN:1990} for details.  Thus, it is 
immediate that if $\widehat{\alpha}$ is the ML estimator of $\alpha$, then the asymptotic variance-covariance 
matrix of $(\widehat{\ul \theta}^{\top}, \widehat{\alpha})$ can be estimated by
$$
\left [ \matrix{\frac{4}{C(\widehat{\alpha})} \sum_{i=1}^n {\ul x}_i {\ul x}_I^{\top} & {\bf 0} \cr  \cr
{\bf 0}^{\top} & \frac{\widehat{\alpha}^2}{2n} \cr} \right ],
$$
which can be used for constructing confidence intervals.

Although the ML estimator of ${\ul \theta}$ cannot be obtained in closed-form, the least squares estimator of
${\ul \theta}$ can be obtained in closed-form as  
$$
\widehat{\ul \theta} = ({\ul X}^{\top}{\ul X})^{-1} {\ul X}^{\top} {\ul Y},
$$
where ${\ul Y}$ is the column vector of the $y_i$'s and ${\ve X}^{\top} = ({\ul x}_1, \cdots, {\ul x}_n)$.   But, it has
been shown, using Monte Carlo simulations, by Rieck and Nedelman \cite{RN:1991} that the LSE of ${\ul \theta}$ is 
not as efficient as the ML estimator of ${\ul \theta}$.  Yet, it is an unbiased estimator of ${\ul \theta}$ and  
is quite efficient for small values of $\alpha$.

\subsection{\sc Testing of Hypotheses}

Lemonte et al. \cite{LFSC:2010} considered the following testing problem for log-BS model:
$$
\hbox{H}_{a0}: \beta_r = \beta_r^0 \ \ vs. \ \ \hbox{H}_{a1}: \beta_r \ne \beta_r^0,
$$
$$ 
\hbox{H}_{b0}: \beta_r \ge \beta_r^0 \ \ vs. \ \ \hbox{H}_{b1}: \beta_r < \beta_r^0,
$$ 
$$
\hbox{H}_{c0}: \beta_r \le \beta_r^0 \ \ vs. \ \ \hbox{H}_{c1}: \beta_r > \beta_r^0.
$$ 
They then discussed test procedures for the above hypotheses and their performance characteristics.

\section{\sc Generalized BS Distribution}

The univariate, bivariate and multivariate BS distributions have been derived by making suitable transformations on the univariate,
bivariate and multivariate normal distributions, respectively.  Attempts have been made to generalize BS distributions by replacing
normal distribution with the elliptical distribution.  In this section, we introduce univariate, multivariate and 
matrix-variate 
generalized BS distributions, discuss their properties and also address some inferential issues.

\subsection{\sc Univariate Generalized BS Distribution}

D{\'i}az-Garc{\'i}a and Leiva-S{\'a}nchez \cite{DL:2005} generalized the univariate BS distribution by 
using an elliptical distribution in place of normal distribution.  A random variable $X$ is said
to have an elliptical distribution if the PDF of $X$ is given by
\be
f_X(x) = c g \left [ \frac{(x-\mu)^2}{\sigma^2} \right ], \ \ \ x \in \mr,   \label{ec-pdf}
\ee
where $g(u)$, with $u > 0$, is a real-valued function and corresponds to the kernel of the PDF
of $X$, and $c$ is the normalizing constant, such that $f_X(x)$ is a valid PDF.  If the random variable
$X$ has the PDF in (\ref{ec-pdf}), it will be denoted by EC$(\mu, \sigma^2, g)$.  Here, $\mu$ and 
$\sigma$ are the location and scale parameters, respectively, and $g$ is the corresponding kernel.
Several well-known 
distributions such as the normal distribution, $t$-distribution, Cauchy distribution, Pearson
VII distribution, Laplace distribution, Kotz type distribution and logistic distribution are all  members of the family 
of elliptically contoured distributions.

Based on the elliptically contoured distribution, D{\'i}az-Garc{\'i}a and Leiva-S{\'a}nchez \cite{DL:2005} defined the 
generalized BS (GBS) distribution as follows.  The random variable $T$ is said to have a 
GBS distribution, with parameters $\alpha$, $\beta$ and kernel $g$, if 
$$
U = \frac{1}{\alpha} \left [ \sqrt{\frac{T}{\beta}} - \sqrt{\frac{\beta}{T}} \right ] \sim
\hbox{EC}(0, 1, g),   %\label{gbs-pdf}
$$
and is denoted by $\ds T \sim \hbox{GBS}(\alpha, \beta, g)$.  It can be shown, using the standard 
transformation technique, that if $\ds T \sim \hbox{GBS}(\alpha, \beta, g)$, then the PDF of $T$ is
as follows:
$$
f_T(t) = \frac{c}{2 \alpha \sqrt{\beta}} t^{-3/2} (t + \beta) g \left ( \frac{1}{\alpha^2} 
\left [ \frac{t}{\beta} + \frac{\beta}{t} - 2 \right ] \right ), \ \ \ t > 0, 
$$
where $c$ is the normalizing constant as mentioned in (\ref{ec-pdf}).  For different special cases,
$c$ can be explicitly obtained and these can be found in D{\'i}az-Garc{\'i}a and Leiva-S{\'a}nchez \cite{DL:2005}.  These
authors have derived the moments of the GBS distribution in terms of the moments of the corresponding 
elliptically contoured distribution.  Inferential procedures have not yet been developed in general, but 
only in some special cases.  It will, therefore,  be of interest to develop efficient inferential procedures for different kernel functions.  In fact, 
in practice, choosing a proper kernel is quite an important issue that  has not been addressed in detail.

It is important to mention here that Fang and Balakrishnan \cite{FB:2017} recently extended some of the stochastic
ordering results for the univariate BS model detailed earlier in Section 8 to the case of generalized BS models with associated
random shocks.  There seems to be a lot of potential for further exploration in this direction!

\subsection{\sc Multivariate Generalized BS Distribution}

Along the same lines as the univariate GBS distribution, the multivariate GBS distribution can be defined.  First,
let us recall [see Fang and Zhang \cite{FZ:1990}] the definition of the multivariate elliptically 
symmetric distribution.  A $p$-dimensional
random vector ${\ul X}$ is said to have an elliptically symmetric distribution with $p$-dimensional
location vector ${\ul \mu}$, a $p \times p$ positive definite dispersion matrix ${\ve \Sigma}$ and the
density generator (kernel) $\ds h^{(p)}: \mr_+ \rightarrow \mr_+$, if the PDF of ${\ul X}$ is of 
the form
\be
f_{EC_p}({\ul x}; {\ul \mu}, {\ve \Sigma}, h^{(p)}) = |{\ve \Sigma}|^{-1/2} 
h^{(p)}(w({\ul x} - {\ul \mu})^{\top}{\ve \Sigma}^{-1} ({\ul x} - {\ul \mu})),    \label{pdf-ellip}
\ee
where $w(\ul{x}): \mr^p \rightarrow \mr_+$ with $w(\ul{x}) =
({\ul x} - {\ul \mu})^{\top} \ve{\Sigma}^{-1}({\ul x} - {\ul \mu})$, $h:\mr_+ \rightarrow \mr_+$,
and
$$
\int_{\mr^p}f_{EC_p}({\ul x}; \ul{\mu}, \ve{\Sigma}, h^{(p)}) \hbox{d}\ul{x} = 1.
$$
In what follows, we shall denote $\ds
f_{EC_p}({\ul x}; \ul{0}, \ve{\Sigma}, h^{(p)})$ simply by
$f_{EC_p}({\ul x}; \ve{\Sigma}, h^{(p)})$
 and the corresponding random vector by EC$_p$($\ve{\Sigma}, h^{(p)})$.

Based on the elliptically symmetric distribution, Kundu et al. \cite{KBJ:2013} introduced the generalized
multivariate BS distribution as follows.

\noindent {\sc Definition:} Let $\ul{\alpha},  \ul{\beta} \in \mr^p$, where
 $\ul{\alpha} = (\alpha_1, \cdots, \alpha_p)^{\top}$ and $\ul{\beta} = (\beta_1, \cdots, \beta_p)^{\top}$, with
$\alpha_i > 0, \beta_i > 0$, for $i = 1, \cdots, p$.  Let $\ve{\Gamma}$ be a $p \times p$
positive-definite correlation matrix.  Then, the random vector ${\ul T} = (T_1, \cdots, T_p)^{\top}$ is said to have a
generalized multivariate BS distribution, with parameters 
$(\ul{\alpha}, \ul{\beta}, \ve{\Gamma})$ and density generator $h^{(p)}$, denoted by 
$\ul{T} \sim$ GBS$_p(\ul{\alpha}, \ul{\beta}, \ve{\Gamma}, h^{(p)})$, if the
CDF of $\ul{T}$, i.e., $\ds P(\ul{T} \le \ul{t}) = P(T_1 \le t_1, \cdots, T_p \le t_p)$, is given by
$$
P(\ul{T} \le \ul{t}) = F_{EC_p} \left [ \frac{1}{\alpha_1}
\left ( \sqrt{\frac{t_1}{\beta_1}} - \sqrt{\frac{\beta_1}{t_1}}
\right ), \cdots, \frac{1}{\alpha_p} \left ( \sqrt{\frac{t_p}{\beta_p}} -
\sqrt{\frac{\beta_p}{t_p}} \right ); \ve{\Gamma}, h^{(p)} \right ]    %\label{def-mgbs}
$$
for $\ul{t} > \ul {0}$, where $\ds F_{EC_p}(\cdot;  \ve{\Gamma}, h^{(p)})$ denotes the CDF of
$\ds EC_p(\ve{\Gamma}, h)$.  The corresponding joint PDF of ${\ul T} = (T_1, \cdots, T_p)^{\top}$
is, for $\ul{t} > \ul{0}$,
\beanno
f_{\ul{T}}(\ul{t}; \ul{\alpha}, \ul{\beta}, \ve{\Gamma})
& = & f_{EC_p} \left ( \frac{1}{\alpha_1} \left (
\sqrt{\frac{t_1}{\beta_1}} - \sqrt{\frac{\beta_1}{t_1}} \right ), \cdots,
\frac{1}{\alpha_p} \left ( \sqrt{\frac{t_p}{\beta_p}} -
\sqrt{\frac{\beta_p}{t_p}} \right ); \ve{\Gamma}, h^{(p)}
\right )   \nonumber  \\
&  & \ \ \ \  \times \prod_{i=1}^p  \frac{1}{2 \alpha_i \beta_i} \left \{ \left (
\frac{\beta_i}{t_i} \right )^\frac{1}{2} +\left (
\frac{\beta_i}{t_i} \right )^\frac{3}{2} \right \},
     %\label{pdf-mgbs}
\eeanno
where $\ds f_{EC_p}(\cdot)$ is as given in (\ref{pdf-ellip}).

The density generator $h^{(p)}(\cdot)$ can take on different forms resulting in multivariate 
normal distribution, symmetric Kotz type distribution, multivariate $t$-distribution, symmetric 
multivariate Pearson type VII distribution, and so on.  These authors have then discussed different properties of the  
generalized multivariate BS distribution including the distributions of the marginals, 
distributions of reciprocals, total positivity of order two property etc.  Special attention has  
been paid to  
two particular cases, namely, (i) multivariate normal and (ii) multivariate $t$.  Generation of random samples and 
different inferential issues have also been discussed by these authors in detail.  

Recently, Fang et al. \cite{FZB:2017} proceeded along the lines of Section 8 and discussed stochastic comparisons 
of minima and maxima arising from independent and non-identically distributed bivariate BS random vectors with respect to
the usual stochastic order.  Specifically, let $(X_1, X_2) \sim$ BS$_2(\alpha_1, \beta_1, \alpha_2, \beta_2, \rho)$ and 
$(X_1^*, X_2^*) \sim$ BS$_2(\alpha_1^*, \beta_1^*, \alpha_2^*, \beta_2^*, \rho)$ independently, and let further $0 < v \le 2$.
Then, Fang et al. \cite{FZB:2017} established that when $\alpha_1 = \alpha_2 = \alpha_1^* = \alpha_2^*$, 
$\ds (\beta_1^{-1/v}, \beta_2^{-1/v}) \ge_m (\beta_1^{*-1/v}, \beta_2^{*-1/v})$ implies $X_{2:2} \ge_{st} X_{2:2}^*$ and 
$\ds (\beta_1^{1/v}, \beta_2^{1/v}) \ge_m (\beta_1^{* 1/v}, \beta_2^{* 1/v})$ implies $X_{1:2}^* \ge_{st} X_{1:2}$.  In a 
similar manner, they also proved that when $\beta_1 = \beta_2 = \beta_1^* = \beta_2^*$, $\ds \left ( 1/\alpha_1, 
1/\alpha_2 \right ) \ge_m \left ( 1/\alpha_1^*, 1/\alpha_2^* \right )$ implies 
$X_{2:2} \ge_{st} X_{2:2}^*$ and $X_{1:2}^* \ge_{st} X_{1:2}$.  Generalizations of these results to bivariate generalized BS distributions, 
analogous to those mentioned in the last subsection, remain open!

\subsection{\sc Matrix-Variate Generalized BS Distribution}

We have already detailed the univariate and multivariate generalized BS distributions, and they 
are natural extensions of the univariate and multivariate BS distributions.  Along the 
same lines, Caro-Lopera et al. \cite{CLB:2012} defined the matrix-variate generalized BS  
distribution using an elliptic random matrix.  First, we introduce a matrix-variate elliptic distribution.  
Let ${\ve X} = 
(X_{ij})$ be an $n \times k$ random matrix.  It is said to have a matrix-variate elliptic distribution 
with location
matrix ${\ve M} \in \mr^{n \times k}$, scale matrices ${\ve \Omega} \in \mr^{k \times k}$ with 
rank$({\ve \Omega}) = k$, and ${\ve \Sigma} \in \mr^{n \times n}$ with rank$({\ve \Sigma}) = n$, and a 
density generator $g$, if the PDF of ${\ve X}$ is 
$$
f_{\ve X}({\ve X}) = c |{\ve \Omega}|^{-n/2} |{\ve \Sigma}|^{-k/2} g(\hbox{tr}({\ve \Omega}^{-1} ({\ve X} - {\ve M})^{top}
{\ve \Sigma}^{-1} ({\ve X} - {\ve M}))), \ \ \ {\ve X} \in \mr^{n \times k},
$$
where $c$ is the normalizing constant for the density generator $g$.   It will be denoted by EC$_{n \times k}
({\ve M}, {\ve \Omega}, {\ve \Sigma}; g)$.  Now, we are in a position to define the matrix-variate generalized
BS distribution.  

\noindent {\sc Definition:} Let ${\ve Z} = (Z_{ij}) \sim$ EC$_{n \times k}({\ve 0}, {\ve I}_k, 
{\ve I}_n;g)$ and ${\ve T} = (T_{ij})$, where
$$
T_{ij} = \beta_{ij} \left [ \frac{\alpha_{ij} Z_{ij}}{2} + \sqrt{ \left \{ \frac{\alpha_{ij} Z_{ij}}{2} 
\right \}^2 + 1} \right ]^2, \ \ \ \alpha_{ij} > 0, \ \beta_{ij} > 0, \ i = 1, \dots, n, j = 1, \dots, k. 
$$
Then, the random matrix ${\ve T}$ is said to have a generalized matrix-variate BS distribution,
denoted by GBS$_{n \times k}({\ve A}, {\ve B}, g)$, where ${\ve A} = (\alpha_{ij})$ and 
${\ve B} = (\beta_{ij})$.  It has been shown by Caro-Lopera et al. \cite{CLB:2012} that if $T \sim$ 
GBS$_{n \times k}({\ve A}, {\ve B}, g)$, then the PDF of ${\ve T}$ is given by
$$
f_{\ve T}({\ve T}) = \frac{c}{2^{n+k}} g \left ( \sum_{i=1}^n \sum_{j=1}^k \frac{1}{\alpha_{ij}^2} 
\left [ \frac{T_{ij}}{\beta_{ij}} + \frac{\beta_{ij}}{T_{ij}} - 2 \right ] \right ) 
\prod_{i=1}^n \prod_{j=1}^k \frac{T_{ij}^{-3/2}[T_{ij} + \beta_{ij}]}{\alpha_{ij} \sqrt{\beta_{ij}}}, \ \  
T_{ij} > 0,
$$
for $i = 1, \dots, n$ and $j = 1, \dots, k$.

Note that if we take
$$
g(u) = \exp \left \{-\frac{u}{2} \right \}, \ \  u > 0, \ \ \ \hbox{and} \ \ \ c = \frac{2^{nk/2}}{\pi^{nk/2}},
$$
then we obtain the matrix-variate BS distribution.  Therefore, a random matrix ${\ve T} \in
\mr^{n \times k}$ is said to have a matrix-variate BS distribution if the PDF of ${\ve T}$ is
$$
f_{\ve T}({\ve T}) = \frac{c}{2^{n+k}} \exp \left ( - \frac{1}{2}\sum_{i=1}^n \sum_{j=1}^k \frac{1}{\alpha_{ij}^2} 
\left [ \frac{T_{ij}}{\beta_{ij}} + \frac{\beta_{ij}}{T_{ij}} - 2 \right ] \right ) 
\prod_{i=1}^n \prod_{j=1}^k \frac{T_{ij}^{-3/2}[T_{ij} + \beta_{ij}]}{\alpha_{ij} \sqrt{\beta_{ij}}}, \ \  
T_{ij} > 0,
$$
for $i = 1, \dots, n$ and $j = 1, \dots, k$.

These authors have provided an alternate representation of the PDF of the matrix-variate generalized 
BS distribution in terms of Hadamard matrix products, and also indicated some open problems.
Interested readers may refer to their article.  An important open problem is the inference
for the model parameters in this case.

\section{\sc Illustrative Examples}

In this section, we present some numerical examples in order to illustrate some of the inferential results 
described in the preceding sections as well as the usefulness of the models introduced.

\subsection {\sc Example 1: Fatigue Data}  

This data set was originally analyzed by Birnbaum and Saunders \cite{BS:1969b} and 
it represents the fatigue life of 6061-T6 aluminum coupons cut parallel to the direction of rolling 
and oscillated at 18 cycles per seconds (cps). It consists of 101 observations with maximum stress per 
cycle 31,000 psi.  The data are as follows: 

 70  90  96  97  99 100 103 104 104 105 107 108 108 108 109
109 112 112 113 114 114 114 116 119 120 120 120 121 121 123
124 124 124 124 124 128 128 129 129 130 130 130 131 131 131
131 131 132 132 132 133 134 134 134 134 134 136 136 137 138
138 138 139 139 141 141 142 142 142 142 142 142 144 144 145
146 148 148 149 151 151 152 155 156 157 157 157 157 158 159
162 163 163 164 166 166 168 170 174 196 212

\noindent In summary, we have in this case $n$ = 101, $\ds s = (1/n) \sum_{i=1}^n t_i$ = 133.73267, and 
$\ds r = n /\sum_{i=1}^n t_i^{-1}$ = 129.93321.  The ML, MM, UML and UMM estimates 
are presented in Table \ref{result-pe-1}; see Ng et al. \cite{NKB:2003}.
\begin{table}[h]
\bc
\caption{Point estimates of $\alpha$ and $\beta$ for Example 1  \label{result-pe-1}}
\begin{tabular}{|c|c|c|}  \cline{1-3}
Estimate & $\alpha$ & $\beta$  \\  \hline
ML  &      0.170385  & 131.818792  \\
MM  &      0.170385  & 131.819255  \\
UML &      0.172089  & 131.809130  \\
UMM &      0.172089  & 131.809593  \\  \hline
\end{tabular}
\ec
\end{table}
The associated 90\% and 95\% confidence intervals are presented in Table \ref{result-ci-1}.
\begin{table}[h]
\bc
\caption{90\% and 95\% confidence intervals for $\alpha$ and $\beta$ for Example 1  \label{result-ci-1}}
\begin{tabular}{|c|c|c|c|c|}  \hline  
Estimate & \multicolumn{2}{|c|}{$\alpha$} & \multicolumn{2}{|c|}{$\beta$}  \\  \cline{1-5} \hline
    &     90\%  &     95\%  &     90\%  &       95\%    \\
ML  & (0.1527,0.1927) & (0.1497,0.1976) & (128.2552,135.5861) &  (127.5944,136.3325)  \\
MM  & (0.1527,0.1927) & (0.1497,0.1976) & (128.2556,135.5866) &  (127.5948,136.3330)  \\
UML & (0.1541,0.1949) & (0.1511,0.1999) & (128.2116,135.6143) &  (127.5448,136.3685)  \\
UMM & (0.1541,0.1949) & (0.1511,0.1999) & (128.2121,135.6148) &  (127.5452,136.3690)  \\  \hline

\end{tabular}
\ec
\end{table}

%\newpage

\subsection {\sc Example 2: Insurance Data} The following data represent Swedish third party motor insurance for 1977 for one of 
several geographical zones.  The data were compiled by a Swedish committee on the analysis of risk premium in motor 
insurance.  The data points are the aggregate payments by the insurer in thousand Skr (Swedish currency).  The data set
was originally reported in Andrews and Herzberg \cite{AH:1985}, and is as as follows:

\noindent   5014 5855  6486 6540  6656 6656 7212 7541 7558 7797  8546  9345  11762  12478 13624  14451 14940 
 14963 15092  16203      16229   16730 
       18027 18343  19365 
  21782  24248        29069     34267 38993

We would like to see whether the two-parameter BS distribution fits these data well or not.  We 
divide all the data points by 10000 and obtain   
the ML, MM, UML and UMM estimates, and these results are presented in Table \ref{result-pe-insurance}.
\begin{table}[h]
\bc
\caption{Point estimates of $\alpha$ and $\beta$ for Example 2  \label{result-pe-insurance}}
\begin{tabular}{|c|c|c|}  \cline{1-3}
Estimate & $\alpha$ & $\beta$  \\  \hline
ML  &      0.559551  & 1.255955   \\
MM  &      0.559551  & 1.256602   \\
UML &      0.540899  & 1.253993   \\
UMM &      0.540899  & 1.253346   \\  \hline
\end{tabular}
\ec
\end{table}
The associated 90\% and 95\% confidence intervals are presented in Table \ref{result-ci-insurance}.
\begin{table}[h]
\bc
\caption{90\% and 95\% confidence intervals for $\alpha$ and $\beta$ for Example 2  \label{result-ci-insurance}}
\begin{tabular}{|c|c|c|c|c|}  \hline  
Estimate & \multicolumn{2}{|c|}{$\alpha$} & \multicolumn{2}{|c|}{$\beta$}  \\  \cline{1-5} \hline
    &     90\%  &     95\%  &     90\%  &       95\%    \\
ML  & (0.4407,0.6783) & (0.4176,0.7014) & (0.9854,1.5278) &  (0.9326,1.5806)  \\
MM  & (0.4407,0.6783) & (0.4176,0.7014) & (0.9848,1.5271) &  (0.9321,1.5798)  \\
UML & (0.4221,0.6597) & (0.3989,0.6828) & (0.9827,1.5252) &  (0.9299,1.5780)  \\
UMM & (0.4220,0.6597) & (0.3989,0.6828) & (0.9822,1.5244) &  (0.9295,1.5771)  \\  \hline

\end{tabular}
\ec
\end{table}
Now to check whether the BS distribution fits the data or not, we have computed the Kolmogorov-Smirnov (KS) distance between the
fitted CDF based on ML, MM, UML and UMM estimates and the empirical CDF.  We have reported these KS distances and the associated
$p$-values in Table \ref{ks-insurance}.  It is clear from the results in Table \ref{ks-insurance} that the BS distribution 
fits the insurance data quite well.
\begin{table}[h]
\bc
\caption{The KS distances between the fitted CDF and the empirical CDF for Example 2  \label{ks-insurance}}
\begin{tabular}{|c|c|c|}  \hline  
Estimate & KS Distance & $p$-value   \\  \cline{1-3} \hline   
ML  & 0.1385 &  0.6130  \\
MM  & 0.1387 &  0.6106  \\
UML & 0.1457 &  0.5470   \\
UMM & 0.1455 &  0.5494   \\  \hline

\end{tabular}
\ec
\end{table}

%\newpage

\subsection {\sc Example 3: Ball Bearings Data}

The following data set is from McCool \cite{McCool:1974}, and it provides the fatigue life in hours of ten ball bearings
of a certain type:

152.7 172.0 172.5 173.3 193.0
204.7 216.5 234.9 262.6 422.6

Cohen et al. \cite{CWD:1984} first used this data set as an illustrative example for the fit of a three-parameter Weibull 
distribution.  Ng et al. \cite{NKB:2006} used the first 8 order statistics and fitted the BS  
distribution, based on the assumption that 
it is a Type-II right censored sample with $n$ = 10 and $r$ = 8.  The ML estimates of $\alpha$ and $\beta$ are 
found to be 0.1792 and 200.7262, respectively.  The biased-corrected estimate of $\alpha$ turns out to be 0.2108.  The 90\% and
95\% confidence intervals for $\alpha$ and $\beta$  based on ML and UML estimates are 
presented in Table \ref{result-ci-2}.
\begin{table}[h]
\bc
\caption{90\% and 95\% confidence intervals  for $\alpha$ and $\beta$ for Example 3  \label{result-ci-2}}
\begin{tabular}{|c|c|c|c|c|}    \hline
Estimate & \multicolumn{2}{|c|}{$\alpha$} & \multicolumn{2}{|c|}{$\beta$}  \\  \cline{1-5}  \hline
    &     90\%  &     95\%  &     90\%  &       95\%    \\
ML  & (0.1017,0.2566) & (0.0868,0.2715) & (183.1828,221.9857) &  (180.1662,226.5831)  \\
UML & (0.0925,0.3290) & (0.0698,0.3517) & (180.4109,226.1973) &  (176.9795,231.8331)  \\  \hline

\end{tabular}
\ec
\end{table}

\subsection {\sc Example 4: Progressively Censored Data}

Pradhan and Kundu \cite{PK:2013} considered the same data set as in Example 3, and generated three different Type-II
progressively censored samples as follows:
 
\noindent MCS-1: $n$ = 10, $m=6$, $R_1=4$, $R_2= \cdots = R_{6}=0$; \\
\noindent MCS-2: $n$ = 10, $m=6$, $R_1= \cdots = R_{5}=0$, $R_{6}=4$; \\
\noindent MCS-3: $n$ = 10, $m=7$, $R_1=2$, $R_2=1$, $R_3 \cdots = R_{7}=0$.

The ML estimates of $\alpha$ and $\beta$ were obtained using the EM algorithm.  The number of iterations needed for the 
convergence of the EM algorithm for the three schemes are 30, 44 and 24, respectively.  The ML estimates of $\alpha$ and 
$\beta$, along with their standard errors, and associated 95\% 
confidence intervals are presented in Table \ref{data-table-4}.  
\begin{table}
\begin{center}
\caption{Estimates of $\alpha$ and $\beta$, along with their standard errors, and 95\% confidence intervals for the data set of McCool 
\cite{McCool:1974}}
\label{data-table-4}
\vspace{4mm}
\begin{tabular}{|c| c| c |c |c |c |c| } \hline
Censoring  & \multicolumn{3}{|c}{$\alpha$} &\multicolumn{3}{|c|}{$\beta$}  \\ \cline{2-7}
schemes           & Estimate & s.e. & 95\% CI & Estimate & s.e. & 95\% CI \\ \hline
MCS-1    & 0.1639 &0.0367 &[0.0921, 0.2358]&194.0795&9.9935&[174.4922, 213.6669] \\ \hline
MCS-2    & 0.1484 &0.0332 &[0.0833, 0.2134]&195.4253&9.1186&[177.5528, 213.2978] \\ \hline
MCS-3    & 0.1570 &0.0351 &[0.0882, 0.2258]&195.8228&9.6617&[176.8859, 214.7597] \\ \hline            
\end{tabular}
\end{center}
\end{table}

\subsection {\sc Example 5: Bivariate Bone Mineral Data}  

We now provide the data analysis of a bivariate data set. The data in this case, obtained from 
Johnson and Wichern \cite{JW:1999}, represent the bone mineral
density (BMD) measured in gm/cm$^2$ for 24 individuals. The first figure represents
the BMD of the bone Dominant Radius before starting the study and
the second figure represents the BMD of the same bone after one
year:

\noindent (1.103  1.027), (0.842  0.857), (0.925
0.875), (0.857  0.873), (0.795  0.811), (0.787  0.640), (0.933
0.947), (0.799  0.886), (0.945  0.991), (0.921  0.977), (0.792
0.825), (0.815  0.851), (0.755  0.770), (0.880  0.912), (0.900
0.905), (0.764  0.756), (0.733  0.765), (0.932  0.932), (0.856
0.843), (0.890  0.879), (0.688  0.673), (0.940  0.949 ), (0.493
0.463), (0.835  0.776).

\noindent The sample means and sample variances of the two components are
(0.8408, 0.8410) and (0.0128, 0.0149), respectively.  The sample
correlation coefficient is 0.9222.   
Kundu et al. \cite{KBJ:2010} used the BVBS distribution to model these bivariate data.
From the observations, it is observed that $s_1$ = 0.8408, $s_2$ = 0.8410, $r_1$
= 0.8225, $r_2$ = 0.8179, and so the modified moment estimates are
$$
\widetilde{\alpha}_1 = 0.1491, \ \   \widetilde{\alpha}_2 = 0.1674, \ \
\widetilde{\beta}_1 = 0.8316, \ \ \widetilde{\beta}_2 = 0.8294,  \ \ \widetilde{\rho} = 0.9343.
$$
Using these as initial values, the ML estimates are determined as 
$$
\widehat{\alpha}_1 = 0.1491, \ \ \widehat{\alpha}_2 = 0.1674,  \ \  
\widehat{\beta}_1 = 0.8312, \ \ \widehat{\beta}_2 = 0.8292, \ \ \widetilde{\rho} = 0.9343.
$$
The 95\%
confidence intervals for $\alpha_1$, $\alpha_2$, $\beta_1$,
$\beta_2$ and $\rho$, based on the empirical Fisher information
matrix, become (0.1069, 0.1913), (0.1200, 0.2148), (0.7818, 0.8806),
(0.7739, 0.8845), (0.8885, 0.9801), respectively.

\subsection {\sc Example 6: Multivariate Bone Mineral Data} 

We now provide the analysis of a multivariate data, taken from 
Johnson and Wichern (\cite{JW:1999}, page 34),  representing the mineral contents of four major
bones of 25 new born babies.  Here, $T_1$, $T_2$, $T_3$ and $T_4$ represent dominant radius, radius, dominant 
ulna and ulna, respectively.  The data are not presented here, but the sample mean, variance and skewness of
the individual $T_i$'s and their reciprocals are all presented in Table \ref{data-basic}.

\begin{table}
{\small
\caption{The sample mean, variance and coefficient of skewness of $T_i$ and $T_i^{-1}$, for $i = 1, \cdots, 4$, 
for Example 6. \label{data-basic}}
\bc
\begin{tabular}{|c|c|c|c|c|c|c|c|c|}  \cline{1-9}
Variables $\rightarrow$ & $T_1$ & $T_1^{-1}$ & $T_2$ & $T_2^{-1}$ & $T_3$ & $T_3^{-1}$ & $T_4$ & $T_4^{-1}$   \\
Statistics $\downarrow$ &       &            &       &            &       &            &       &          \\  \hline
Mean & 0.844 &  1.211 & 0.818  & 1.245 & 0.704 & 1.452 & 0.694 & 1.474 \\
Variance & 0.012 & 0.041 & 0.011 & 0.034 & 0.011 & 0.050 & 0.010 & 0.052 \\
Skewness & -0.793 & 2.468 & -0.543 & 1.679 & -0.022 & 0.381 & -0.133 & 0.755  \\
\hline
\end{tabular}
\ec}
\end{table}
Kundu et al. \cite{KBJ:2013} fitted a 4-variate BS distribution to this data set.  First, the 
MM estimates are obtained from the marginals and they are provided in 
Table \ref{ks-data}.   
\begin{table}[h]
\caption{The MM estimates of $\alpha_i$ and $\beta_i$, the KS distance between the empirical
distribution function and the fitted distribution function, and the corresponding $p$ values.
\label{ks-data}}
\bc
\begin{tabular}{|c|c|c|c|c|}  \cline{1-5}
      & $\alpha$ & $\beta$ & KS & $p$    \\
      &          &  distance &    &           \\  \hline
$T_1$ & 0.1473 & 0.8347 & 0.161 & 0.537   \\
$T_2$ & 0.1372 & 0.8107 & 0.145 & 0.671   \\
$T_3$ & 0.1525 & 0.6963 & 0.109 & 0.929   \\
$T_4$ & 0.1503 & 0.6861 & 0.094 & 0.979  \\
\hline
\end{tabular}
\ec
\end{table}
Using MM estimates as initial guess, the ML estimates of $\beta_1$, $\beta_2$, $\beta_3$ and $\beta_4$ 
are obtained as 0.8547, 0.7907, 0.7363  and 0.8161, respectively, and the corresponding maximized  
log-likelihood value (without the additive constant) is 402.61.  Finally, the corresponding ML estimates of $\alpha_1$, $\alpha_2$, $\alpha_3$ and $\alpha_4$
are obtained as 0.1491, 0.1393, 0.1625 and 0.2304, respectively. The 95\% non-parametric 
bootstrap confidence intervals of
$\beta_1$, $\beta_2$, $\beta_3$ and $\beta_4$ are then obtained as (0.8069, 0.9025), 
(0.7475, 0.8339), (0.6950, 0.7776) and
(0.7760, 0.8562), respectively.  Similarly,  the 95\% non-parametric bootstrap confidence intervals of
$\alpha_1$, $\alpha_2$, $\alpha_3$ and $\alpha_4$ are obtained as (0.1085, 0.1897), (0.1015, 0.1771), (0.1204, 0.2046) and
(0.1890, 0.2718), respectively.  The ML estimate of $\ve{\Gamma}$ is obtained as 
\be
\widehat{\ve \Gamma} = \left [ \matrix{1.000 & 0.767 & 0.715 & 0.515 \cr
0.767 & 1.000 & 0.612 & 0.381 \cr
0.715 & 0.612 & 1.000 & 0.693 \cr
0.515 & 0.381 & 0.693 & 1.000 \cr } \right ].
\ee

\subsection {\sc Example 7: Multivariate Bone Mineral Data (Revisited)} 

Kundu et al. \cite{KBJ:2013} analyzed the data set in Example 6 by using 
generalized multivariate BS distribution with multivariate $t$-kernel.  The degrees of 
freedom $\nu$ was varied from 1 to 20 for a 
profile analysis with respect to $\nu$.  The
ML estimates of all the unknown parameters and the corresponding maximized log-likelihood values, for different 
choices of $\nu$, are obtained and are presented in Table \ref{max-ll}.  It is observed that the maximized 
log-likelihood values first increase and then decrease.  
\begin{table}[h]
\caption{The maximized log-likelihood value vs. degrees of freedom $\nu$ = 1(1)20 for Example 7.
\label{max-ll}}
\bc
\begin{tabular}{|c|c||c|c||c|c||c|c||}  \cline{1-8}
 $\nu$ & Maximized  & $\nu$ & Maximized & $\nu$ & Maximized & $\nu$ & Maximized \\
       & log-likelihood  &  & log-likelihood &  & log-likelihood & & log-likelihood \\  \hline
 1 &  428.315491 & 2 &  433.549164 & 3 &  436.112915 & 4 &  436.911774 \\
 5 &  437.847198 & 6 &  438.225861 & 7 &  439.095734 & 8 &  439.184052 \\
 9 &  443.246613 & 10 & 442.009432 & 11 & 441.526855 & 12 &  441.068146 \\
 13 &  439.946442 & 14 &  438.317993 & 15 &  437.837219 & 16 &  437.024994 \\
 17 &  436.234161 & 18 &  435.532867 & 19 &  434.860138 & 20 &  434.289551 \\  \hline
\end{tabular}
\ec
\end{table}
The maximum occurs at $\nu$ = 9, with the associated log-likelihood value (without
the additive constant) being 443.2466.  It is important to mention here that the selection of the best 
$t$-kernel function through the maximized log-likelihood value is equivalent to selecting by the 
Akaike Information Criterion since the number of model parameters remains the same when $\nu$ varies.
Furthermore, this maximized log-likelihood value of 443.246 for the multivariate $t$ kernel with 
$\nu$ = 9 degrees of freedom is significantly larger than the corresponding value of 402.61 for the 
multivariate normal kernel, which does provide a strong evidence to the fact that the multivariate $t$ kernel
provides a much better fit for these data.  

Now, we provide detailed results for the case $\nu$ = 9.  In this
case, the ML estimates of $\beta_1$, $\beta_2$, $\beta_3$ and $\beta_4$ are found to be 
35.1756, 30.1062, 31.9564 and 40.1928, respectively.  The corresponding 95\% confidence intervals, 
 obtained by the use of 
non-parametric bootstrap method, are (25.216, 45.134), (20.155, 40.057), (22.740, 41.172), and (28.822, 51.562), 
respectively. The ML estimates of $\alpha_1$, $\alpha_2$, $\alpha_3$ and $\alpha_4$ are 0.7746, 1.0585, 0.9457 and 
0.9193, and the associated 95\% non-parametric bootstrap confidence intervals  
are (0.6940, 0.8551), (0.9538, 1.1632), (0.8488, 1.0425), and 
(0.8418, 0.9968), respectively.  Finally, the ML estimate of $\ve{\Gamma}$ is obtained as 
\be
\widehat{\ve \Gamma} = \left [ \matrix{1.000 & 0.796 & 0.696 & 0.583 \cr
0.796 & 1.000 & 0.735 & 0.813 \cr
0.696 & 0.735 & 1.000 & 0.693 \cr
0.583 & 0.813 & 0.693 & 1.000 \cr } \right ].
\ee

\section{\sc Concluding Remarks and Further Reading}

In this paper, we have considered the two-parameter BS distribution, which was introduced 
almost fifty years ago.  The BS model has received considerable attention since then 
for various reasons.  The two-parameter BS distribution has a shape and a scale parameter.  
Due to the presence of the shape parameter, the PDF of a BS distribution can take on different shapes.
It has non-monotone HF and it has a nice physical interpretation.  Several generalizations 
of the BS distribution have been proposed in the literature and they have found numerous applications in 
many different fields.  Recently, bivariate, multivariate and matrix-variate BS distributions have also been introduced in 
the literature.  We have provided a detailed review of various models and methods available to date with regard to 
these models, and have also mentioned several open problems for future work.

Extensive work has been done on different issues with regard to BS  distribution during the 
last 15 years.  Due to limited space, we are unable to provide detailed description of all the work.  Efficient
$R$ packages have been developed by Leiva et al. \cite{LHR:2006} and Barros et al. \cite{BPL:2009}.
Interested readers are referred to the 
following articles for further reading.  For different applications of the univariate BS, multivariate BS and related 
distributions, one may look at Ahmed et al. \cite{ACFLS:2010},  Aslam et al. \cite{AJA:2011}, 
Aslam and Kantam \cite{AK:2008}, Baklizi and El Masri \cite{BE:2004}, 
Balakrishnan et al. \cite{BLL:2007}, Balamurali et al. \cite{BJU:2018},
Castillo et al. \cite{CGB:2011}, Desousa et al. \cite{DSLS:2017}, Garcia-Papani et al. \cite{GULR:2017}, 
Gomes et al. \cite{GFL:2012},  Ismail \cite{Ismail:2005}, 
Kotz \cite{KLS:2010}, 
Leiva and Marchant \cite{LM:2017}, Leiva and Saulo \cite{LS:2017}, 
Leiva et al. \cite{LBPS:2008b, LSKA:2010, LVBS:2010, LAAM:2011, LSCC:2011, LPMB:2012, LRGS:2014, LSCB:2014, LSLM:2014, 
LMSAR:2014, LMRS:2015, LTGSOM:2015, LSCB:2016, LFGL:2016}, Lio and Park \cite{LP:2008}, Lio et al. \cite{LTW:2010},
Marchant et al. \cite{MBLS:2013, MLCS:2013, MLCL:2017}, Paula et al. \cite{PLBL:2012}, Podlaski \cite{Podlaski:2008},
Rojas et al. \cite{RLWM:2015}, 
Saulo et al. \cite{SLZM:2013, SLR:2015, SLLA:2017}, Upadhyay et al. \cite{UMG:2009}, Vilca et al. \cite{VSLC:2010},
Villegas et al. \cite{VPL:2011}, Wanke et al. \cite{WELR:2016},
Wanke and Leiva \cite{WL:2015}, Wu and Tsai \cite{WT:2005},
Zhang et al. \cite{ZLD:2016, ZSBF:2017}, and the references cited
therein.  In a recent paper, Mohammadi et al. \cite{MOM:2017} have discussed the modeling of wind speed and wind power
distributions by BS distribution.  Also, in the reliability analysis of nano-materials, Leiva et al. \cite{LRSV:2017}
have recently made use of the BS lifetime distribution.  Garcia-Papani et al. \cite{GULA:2017, GULA:2017a} have 
considered a spatial modelling 
of the BS distribution and applied it to agricultural engineering data.

For different inference related issues, one may refer to Ahmed et al. \cite{ABLV:2008}, Arellano-Valle et al. \cite{AGQ:2005},
Athayde \cite{AABL:2017}, Audrey et al. \cite{AFC:2008}, Azevedo et al. \cite{ALAB:2012}, 
Balakrishnan et al. \cite{BLSV:2009, BSL:2017, BSBZ:2017},  Balakrishnan and Zhu \cite{BZ:2013, BZ:2015a, BZ:2015b}, Barros et al. \cite{BLOT:2014}, 
Chang and Tang \cite{CT:1994c}, 
Cordeiro et al. \cite{CLCP:2015}, Cysneiros et al. \cite{CCA:2008}, Desmond and Yang \cite{DY:1996}, Farias and Lemonte \cite{FL:2011}, Guo et al. \cite{GWLL:2017}, Jeng \cite{Jeng:2003}, 
 Lachos et al. \cite{LDCL:2017}, Lemonte \cite{Lemonte:2011, Lemonte:2012, Lemonte:2013a, Lemonte:2013b, Lemonte:2013c, Lemonte:2016},
Lemonte and Cordeiro \cite{LC:2009},  
Lemonte et al. \cite{LFSC:2010, LCM:2012, LMM:2015}, Lemonte and Ferrari \cite{LF:2011a, LF:2011b, LF:2011c,LF:2011d},  
Lemonte and Patriota \cite{LP:2011}, Li et al. \cite{LCX:2012}, Li and Xu \cite{LX:2014}, Lillo et al. \cite{LLNA:2017},
Lu and Chang \cite{LC:1997}, Meintanis \cite{Meintanis:2010}, 
Moala et al. \cite{MAG:2015}, Niu et al. \cite{NGXZ:2013}, Padgett and Tomlinson \cite{PT:2003}, P{\'e}rez and Correa \cite{PC:2008}, Qu and Xie \cite{QX:2011}, 
Riquelme \cite{RLGS:2011}, 
S{\'a}nchez et al. \cite{SLCC:2015}, 
Santana et al. \cite{SVL:2011}, Santos-Neto et al. \cite{SCLA:2012, SCLB:2014}, 
Saulo et al. \cite{SBZGL:2017}, Sha and Ng \cite{SN:2017}, Teimouri et al. \cite{THN:2013}, Tsionas \cite{Tsionas:2001}, 
Upadhyay and Mukherjee \cite{UM:2010}, Vanegas and Paula \cite{VP:2017}, 
Vanegas et al. \cite{VRC:2012}, Vilca et al. \cite{VSLB:2011, VAB:2016, VBZ:2014a, VBZ:2014b, VBZ:2015, VRB:2016}, Wang \cite{Wang:2012}, Wang and Fei \cite{WF:2004, WF:2006}, Wang et al. \cite{WPS:2015, WZSP:2013},
Xiao et al. \cite{XLBL:2010}, Xie and Wei \cite{XW:2007}, Xu and Tang \cite{XT:2011}, 
Xu et al. \cite{XYL:2016}, Zhu and Balakrishnan \cite{ZB:2015},
 and the references cited therein.

Several different models relating to univariate and multivariate BS distributions can be found in Athayde et al. \cite{AALS:2012}, Balakrishnan and Saulo \cite{BS:2016}, 
Barros et al. \cite{BPL:2008}, Bhatti \cite{Bhatti:2010}, Cancho et al. 
\cite{CEP:2010}, Cordeiro et al.
 \cite{CCOB:2016},  
Cordeiro and Lemonte \cite{CL:2011, CL:2014}, Desmond et al. \cite{DCSL:2012}, 
D{\'i}az-Garc{\'i}a and Dom{\'i}nguez-Molina \cite{DD:2006a, DD:2006b}, D{\'i}az-Garc{\'i}a and Leiva \cite{DL:2002}, Ferreira et al.
 \cite{FGL:2012}, Fierro et al. \cite{FLRS:2013}, Fonseca and Cribari-Neto \cite{FC:2018},
Gen{\c c} \cite{Genc:2013}, 
Gomes et al. \cite{GOB:2009}, Guiraud et al. \cite{GLF:2009}, Hashimoto \cite{HOCC:2014},
Jamalizadeh and Kundu \cite{JK:2015}, Khosravi et al.
 \cite{KKJ:2015, KLJE:2016}, 
Kundu \cite{Kundu:2015a, Kundu:2015b, Kundu:2016}, Leiva et al. \cite{LSSG:2008}, 
Mart{\'i}nez-Fl{\'o}rez et al. \cite{MBG:2014, MBG:2017}, Marchant et al. \cite{MLC:2016, MLCV:2016}, 
Olmos et al. \cite{OMB:2017}, 
Onar and Padgett \cite{OnP:2000}, Ortega et al. \cite{OCL:2012}, Owen \cite{Owen:2006}, Owen and Padgett \cite{OP:1998, OP:1999, 
OP:2000}, Park and Padgett \cite{PP:2006}, Patriota \cite{Patriota:2012},
Pescim et al. \cite{PCNDO:2014},  Pourmousa et al. \cite{PJR:2015}, Raaijmakers \cite{Raai:1980, Raai:1981},
Romeiro et al. \cite{RVB:2016}, 
 Reina et al. \cite{RVZB:2016}, Reyes et al. \cite{RVGG:2017},
Sanhueza et al. \cite{SLF:2009},
Vilca and Leiva \cite{VL:2006}, Volodin and Dzhungurova \cite{VD:2000}
and Ziane et al. \cite{ZZA:2018}.  Recently, some survival analytic methods have been developed based on BS and related
models.  For example, one may refer to Le{\~a}o \cite{Leao:2017}, Le{\~a}o et al. \cite{LLST:2017a, LLST:2017b, LLST:2017c} 
and Balakrishnan
and Liu \cite{BL:2017}.  This is one direction in which there is a lot more research work that could be carried out.

\section*{\sc Acknowledgements:} The authors would like to thank the associate editor and three anonymous 
reviewers for their constructive
comments, which have helped to improve the manuscript significantly.

%\newpage

\end{document}